\documentclass{article}

\newcommand{\btheta}{ \mbox{\boldmath $\theta$}}

\newcommand{\bmu}{ \mbox{\boldmath $\mu$}}

\newcommand{\bzeta}{ \mbox{\boldmath $\zeta$}}

\newcommand{\bepsilon}{ \mbox{\boldmath $\epsilon$}}

\newcommand{\bkappa}{ \mbox{\boldmath $\kappa$}}
\newcommand{\btau}{ \mbox{\boldmath $\tau$}}

\newcommand{\bA}{ \mbox{\bf A}}

\newcommand{\bE}{ \mbox{\bf E}}

\newcommand{\bx}{ \mbox{\bf x}}
\newcommand{\bX}{ \mbox{\bf X}}

\newcommand{\bu}{ \mbox{\bf u}}
\newcommand{\bU}{ \mbox{\bf U}}

\newcommand{\bI}{ \mbox{\bf I}}

\newcommand{\bzero}{ \mbox{\bf 0}}
\newcommand{\bone}{ \mbox{\bf 1}}

\newcommand{\beq}{ \begin{equation}}
\newcommand{\eeq}{ \end{equation}}
\newcommand{\beqn}{ \begin{eqnarray}}
\newcommand{\eeqn}{ \end{eqnarray}}

\usepackage{amsmath, amsthm, amssymb, amsfonts}

\usepackage{arxiv}

\usepackage[utf8]{inputenc} 
\usepackage[T1]{fontenc}    
\usepackage{xcolor}		
\usepackage{hyperref}       
\usepackage{url}            
\usepackage{booktabs}       
\usepackage{nicefrac}       
\usepackage{microtype}      
\usepackage{lipsum}		
\usepackage{graphicx}
\usepackage{natbib}
\usepackage{doi}

\usepackage{float}			
\usepackage{arydshln}
\usepackage{subfig}

\usepackage{xr}

\makeatletter
\newcommand*{\addFileDependency}[1]{
  \typeout{(#1)}
  \@addtofilelist{#1}
  \IfFileExists{#1}{}{\typeout{No file #1.}}
}
\makeatother

\newcommand*{\myexternaldocument}[1]{%
    \externaldocument{#1}%
    \addFileDependency{#1.tex}%
    \addFileDependency{#1.aux}%
}

\myexternaldocument{supplementary_materials_aux}

\title{BayesClint: Bayesian multi-scale clustering and multi-sample integration with feature selection for spatial transcriptomics data}

\date{} 					

\author{\href{https://orcid.org/0000-0003-1396-256X}{\includegraphics[scale=0.06]{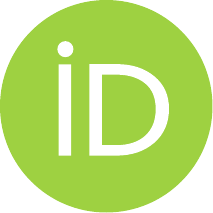}\hspace{1mm}Alvin Sheng} \\
	Division of Biostatistics and Health Data Science\\
	University of Minnesota\\
	Minneapolis, MN 55414 \\
	\texttt{sheng123@umn.edu} \\
\And
\href{https://orcid.org/0000-0001-9593-4778}{\includegraphics[scale=0.06]{orcid.pdf}\hspace{1mm}Sandra E.~Safo} \\
	Division of Biostatistics and Health Data Science\\
	University of Minnesota\\
	Minneapolis, MN 55414 \\
	\texttt{ssafo@umn.edu} \\
\And
\href{https://orcid.org/0000-0002-3594-9182}{\includegraphics[scale=0.06]{orcid.pdf}\hspace{1mm}Thierry Chekouo}\thanks{Corresponding author.} \\
	Division of Biostatistics and Health Data Science\\
	University of Minnesota\\
	Minneapolis, MN 55414 \\
	\texttt{tchekouo@umn.edu} \\
}

\renewcommand{\shorttitle}{\textit{arXiv} Template}

\hypersetup{
pdftitle={A template for the arxiv style},
pdfsubject={q-bio.NC, q-bio.QM},
pdfauthor={David S.~Hippocampus, Elias D.~Striatum},
pdfkeywords={First keyword, Second keyword, More},
}

\begin{document}
\maketitle

\begin{abstract}
	Recent advances in spatial transcriptomics have enabled researchers to profile gene expression at the single-cell spatial resolution, often for multiple tissue samples in a single study. This high-dimensional molecular profile for each cell can be used to sort cells into cell types with distinct functions, or segment the tissue into biologically relevant spatial domains. Although many non-spatial and spatial clustering methods have been developed to cluster these cells into cell types or spatial domains, most have two main limitations: first, they perform dimension reduction and clustering separately; second, they cluster cells at a single scale, rather than treating cell type and spatial domain clustering as distinct tasks at two different scales. To overcome these limitations, we propose BayesClint, a Bayesian method that simultaneously performs factor analysis and spatial clustering on multiple samples, where the clustering is done jointly at the single-cell and tissue regional scale. To increase interpretability, we employ a feature selection mechanism within the estimation of the sparse factor loadings matrix, which detects active genes and differentially expressed genes that discriminate between cell type clusters. We illustrate the advantages of the method over alternative state-of-the-art approaches through simulation studies and two real data applications.
\end{abstract}


\section{Introduction}

Spatially resolved transcriptomics (SRT) refers to technologies that enable gene expression profiling in tissue, while retaining the spatial information of the measurements. These technologies are categorized into next-generation sequencing (NGS)- and imaging-based methods \citep{heumos_best_2023}. NGS-based methods measure gene expression on an array of capture sites known as spots, each of which may encompass multiple cells. For example, for spatial transcriptomics (ST) \citep{stahl_visualization_2016} technology, each spot assays 10-40 cells, while for 10x Visium \citep{noauthor_mastering_nodate}, each spot assays 1-10 cells. Other NGS-based technologies, Slide-seq \citep{rodriques_slide-seq_2019} and the subsequent Slide-seq V2 \citep{stickels_highly_2021}, approach cellular resolution with each spot containing 1-3 cells. 
On the other hand, imaging-based methods may involve single-molecule fluorescence in situ hybridization, e.g., MERFISH \citep{chen_spatially_2015} and 10x Xenium \citep{janesick_high_2023}, or in situ sequencing, e.g., spatially resolved transcript amplicon readout mapping (STARmap) \citep{wang_three-dimensional_2018}. In either case, imaging-based methods can directly measure gene expression at single-cell resolution.

Quantifying gene expression at cellular or near-cellular resolution enables scientists to analyze tissue samples at two anatomical scales: the single-cell scale and the spatial domain scale. In the rest of this paper, we denote the joint analysis of two scales as ``multi-scale'' analysis. At the single-cell scale, transcriptomic heterogeneity can be used to cluster cells into distinct cell types \citep{wang_three-dimensional_2018,moffitt_molecular_2018}.
At the spatial domain scale, transcriptomic heterogeneity can be used to determine the spatial organization of complex tissues. Detecting the spatial domains that make up the tissue can improve our understanding of the spatial structure of healthy tissue, in contrast to diseased tissue
\citep{ma_accurate_2024}.
In SRT research, it is common to collect several adjacent tissue samples from a single individual or to obtain tissue samples from the same anatomical site across multiple individuals. Because samples from comparable tissues usually have the same cell type compositions and spatial patterns, combining data from different tissue specimens can enhance the identification of these domains and cell types \citep{li_bass_2022}.

In this paper, we focus on two SRT datasets of the mouse brain from \citep{wang_three-dimensional_2018}. A current challenge in neuroscience is detecting and classifying cell types and spatial domains in the neocortex of the adult mouse brain \citep{wang_three-dimensional_2018}. The authors use STARmap to profile the gene expression and location of cells in two parts of the neocortex: the mouse primary visual neocortex (MVC) and medial prefrontal cortex (mPFC). The MVC is an important cortical region of study, because it serves as a useful model to explore how sensory inputs are transformed into goal-directed perceptions and actions \citep{glickfeld_mouse_2014}. On the other hand, the mPFC is also a crucial cortical region due to its role in high-level cognitive functions like decision-making, memory, attention, and emotion \citep{carlen_what_2017}. Dysfunction in this cortical region has been linked to various neurological and psychiatric disorders like depression and Alzheimer's Disease \citep{xu_medial_2019}. In these and other SRT datasets, the usual scientific objectives are to cluster cells into different cell types or spatial domains that may play important roles in tissue function and disease. In addition, researchers may be interested in determining the main genes driving the variation in the data and the differentiation into cell types. 

For the detection of clusters in SRT data, proposed methods include the hidden Markov random field (HMRF) \citep{zhu_identification_2018}, BayesSpace \citep{zhao_spatial_2021}, SpaGCN \citep{hu_spagcn_2021}, PRECAST \citep{liu_probabilistic_2023}, BayesCafe \citep{li_interpretable_2024}, and BASS \citep{li_bass_2022}. Because of the high number of features (genes) in SRT data, all methods except PRECAST and BayesCafe are applied to low-dimensional representations of the gene expression data, e.g., those returned by performing principal components analysis (PCA) on the gene expression data. These clustering methods allocate spots or cells into spatial domains while accounting for the spatial correlation among nearby spots or cells. Most methods implement this spatial dependency structure via a Potts model, while SpaGCN implements it via a graph convolutional network. PRECAST and BASS are both able to integrate information across multiple tissue samples, while the other methods are designed to analyze just one tissue sample. BASS is the only one of these methods that can jointly perform cell type and spatial domain clustering, and BayesCafe is the only one that can jointly perform spatial domain clustering and differentiating gene selection.

There are two main drawbacks of existing methods for clustering SRT data. The first drawback is that most methods treat dimension reduction (e.g., PCA) as a separate analysis step from clustering. Performing dimension reduction and clustering sequentially, rather than jointly, is not ideal for two main reasons. First, dimension reduction and clustering would each optimize different objective criteria that may not be consistent with each other when aiming to achieve optimal clustering performance \citep{markos_beyond_2019}. Second, the dimension reduction step does not consider the uncertainty in obtaining the low-dimensional representations. Consequently, the low-dimensional representations would be treated as error-free in the clustering analysis step, which would make the statistical inference of clustering parameters less accurate \citep{liu_joint_2022}.

The other drawback is that all of the aforementioned methods, except BASS, collapse the tasks of clustering cell types and clustering spatial domains into one clustering task, instead of conducting a multi-scale analysis. Thus, these clustering methods determine the spatial domains directly from gene expression heterogeneity rather than from the unique cell type compositions underlying each spatial domain. This approach makes it difficult to characterize the cellular landscape of the tissue and reduces the accuracy and interpretability of the detected spatial domains \citep{ma_accurate_2024}.

In this paper, we propose the novel algorithm \underline{Bayes}ian \underline{cl}ustering and \underline{int}egration (BayesClint), which builds upon the cutting-edge methods and overcomes their limitations. BayesClint overcomes the first main drawback by jointly performing dimension reduction and clustering via a Bayesian factor analysis model. BayesClint overcomes the second main drawback by performing multi-scale analysis that jointly clusters cells into cell types and spatial domains. In addition, BayesClint is able to integrate multiple tissue samples like the cutting-edge methods PRECAST and BASS, and is unique in that it performs nested feature selection of active genes and differentiating genes. That is, BayesClint selects genes that contribute to the covariance of the gene expression matrix (active genes) and genes that differentiate between the cell types (differentiating genes). BayesClint is able to jointly perform all of these tasks, i.e., dimension reduction, multi-scale clustering, multi-sample integration, and nested feature selection, through a Bayesian hierarchical model. Web Table \ref{tab:bayesclint_competing_comparison} compares the overall features of BayesClint and the cutting-edge SRT clustering algorithms BayesCafe, PRECAST, and BASS.

When we applied BayesClint separately to the two SRT datasets of the mouse brain from \cite{wang_three-dimensional_2018}, we achieved a cell type and spatial domain clustering that more closely aligned with the ground truth than the results of most other current approaches. In addition, we identified specific genes critical for distinguishing cell type clusters, offering a valuable starting point for subsequent studies.

The rest of the article is organized as follows. In Section \ref{sec:model_framework}, we present the statistical framework of the model. In Section \ref{sec:posterior_inference}, we describe the computational details, such as the Markov Chain Monte Carlo algorithm and posterior inference. Then, we evaluate and compare our methodology with existing methods on simulated datasets in Section \ref{sec:simulation_study}, and two real datasets in Section \ref{sec:real_data_analysis}. Finally, we conclude our paper in Section \ref{sec:discussion}.

\section{Model Framework}
\label{sec:model_framework}

\begin{figure}

\begin{minipage}{0.9\textwidth}
    \centering

    \subfloat[Tissue Sample 1\label{fig:sim_study_repsA}]{\includegraphics[width=0.50\textwidth]{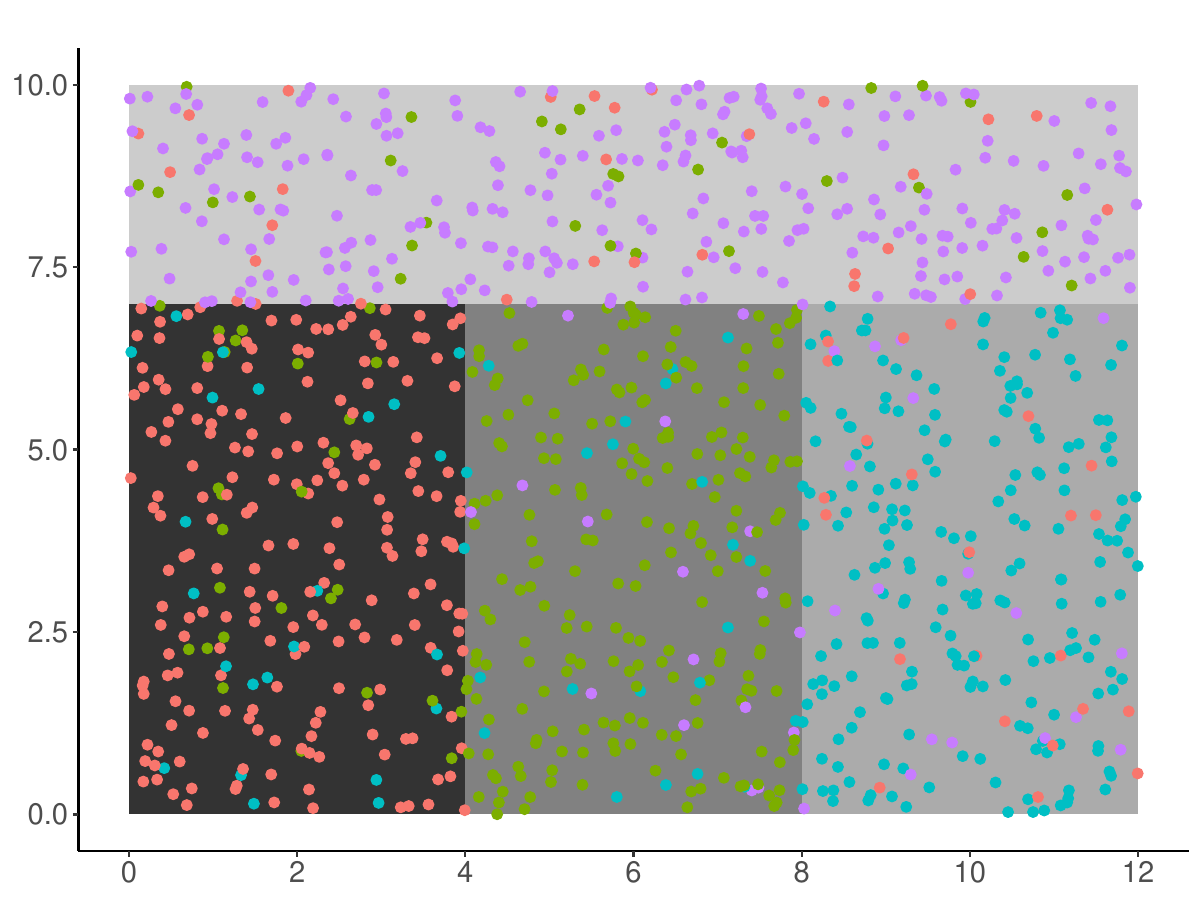}} 
    \subfloat[Tissue Sample 2\label{fig:sim_study_repsB}]{\includegraphics[width=0.50\textwidth]{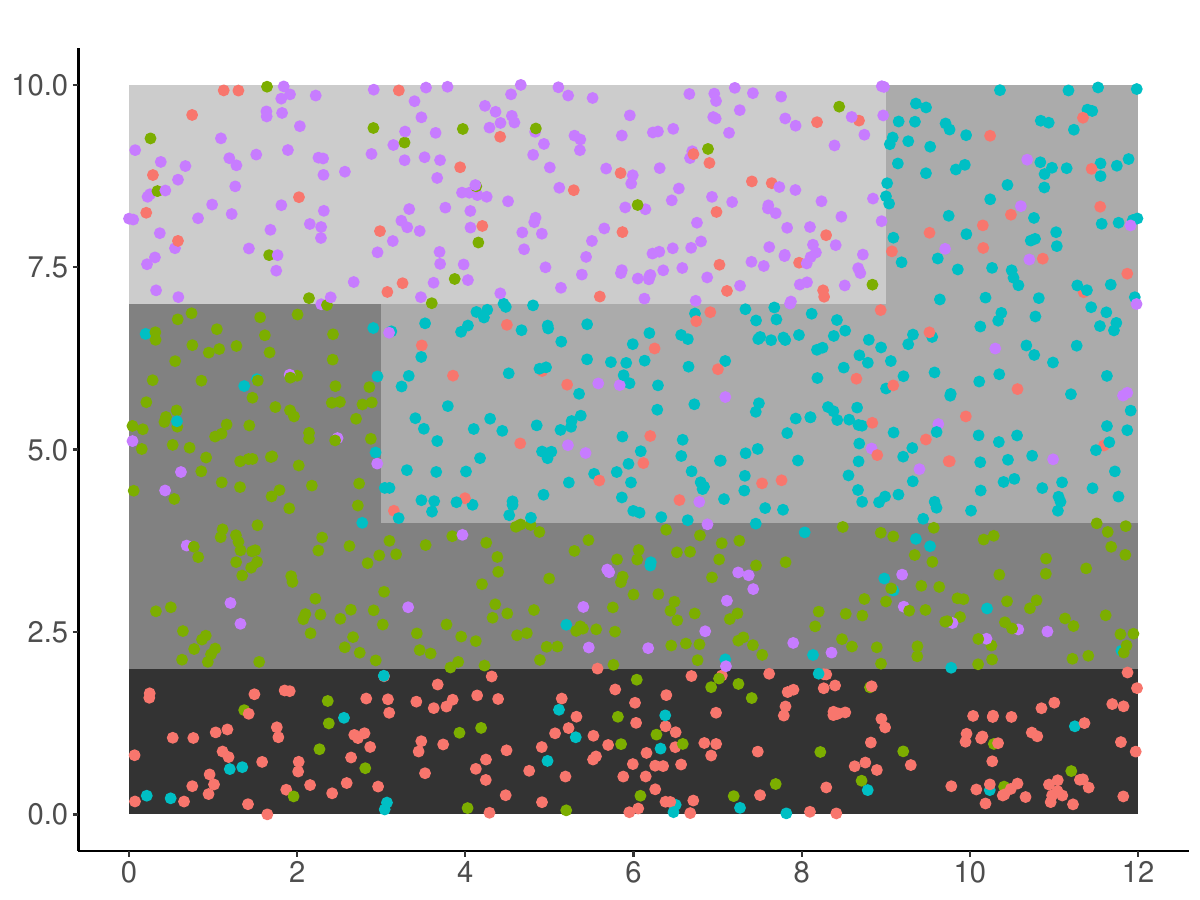}} 

    \subfloat[Tissue Sample 3\label{fig:sim_study_repsC}]{\includegraphics[width=0.50\textwidth]{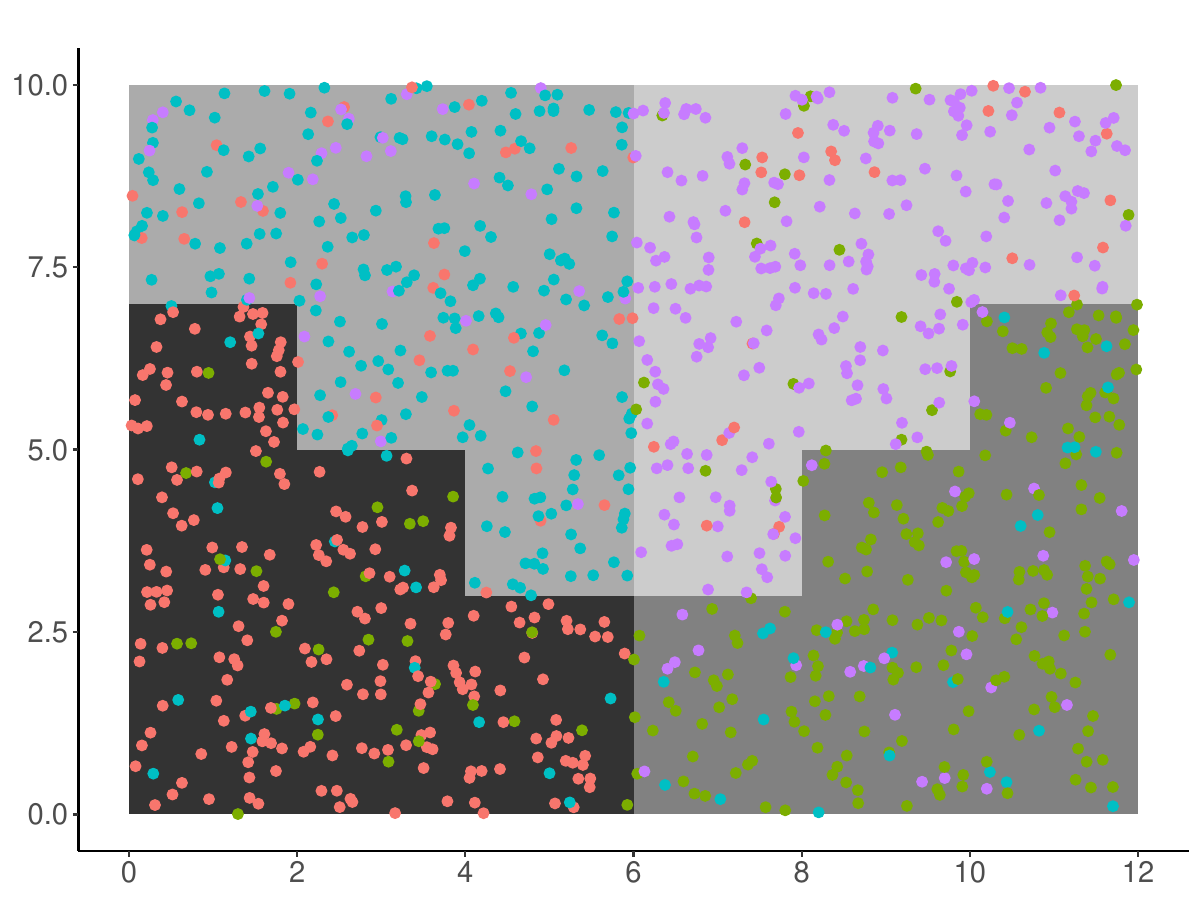}} 
    \end{minipage}
\begin{minipage}{0.1\textwidth}
\includegraphics[width=\textwidth]{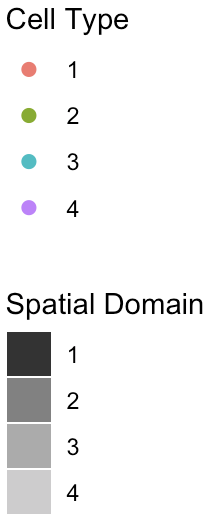}
\end{minipage}
    \caption{Cell types (indicated by colored points) and spatial domains (indicated in grayscale) that will be used in the simulation study. The cell types in this figure are assigned according to the ``regular'' cell type composition, where for each spatial domain, 80\% of cells are assigned to a given type, 10\% are assigned to another type, and 10\% are assigned to a third type.}
    \label{fig:sim_study_reps}
\end{figure}

Our statistical objective is to integrate multiple tissue samples to jointly perform dimension reduction, multi-scale clustering and nested feature selection. Figure \ref{fig:sim_study_reps} shows an example of what an SRT dataset may look like, where each point represents a cell for which we have gene expression counts (i.e., counts of RNA transcripts for a set of genes) and location information. In Section \ref{sec:factor_model}, we introduce a factor analysis model that relates the different tissue samples. In Section \ref{sec:multi_scale_clustering_layer}, we propose a hierarchical clustering of cells based on their latent factors, first into cell types and then into spatial domains. We then adopt a Bayesian feature selection technique in Section \ref{sec:sparse_ft_select} to select important genes. Web Figure \ref{fig:BayesClint_DAG} summarizes the proposed Bayesian hierarchical model via a directed acyclic graph, and Web Table \ref{tab:bayesclint_competing_comparison} itemizes the advantages of the BayesClint model, in comparison with three other cutting-edge algorithms.

\subsection{Factor analysis model for integrating multiple tissue samples}
\label{sec:factor_model}

Let $\bx_{mi}, m = 1, \ldots, N,$ be the normalized, centered, and scaled $P$-dimensional expression vector for cell $i$ in sample $m$ at location $s_{mi}$ for the $N$ tissue samples, after the pre-processing of the original RNA transcript counts. Denote $n_m$ as the number of cells in sample $m$. The vector $\bx_{mi}$ follows the following factor model:
    \begin{align} \label{eq:factor_model}
        \bx_{mi} = \bA^\intercal \bu_{mi} + \bepsilon_{mi},
    \end{align}
    
\noindent where $\bu_{mi} \in \mathbb{R}^r$ is a vector of latent representations ($r \ll P$) that captures the biological effects across the $N$ tissue samples, 
$\mathbf{A} \in \mathbb{R}^{r \times P}$ is the factor loadings matrix, and $\boldsymbol{\epsilon}_{mi}$ is the idiosyncratic error term.

To facilitate joint estimation across all cells within a sample, we express \eqref{eq:factor_model} in matrix form. Let $\mathbf{X}_m = (\mathbf{x}_{m1}, \dots, \mathbf{x}_{mn_m})^\top$ be the $n_m \times P$ observed expression matrix for sample $m$, and let $\mathbf{U}_m = (\mathbf{u}_{m1}, \dots, \mathbf{u}_{mn_m})^\top$ be the $n_m \times r$ matrix of latent factors. Defining the error matrix $\mathbf{E}_m = (\boldsymbol{\epsilon}_{m1}, \dots, \boldsymbol{\epsilon}_{mn_m})^\top$, the model for the $m$-th sample is:
\begin{equation} \label{eq:matrix_form}
    \mathbf{X}_m = \mathbf{U}_m \mathbf{A} + \mathbf{E}_m.
\end{equation}

\noindent We assume a common factor loadings matrix $\mathbf{A}$ for all $N$ samples; this global structure facilitates the integration of tissue-specific data by capturing the fundamental gene-to-factor dependencies shared across the replicates.

The residual $\bepsilon_{mi}$ follows the distribution $\bepsilon_{mi} \sim \mathcal{N}(\bzero, \textrm{diag}(\upsilon^2_{m1}, \cdots, \upsilon^2_{mP}))$, and the residual matrix $\bE_{m}$ follows the distribution $\textrm{vec}(\bE_{m}^\intercal) \sim \mathcal{N}(\bzero, \Upsilon_m),$ where $\Upsilon_m = I_{n_m} \otimes \textrm{diag}(\upsilon^2_{m1}, \cdots, \upsilon^2_{mP})$ is a diagonal matrix of the residual variances for sample $m$. The prior distribution of the residual variance $\upsilon^{2}_{mj}$ is $\textrm{Inverse-Gamma}(0.01, 0.01)$. The hyperparameters of the \textrm{Inverse-Gamma} distribution were selected to make the prior weakly informative, allowing the data to drive the Bayesian estimation of $\upsilon^{2}_{mj}$ in the absence of prior information.

\subsection{Multi-scale clustering layer}
\label{sec:multi_scale_clustering_layer}

The latent factors $\bu_{mi}$, which provide a low-dimensional representation of the gene expression  for cell $i$ in sample $m$, are modeled via the multi-scale clustering layer, which jointly clusters cells into cell types and spatial domains:
\begin{align}
    &\bu_{mi} \mid \zeta_{mi} = c \sim \mathcal{N}(\bmu_{c}, \Sigma) \label{eq:factor_u_mi} \\
    &\zeta_{mi} \mid \kappa_{mi} = k \sim \text{Categorical}(\btheta_{k}) \\
    &\bkappa_{m} \sim \text{Potts}(B_m, \beta_m).
\end{align}

\noindent Here, $\zeta_{mi}=c \in \{ 1, \ldots, C\}$ is the cell type label, $\kappa_{mi} =k\in \{ 1, \ldots, K\}$ is the spatial domain label, and $\bzeta_m, \bkappa_m$ are $n_m$-length vectors of the cell type labels and spatial domain labels, respectively. The parameter $\boldsymbol{\theta}_k = (\theta_{1k}, \ldots, \theta_{Ck})^\top$ represents the spatial-domain-specific cell type proportions, organized into the matrix $\boldsymbol{\Theta} = (\boldsymbol{\theta}_1, \ldots, \boldsymbol{\theta}_K) \in \mathbb{R}^{C \times K}$.
We assume a common covariance matrix, $\boldsymbol{\Sigma}$, across all cell type clusters. This approach balances computational tractability with model parsimony, a strategy successfully employed in similar high-dimensional contexts by \cite{li_bass_2022} and \cite{zhu_generalized_2024}. While the latent factors $\mathbf{u}_{mi}$ are generally identifiable only up to an orthogonal rotation, the proposed formulation remains well-posed; specifically, the mixture-of-Gaussians prior, coupled with the shared covariance structure, ensures that the likelihood and resulting posterior cluster assignments are invariant to such rotations.

To account for the spatial architecture within each tissue sample, we assume that neighboring cells are more likely to belong to the same spatial domain, thereby inducing spatially homogeneous clusters. We model the spatial domain labels, $\bkappa_m$, using a  Potts model \citep{potts_generalized_1952,feng_mri_2012}, denoted as $\text{Potts}(B_m, \beta_m)$. The joint probability mass function is given by:
\begin{align}
    &p(\bkappa_m \mid B_m, \beta_m) = \frac{1}{d(B_m, \beta_m)} \exp \left\{\beta_m \sum_{i \sim i'} \bI(\kappa_{mi} = \kappa_{mi'}) \right\} 
    \label{eq:potts_model}
\end{align}

\noindent where $B_m$ represents a $k$-nearest neighbor graph, and $\beta_m \geq 0$ is the smoothing parameter determining the strength of spatial dependence in sample $m$. Unless specified otherwise, $B_m$ is constructed symmetrically such that each cell is connected to at least its 4 nearest neighbors. Omitting the index $m$ to simplify notation, $i \sim i'$ denotes all neighboring pairs of cells in the graph $B_m$.

The normalizing constant, or partition function, $d(B_m, \beta_m)=\sum_{\bkappa_m}\exp \left\{\beta_m \sum_{i \sim i'} \bI(\kappa_{mi} = \kappa_{mi'}) \right\}$ is computationally intractable as it involves $K^{n_m}$ terms.  Since evaluating this constant is essential for posterior inference via Markov chain Monte Carlo (MCMC), we employ thermodynamic integration as proposed by \citet{reich_spatial_2014}. Further details regarding the implementation of this approach are provided in Web Appendix Section \ref{sec:betam_sampling}.

For the rest of the parameters, we specify the following priors:
\begin{align}
    &\bmu_{c} \sim \mathcal{N}(\bzero, I_r) \label{eq:mu_prior} \\
    &\Sigma \sim \textrm{Inverse-Wishart}(df_0, \Omega_0) \label{eq:Sigma_prior} \\
    &\btheta_k \sim Dir(c_0 \bone^\intercal_C) \label{eq:theta_prior} \\
    &\beta_m \sim Uniform(0, \beta_{max}), \label{eq:beta_prior}
\end{align}

\noindent where $k = 1, \ldots, K$ and $m = 1, \ldots, N$. Here, we place a Normal prior on the cell type means $\bmu_c$ with mean $\bzero$ and variance-covariance matrix $I_r$; an Inverse-Wishart prior on $\Sigma$ with $df_0$ degrees of freedom and a symmetric positive definite scale matrix $\Omega_0$ of size $r \times r$; a Dirichlet prior on $\btheta_k$ with concentration parameter $c_0$; and a uniform prior on $\beta_m$ with a lower bound of 0 and an upper bound of $\beta_{max}$. We set $df_0$ to be $r + 1$ and $\Omega_0$ to be $I_{r}$ to provide a weak prior on the covariance matrix; set $c_0$ to be 1 to encode equal prior probabilities for all possible cell type compositions; and set $\beta_{max}$ to be 4 to represent the extreme case where the spatial domain boundaries are extremely smooth. 

These prior distributions and associated hyperparameters were selected to be weakly informative, allowing the data to drive the Bayesian estimation in the absence of prior information. The first three priors (Equations \ref{eq:mu_prior}, \ref{eq:Sigma_prior}, and \ref{eq:theta_prior}) were selected to be conjugate priors to facilitate the computation via Gibbs sampling (Web Appendix Section \ref{sec:mcmc_impl}).

\subsection{Bayesian feature selection using latent binary indicators}
\label{sec:sparse_ft_select}

To increase the interpretability of our model, we employ a feature selection mechanism to select active genes, i.e., genes that contribute to the covariance of the gene expression matrix, as opposed to genes that only contribute to the random noise. We perform feature selection by selecting a subset of factor loadings in $\bA \in \mathbb{R}^{r \times P}$ to be non-zero, setting the rest of the loadings at zero. We select non-zero factor loadings, indicating whether gene $j = 1, \ldots, P$ is active (corresponding to the columns of $\bA$), in each component $l = 1, \ldots, r$ (corresponding to the rows of $\bA$) through the binary indicator $\gamma_{lj}$. For each component $l$ and gene $j$, the prior distribution of $\gamma_{lj}$ is $\gamma_{lj} \sim \textrm{Bernoulli}(q_\gamma),$ where $q_\gamma$ is the prior probability of selecting features for a given component. We set $q_\gamma$ to 0.05, as we expect that a small number of features are active.

To achieve a parsimonious representation and identify relevant biological markers, we impose sparsity on the factor loadings matrix $\mathbf{A}$ using a spike-and-slab prior. Specifically, for each entry $a_{lj}$, we assume:
\begin{equation}
       a_{lj} \mid \gamma_{lj} \sim (1 - \gamma_{lj}) \delta_0 + \gamma_{lj} \mathcal{N}(0, 1),
\end{equation}

\noindent where $\delta_{0}$ denotes a Dirac delta mass at $0$. As detailed in Section \ref{sec:select_active_differentiating}, the binary indicators $\gamma_{lj}$ will allow us to identify ``active'' genes contributing to the latent space and, among those, distinguish genes that drive the differentiation between cell types. Hence, BayesClint can perform nested feature selection. By sharing $\bA$ across all $M$ tissue samples, we enforce the biologically motivated assumption that the fundamental gene-factor regulatory structures are preserved across samples.

Furthermore, this framework naturally accommodates component-level selection through the introduction of indicators $\rho_l$ for $l = 1, \dots, r$. This allows for a multi-stage selection process where active components are first identified, followed by the selection of specific gene loadings within those components/factors. Implementation details for this extended selection mechanism are provided in Web Appendix Section \ref{sec:gamma_eta_sampling}.

\section{Computational Details and Posterior Inference}
\label{sec:posterior_inference}

For posterior inference, our primary interest is in the estimation of the feature selection indicators $\gamma_{lj}$, which allow us to select the active and differentiating genes, and the estimation of the cell type labels $\zeta_{mi}$, spatial domain labels $\kappa_{mi}$, and cell type proportions $\theta_{ck}$. 

\subsection{Estimation of the factor loadings matrix}
\label{sec:estimation_factor_loadings}

We implement a Markov Chain Monte Carlo (MCMC) algorithm for posterior inference that employs a partially collapsed Gibbs sampling \citep{van_dyk_partially_2008} to sample the feature selection indicators $\gamma_{lj}$. That is, to sample $\gamma_{lj}$ given the other parameters, we integrate out all the loadings $\bA$ and use a Metropolis-Hastings step. We then sample the loadings of selected variables (according to $\gamma_{lj}$) from their full conditional distributions. The other parameters are updated using Gibbs or Metropolis steps (see Web Appendix Section \ref{sec:mcmc_impl}).

We set $r$, the maximum number of active components, according to the scree plot of the eigenvalues of the correlation matrix of $\bX = (\bX_1^\intercal, \ldots, \bX_N^\intercal)^\intercal$. We select the number of components using the elbow method, i.e., we select the value $r$ just before the point where eigenvalues start leveling off. Sensitivity analyses of the choice of $r$, as well as the prior probability of selection for $\gamma_{lj}$, $q_\gamma$, are shown in Web Appendix Sections \ref{sec:r_sensitan} and \ref{sec:qv_sensitan}, respectively. These sensitivity analyses validate our selection of $r$ through the elbow method and setting of $q_\gamma$ to 0.05.

\subsection{Selection of active and differentiating genes}
\label{sec:select_active_differentiating}

We define a gene to be ``active'' if it is selected in at least one component in $\bA$. We summarize the feature selection via the posterior probabilities of inclusion (\textrm{PPI}) as follows, for genes $j = 1, \ldots, P$: 
\begin{align}
    \textrm{PPI}_j = \frac{1}{T} \sum_{t = 1}^T \bI\left( \sum_{l = 1}^r \gamma_{lj}^{(t)} > 0 \right),
\end{align}

\noindent where $t = 1, \ldots, T$ indicates the MCMC iterations after burn-in, and $\gamma_{lj}^{(t)}$ is the estimated $\gamma_{lj}$ at MCMC iteration $t$. Then, the active genes are identified if their \textrm{PPI} values exceed a given threshold $\tau$; the set of active genes would be $\{ \bI(\textrm{PPI}_1 \geq \tau), \ldots, \bI(\textrm{PPI}_P \geq \tau) \}$. We can use the threshold $\tau = 0.5$, which is commonly referred to as the median model. 

Alternatively, we can use a threshold controlling for multiplicity \citep{newton_detecting_2004}, which ensures that the expected Bayesian false discovery rate (BFDR) is less than a specified value. The BFDR($\tau'$) is calculated as follows:
\begin{align}
\label{eq:bfdr_post_prob_inc}
    \textrm{BFDR}(\tau') := \frac{\sum_{j = 1}^P (1 - \textrm{PPI}_j) \bI(1 - \textrm{PPI}_j < \tau')}{\sum_{j = 1}^{P} \bI(1 - \textrm{PPI}_j < \tau')},
\end{align}

\noindent where $\textrm{BFDR}(\tau')$ is the expected false discovery rate. Then, the set of active genes would be $\{ \bI(1 - \textrm{PPI}_1 < \tau'), \ldots, \bI(1 - \textrm{PPI}_P < \tau') \}$. For the simulation study and real data analyses, we set the BFDR to be equal to or less than the significance level of 0.05.

To assess the MCMC convergence of our method, we fitted four MCMC chains with different starting points. As is commonly done in Bayesian feature selection (\citealp{chekouo_mirnatarget_2015,chekouo_bayesian_2017,stingo_incorporating_2011,chekouo_bayesian_2020,li_interpretable_2024}), we assessed the agreement (in particular for binary variables) of the results among the four chains by evaluating the pairwise correlation coefficients between the \textrm{PPI}s of all $P$ genes.

To select genes that are not only active but also able to differentiate among cell types, we first determine the components that are differentiating. We define a differentiating component as one where at least one cell type mean for that component is significantly nonzero (i.e., the null hypothesis that the cell type mean is zero would be rejected). From the MCMC samples for cell type mean $\mu_{lc}$, we compute the Bayesian $p$-value as 
\begin{align}
\label{eq:cell_type_mean_pval}
    p_{lc} = 2 \times \min\left( \frac{1}{T} \sum_{t = 1}^T \bI \left(\mu_{lc}^{(t)} <= 0 \right) , \frac{1}{T} \sum_{t = 1}^T \bI\left(\mu_{lc}^{(t)} >= 0 \right) \right).
\end{align}

\noindent We determine a cell type mean $\mu_{lc}$ to be significantly nonzero if $p_{lc}$ is less than the desired significance level. For the simulation study, we use the significance level of 0.05. To correct for the multiplicity of $C$ hypothesis tests and keep the family-wise error rate at the desired significance level, we adjust the Bayesian $p$-values by Holm's correction \citep{holm_simple_1979}. Web Figure \ref{fig:diff_vs_nondiff_component} shows an example of a differentiating component and a non-differentiating component from a simulation study scenario. Let $L_d$ be the set of indices $l$ corresponding to differentiating components.

Then, we examine the \textrm{PPI} for each entry of the factor loadings matrix, $\textrm{PPI}_{lj} = \sum_{t = 1}^T \gamma_{lj}^{(t)} / T$, to get the \textrm{PPI} of gene $j$ into the set of differentiating genes:
\begin{align}
\label{eq:diff_post_prob_inc}
    \textrm{PPI}_{d,j} = \max_{l \in L_d}\{ \textrm{PPI}_{lj} \}.
\end{align}

\noindent Alternatively, one can define $\textrm{PPI}_{d,j}$ as $\frac{1}{T} \sum_{t = 1}^T \bI\left( \sum_{l \in L_d}^r \gamma_{lj} > 0 \right)$. However, according to our simulation study, defining the $\textrm{PPI}_{d,j}$ as in (\ref{eq:diff_post_prob_inc}) leads to better selection performance. After calculating the $\textrm{PPI}_{d,j}$, we can use the threshold that would bound the BFDR at the desired significance level according to Equation \ref{eq:bfdr_post_prob_inc}, to identify the differentiating genes. The set of differentiating genes would be $\{ \bI(\textrm{PPI}_{d,1} \geq \tau), \ldots, \bI(\textrm{PPI}_{d,P} \geq \tau) \}$ using the threshold $\tau$ that does not control for multiplicity; the set of differentiating genes would be $\{ \bI(1 - \textrm{PPI}_{d,1} < \tau'), \ldots, \bI(1 - \textrm{PPI}_{d,P} < \tau') \}$ using the threshold $\tau'$ that ensures a BFDR less than or equal to a pre-set significance level.

Rather than using the binary procedure of categorizing each component as differentiating or non-differentiating, we can instead assign each component a $p$-value indicating the probability that the component accepts the null hypothesis that it is non-differentiating, i.e., that all of the cell type means associated with it are zero. We do this by combining the $p$-values for component $l$, as defined in (\ref{eq:cell_type_mean_pval}), into one $p$-value through the Aggregated Cauchy Association Test (ACAT) \citep{liu_acat_2019,liu_cauchy_2020}. Because the Bayesian $p$-values for the cell type means are computed from $T$ MCMC iterations, some Bayesian $p$-values may be exactly zero. If one of the Bayesian $p$-values to be combined by ACAT is zero, then the combined $p$-value is also zero. Therefore, we correct the zero $p$-values by replacing them with $1 / T$, according to the recommendation of \cite{liu_cauchy_2020}. The $r$ combined $p$-values would allow us to rank the components by how much they reject the null hypothesis of being non-differentiating. We demonstrate this approach in the real data analyses.

\subsection{Estimating the cluster labels and determining the numbers of cell types and spatial domains}
\label{sec:estimation_cluster_labels}
The posterior distribution of our Bayesian hierarchical model is invariant to permutations of the cluster indices (for either cell types or spatial domains), leading to the well-documented ``label switching'' phenomenon \citep{stephens_dealing_2000}. To resolve this and ensure parameter identifiability, we post-process the MCMC chains using the iterative version of the Equivalence Class Representation (ECR-1) algorithm \citep{rodriguez_label_2014}, as implemented in the \texttt{label.switching} R package \citep{papastamoulis_labelswitching_2016}. This procedure yields a consistent ordering for the cluster labels $\zeta_{mi}$, $\kappa_{mi}$, and the cell type centroids $\boldsymbol{\mu}_c$. 

Post-alignment, we obtain point estimates for the labels $\zeta_{mi}$ and $\kappa_{mi}$ via their respective posterior modes, $\hat{\zeta}_{mi}$ and $\hat{\kappa}_{mi}$. Given these estimated labels, the cell-type-to-spatial-domain proportions $\theta_{ck}$ are estimated empirically as:
\begin{equation}
	\hat{\theta}_{ck} = \frac{\sum_{m = 1}^M \sum_{i = 1}^{n_m} \bI(\hat{\zeta}_{mi} = c, \hat{\kappa}_{mi} = k)}{\sum_{m = 1}^M \sum_{i = 1}^{n_m} \bI(\hat{\kappa}_{mi} = k)},
\end{equation}
where $\hat{\theta}_{ck} = 0$ if the denominator is zero. We do not estimate $\theta_{ck}$ by summarizing its posterior samples, because the label switching issue could not be fully resolved by the ECR-1 algorithm, and so the posterior mean estimates of $\theta_{ck}$ could be sensitive to the label switching issue. On the other hand, the posterior estimates of $\zeta_{mi}, \kappa_{mi}$ are less sensitive to the label switching issue, allowing us to estimate $\theta_{ck}$ more accurately.

In practice, the number of cell types $C$ and spatial domains $K$ is often guided by expert pathological knowledge. When such information is unavailable, we propose a data-driven heuristic informed by our sensitivity analyses (Web Appendix Section \ref{sec:CorK_sensitan}). We recommend initially overspecifying the number of cell types and prioritizing calibration of the number of spatial domains, as cell type clustering is more stable under misspecification. Specifically, we fix $C$ to a high value, and then fit the BayesClint model under an increasing number of spatial domains $K$ until the smallest spatial domain constitutes less than 6\% of the total cell population. Note that changing the number of spatial domains would require recomputation of the Potts model statistics involved in the estimation of the Potts normalizing constant (see Section \ref{sec:multi_scale_clustering_layer} and Web Appendix Section \ref{sec:eval_d}). Subsequently, we fix the number of spatial domains at this value, and increase the number of cell types until the rarest type represents less than 2.5\% of the dataset.

\section{Simulation Study}
\label{sec:simulation_study}

In this section, we evaluate the performance of the proposed method with a simulation study and sensitivity analyses. We first describe the simulation scenarios, introduce competing methods and metrics, and finally present the results.

\subsection{Data generation}
\label{sec:data_generation}

The simulation study examines the performance of BayesClint against competing methods across 16 scenarios determined by four factors, under which SRT data are generated. The factors listed, with their respective levels, are i) the number of tissue samples $N\in\{1,3\}$, ii) cell type composition $\theta$ evaluated under both \textit{irregular} and \textit{regular} spatial distributions to assess robustness to tissue architecture, iii) total number of genes $P\in\{200, 1000\}$ representing moderate and high-dimensional datasets, and iv) number of differentially expressed genes $P_d\in\{40,80\}$. 

For each of the 16 scenarios, we generate 50 Monte Carlo (MC) datasets. First, we generate $N$ spatial point patterns to represent cell locations. Locations are generated uniformly within a 10 units $\times$ 12 units window with an intensity of 9 points per unit area. Each cell is assigned a latent spatial domain label $k \in \{1, \dots, 4\}$ based on its coordinate's intersection with the predefined tissue regions. All scenarios utilize the four-domain partition illustrated in Figure \ref{fig:sim_study_reps}. In the single-sample case ($N = 1$), we utilize the partition for Tissue Sample 1 exclusively (Figure \ref{fig:sim_study_repsA}); for the multi-sample setting ($N = 3$), we utilize the distinct regional architectures across all three Tissue Samples (Figures \ref{fig:sim_study_repsA}-\ref{fig:sim_study_repsC}) to assess the model's ability to integrate heterogeneous spatial structures.

Given the number of tissue samples $N$, the cell locations generated for the 50 MC replicates remain invariant across all scenarios and subsequent sensitivity analyses. Within the BayesClint framework, these coordinates serve exclusively to define the spatial topology by constructing a $k$-nearest neighbors graph, which informs the prior distribution of $\bkappa_{m}$ for spatial dependency between adjacent cells. 

The second simulation factor, $\theta$, dictates the cell type composition across four distinct cell types. For each spatial domain $k \in \{1, \dots, 4\}$, cell type labels are drawn from a categorical distribution with domain-specific probability vectors $\btheta_k\in \mathbb{R}^4$.   We evaluate two compositional regimes: i) \textit{Irregular Cell Type Composition}: this regime represents high spatial mixing and overlapping cell type distributions. The probability vectors are defined as $\theta_1 = (0.2, 0.3, 0.3, 0.2)^\intercal$, $\theta_2 = (0.6, 0.1, 0.1, 0.2)^\intercal$, $\theta_3 = (0.05, 0.05, 0.40, 0.50)^\intercal$, and $\theta_4 = (0.0, 0.70, 0.15, 0.15)^\intercal$. ii) \textit{Regular Cell Type Composition}: this regime reflects a more structured tissue architecture where each domain is dominated by a primary cell type. The vectors are specified as $\theta_1 = (0.8, 0.1, 0.1, 0)^\intercal$, $\theta_2 = (0, 0.8, 0.1, 0.1)^\intercal$, $\theta_3 = (0.1, 0, 0.8, 0.1)^\intercal$, and $\theta_4 = (0.1, 0.1, 0, 0.8)^\intercal$.
Web Figure \ref{fig:irreg_vs_reg} shows the probabilities for the two compositional regimes, and Figure \ref{fig:sim_study_reps} shows the tissue samples where the cell types have been assigned according to the ``regular'' cell type composition.

Log-mean gene expression profiles of each cell are generated according to the factor model defined in Equation \ref{eq:factor_model}. The third factor, $P$, specifies the total number of genes and corresponds to the number of columns in $\bA$, the factor loadings matrix. For all scenarios, we fix the number of latent factors at $r = 4$. To reflect biological sparsity, we define ``active genes'' as columns in $\bA$ containing at least one non-zero entry, with the proportion of active genes fixed at 40\% of $P$. We generate the non-zero loadings as independent and identically distributed ($i.i.d.$) random samples from the uniform distribution on the disjoint union set $[-0.5, -0.3] \cup [0.3, 0.5]$.

The latent factors $\bu_{mi}$ for cell $i$ in sample $m$ are generated from a multivariate normal distribution centered at a cell type-specific mean, $\bmu_{\zeta_{mi}}$, following Equation \ref{eq:factor_u_mi}. The variance-covariance matrix $\Sigma$ for this distribution will be drawn from the Inverse-Wishart($2r$, $I_r$) distribution. The means for components 1 to 4, respectively, for cell type 1 are $\bmu_1 = (2.5, -1.0, 1.0,  0.0)^\intercal$. For the other cell types, the mean vectors are $\bmu_2 = (1, 1, 1, 0)^\intercal, \bmu_3 = (-3, -1, 0, 0)^\intercal,$ and $\bmu_4 = (1.5, -0.5, -2.0, 0.0)^\intercal$. Under this configuration, the first three components serve as differentiating components that drive the separation between cell types, whereas the fourth component, characterized by a zero mean across all cell types, serves as a non-differentiating component.

The fourth simulation factor, $P_d$, defines the number of differentially expressed (DE) genes, representing features (genes) that capture distinct expression profiles across the four cell types as governed by $\bmu_c, c = 1, \ldots, 4$. The structure of the factor loadings matrix $\bA$ is partitioned to reflect this biological distinction: i) \textit{Differentially Expressed Genes}: For the $P_d$ columns corresponding to DE genes, all entries in the first three rows (the differentiating components) are non-zero, ensuring these genes manifest the cell-type-specific signals. ii) \textit{Active, Non-DE Genes}: For the remaining active genes ($0.4P - P_d$), the corresponding columns in $\mathbf{A}$ contain non-zero entries only in the fourth row. Because the fourth latent component, $\bU_{\cdot 4}$, is non-differentiating (zero-centered for all cell types), these genes exhibit systematic variation across cells but do not contribute to cell type discrimination. This construction allows us to rigorously evaluate BayesClint's ability to isolate true discriminatory signals from background biological noise.

Finally, we generated the high-dimensional gene expression vectors similar to Equation \ref{eq:factor_model}: $\bx_{mi} = \btau_m + \bA^\intercal \bu_{mi} + \bepsilon_{mi}$, where $\btau_m \sim \mathcal{N}(\bzero, I_P)$ is the vector of intercepts specific to sample $m$ and each gene, $\bepsilon_{mi} \sim \mathcal{N}(\bzero, \textrm{diag}(\upsilon^2_{m1}, \cdots, \upsilon^2_{mP}))$, where $\upsilon^2_{mj} = 0.1 + |z_j|$ and $z_j \sim \mathcal{N}(0, 1)$. The generated raw gene expression counts $\tilde{\bx}_{mi} = (\tilde{x}_{mi1}, \ldots, \tilde{x}_{mip})^\intercal$ are subsequently generated via a Poisson model: $\tilde{x}_{mij} \sim \textrm{Poisson}(\exp(x_{mij}))$.

\subsection{Competing methods and metrics}

For all simulation scenarios, the BayesClint model was fitted using 13,000 MCMC iterations, with the initial 6,500 discarded as burn-in. The number of latent components was set to $r=4$, as selected by the elbow method (Section \ref{sec:estimation_factor_loadings}) applied to a scree plot of a representative generated dataset (Web Figure \ref{fig:sim_study_screeplot}). For the problem of clustering cells into cell types and spatial domains, we compare BayesClint against four established clustering algorithms. The first two are two-stage approaches combining Sparse Principal Components analysis (SPC; \citealp{witten_penalized_2009}) with either $k$-means clustering \citep{hartigan_algorithm_1979} or Gaussian mixture model-based clustering \citep{scrucca_model-based_2023}, denoted as SPC+$k$-means and SPC+mclust, respectively. SPC performs PCA on the data and imposes $l_1$-norm regularization on the principal component loadings to select variables. SPC, $k$-means, and the model-based clustering are implemented in the R package \texttt{PMA} (version 1.2.4) \citep{witten_pma_2024}, \texttt{stats} (version 4.4.1) \citep{r_core_team_r_2021}, and \texttt{mclust} (version 6.1.2) \citep{scrucca_model-based_2023}, respectively. We also compare against PRECAST (v1.8; \citealp{liu_probabilistic_2023}) and BASS (v1.1.0.17; \citealp{li_bass_2022}), which are specifically designed for multi-sample integration.

While PRECAST and BASS natively handle multiple tissue samples, for SPC+$k$-means and SPC+mclust in $N=3$ scenarios, we concatenate the expression matrices across samples as input. Furthermore, only BASS explicitly distinguishes between cell type and spatial domain clustering; for the other three methods, a single clustering partition is evaluated against both ground truths. 
For more details on the implementation of competing clustering algorithms, see Web Appendix Section \ref{sec:additional_competing_details}.

The clustering performance of each method is evaluated by comparing the estimated partition with the true partition for both cell types and spatial domains. The comparison is done using the Adjusted Rand Index (ARI) \citep{hubert_comparing_1985}, which measures the similarity between two partitions by counting pairs of elements that are clustered together or separately in the two partitions. The two partitions may have different numbers of clusters. If the two partitions are almost identical, the ARI is close to one; if one of the partitions was randomly generated, the ARI tends toward zero. The ARI can be negative if the proposed partition is worse than a random partition. For multi-sample scenarios ($N=3$), we calculate a single global ARI for each task by pooling the cluster labels and ground-truth assignments across all tissue samples, thereby evaluating the model's integrative performance across the entire study.  

For the problem of feature selection, BayesClint will select genes according to a threshold that corresponds to a BFDR of 0.05 (Section \ref{sec:select_active_differentiating}). We also compare BayesClint with five established feature selection algorithms. To assess the identification of active genes, we compare BayesClint with two scRNA-seq-based approaches: \textit{modelGeneVar} for identifying highly variable genes (HVGs) and \textit{correlateGenes} for identifying highly correlated genes, both implemented in the R package \texttt{scran} (version 1.32.0; \cite{lun_step-by-step_2016}). For the selection of differentiating genes, we benchmark against SPARK-X (v1.1.1; \citealp{zhu_spark-x_2021}), a method for detecting spatially variable genes, and DESeq2 (v1.44.0; \citealp{love_moderated_2014}), a method for finding differentiating genes in RNA-seq data. Since DESeq2 requires pre-defined cell groups as input to find genes that differentiate between the groups, we use the cell type clusters estimated by BASS as input (referred to as BASS+DESeq2), as BASS demonstrated the second-best cell type clustering performance (after BayesClint) among clustering methods in the simulation study. For these four approaches, $p$-values were adjusted using the Benjamini-Hochberg procedure \citep{benjamini_controlling_1995} to maintain a false discovery rate (FDR) of 0.05. Finally, we include SPC as a baseline; since it does not distinguish between active and differentiating features, its unified selection results are applied to both evaluation tasks. For the multi-sample implementation of all competing gene selection methods, expression matrices were concatenated prior to analysis. Further implementation details are provided in Web Appendix Section \ref{sec:additional_competing_details}.

The performance of each feature selection method is assessed by its ability to correctly identify active and differentiating genes. We primarily quantify this via the classification accuracy, defined as the proportion of genes correctly assigned to their respective latent categories (e.g., active versus inactive). For each simulation scenario and sensitivity analysis, performance metrics are averaged across the 50 Monte Carlo replicates. Additionally, to evaluate the trade-off between sensitivity and specificity across varying selection thresholds, we report the Area Under the Receiver Operating Characteristic Curve (AUC) for all methods in Web Appendix Section \ref{sec:additional_ft_select_results}.

\subsection{Simulation study results}

First, we examined the convergence of our algorithm on an example scenario, with $N = 3$, irregular $\theta$ composition, $P = 200$, and $P_d = 40$. After running 4 MCMC chains, each with 6,500 burn-in and 6,500 MCMC iterations, we examined the pairwise correlation coefficients between the \textrm{PPI}s of all $P$ genes. These indicated good concordance between the four chains, with all pairwise Pearson correlation coefficients falling in the range [0.993, 0.999] (see Web Figure \ref{fig:sim_study_corrplot_4_chains}).

Regarding cell type clustering performance, our method, BayesClint, achieved the highest ARI in ten out of sixteen evaluated scenarios (top of Table \ref{tab:sim_results_zeta_kappa_ari}). In the remaining scenarios, either BASS or PRECAST achieved the highest ARI; however, BayesClint maintained competitive performance with only marginal differences.
Across all methods, an increase in the number of differentially expressed genes from $P_d = 40$ to $P_d = 80$ consistently improved clustering performance. This trend is consistent with expectations, as a higher density of discriminatory features increases signals, thereby facilitating more separation of latent cell type structures.

\begin{table}[]
\centering
\begin{tabular}{lllllllll}
Cell Types \\
\hline
$N$ & $\theta$ & $P$ & $P_d$ & BayesClint & SPC+$k$-means & SPC+mclust & PRECAST & BASS \\ 
\hline
1 & irr & 200 & 40 & \textbf{89} (1.5) & 75.2 (2.1) & 85.2 (1.1) & 87.7 (0.8) & 89.0 (0.8) \\ 
1 & irr & 200 & 80 & \textbf{97} (0.4) & 87.9 (2.2) & 94.7 (0.8) & 95.3 (0.6) & 96.3 (0.6) \\ 
1 & irr & 1000 & 40 & \textbf{92.8} (1) & 53.6 (3.4) & 77.7 (1.7) & 83.0 (1.6) & 79.4 (2.5) \\ 
1 & irr & 1000 & 80 & \textbf{94.6} (1.7) & 74.3 (3.3) & 91.8 (1.7) & 93.1 (1.5) & 91.7 (1.6) \\ 
\hdashline
1 & reg & 200 & 40 & \textbf{92.5} (0.9) & 75.2 (2.2) & 85.1 (1.2) & 91.9 (0.6) & 91.8 (0.6) \\ 
1 & reg & 200 & 80 & 96.2 (0.9) & 84.4 (2.1) & 94.1 (0.8) & 95.8 (0.7) & \textbf{97.1} (0.3) \\ 
1 & reg & 1000 & 40 & \textbf{95.9} (0.2) & 56.4 (3.3) & 79.2 (1.5) & 88.4 (1.2) & 87.2 (0.7) \\ 
1 & reg & 1000 & 80 & \textbf{96.4} (1) & 71.8 (3.4) & 92.2 (1.2) & 94.7 (0.9) & 93.4 (1.4) \\ 
\hdashline
3 & irr & 200 & 40 & \textbf{90.2} (0.7) & 74.3 (2.6) & 86.5 (0.9) & 87.2 (0.9) & 89.4 (0.8) \\ 
3 & irr & 200 & 80 & 93.6 (1.3) & 81.9 (2.3) & 91.2 (1.2) & 91.8 (1.2) & \textbf{94.9} (0.6) \\ 
3 & irr & 1000 & 40 & \textbf{89.7} (1.5) & 55.6 (3.3) & 79.6 (1.4) & 87.1 (1.3) & 82.7 (1.9) \\ 
3 & irr & 1000 & 80 & 93.0 (1.6) & 70.6 (3.1) & 94.6 (0.7) & \textbf{94.9} (0.7) & 94.8 (0.7) \\ 
\hdashline
3 & reg & 200 & 40 & 91.7 (1.3) & 79.1 (2.2) & 88.3 (0.7) & 92.6 (0.5) & \textbf{94.1} (0.3) \\ 
3 & reg & 200 & 80 & 96.0 (1.1) & 82.8 (2.4) & 93.4 (0.9) & 95.7 (0.7) & \textbf{96.8} (0.4) \\ 
3 & reg & 1000 & 40 & \textbf{94.6} (0.9) & 59.3 (3.3) & 82.0 (1.4) & 89.6 (1.2) & 87.1 (1.4) \\ 
3 & reg & 1000 & 80 & 96.0 (1.1) & 73.2 (3.0) & 94.1 (0.9) & \textbf{96.1} (0.7) & 94.5 (1.2) \\
\hline
\\
Spatial Domains \\
\hline
1 & irr & 200 & 40 & 81.1 (1.7) & 14.5 (0.7) & 17.4 (0.4) & 20.5 (0.3) & \textbf{83} (1.1) \\ 
1 & irr & 200 & 80 & 85.8 (0.9) & 19.1 (0.7) & 21.1 (0.3) & 22.1 (0.3) & \textbf{86.2} (0.6) \\ 
1 & irr & 1000 & 40 & \textbf{82.1} (1.4) & 8.4 (0.8) & 14.4 (0.6) & 18.9 (0.6) & 77.8 (2.6) \\ 
1 & irr & 1000 & 80 & \textbf{83.1} (1.6) & 14.2 (0.9) & 19.5 (0.6) & 20.9 (0.4) & 82.2 (1.7) \\ 
\hdashline
1 & reg & 200 & 40 & 93.8 (0.8) & 41.1 (1.2) & 46.7 (0.7) & 55.3 (0.3) & \textbf{94.8} (0.2) \\ 
1 & reg & 200 & 80 & 95.1 (0.5) & 45.9 (1.3) & 51.6 (0.5) & 54.4 (0.5) & \textbf{95.6} (0.2) \\ 
1 & reg & 1000 & 40 & \textbf{95.3} (0.2) & 30.9 (1.8) & 43.5 (0.9) & 53.8 (0.7) & 92.6 (0.7) \\ 
1 & reg & 1000 & 80 & \textbf{95.3} (0.4) & 39.1 (1.9) & 50.4 (0.8) & 54.1 (0.6) & 94.4 (0.7) \\ 
\hdashline
3 & irr & 200 & 40 & 82.6 (1.1) & 14.1 (0.7) & 17.5 (0.3) & 20.3 (0.2) & \textbf{83.4} (0.3) \\ 
3 & irr & 200 & 80 & 82.8 (1.3) & 16.2 (0.7) & 18.8 (0.4) & 20.2 (0.4) & \textbf{85.2} (0.3) \\ 
3 & irr & 1000 & 40 & \textbf{80.4} (1.5) & 8.9 (0.8) & 14.8 (0.5) & 20.0 (0.3) & 78.4 (1.8) \\ 
3 & irr & 1000 & 80 & 79.7 (1.8) & 12.9 (0.8) & 19.8 (0.3) & 20.9 (0.2) & \textbf{83.8} (0.9) \\ 
\hdashline
3 & reg & 200 & 40 & 92.5 (0.8) & 42.8 (1.2) & 48.2 (0.4) & 55.5 (0.2) & \textbf{93.9} (0.1) \\ 
3 & reg & 200 & 80 & 93.9 (0.5) & 44.7 (1.4) & 50.9 (0.5) & 54.3 (0.4) & \textbf{94.3} (0.2) \\ 
3 & reg & 1000 & 40 & \textbf{93.4} (0.7) & 32.0 (1.8) & 44.6 (0.8) & 54.6 (0.7) & 90.9 (0.8) \\ 
3 & reg & 1000 & 80 & 93.1 (0.9) & 39.6 (1.7) & 51.3 (0.6) & 54.6 (0.4) & \textbf{93.5} (0.6) \\
\hline
\end{tabular}
\caption{Results of the clustering of cell types (top) or spatial domains (bottom). The average adjusted Rand index (multiplied by 100) across 50 Monte Carlo iterations is presented for each of the five competing methods under 16 scenarios determined by 4 factors. The standard error of the average is in parentheses. $N$ is the number of simulated tissue samples, $\theta$ refers to the cell type compositions (irregular or regular, abbreviated as `irr' or `reg', respectively), $P$ is the number of genes in total, and $P_d$ is the number of differentially expressed genes. Our proposed algorithm is denoted as BayesClint, whereas the competing methods are SPC+$k$-means, SPC+mclust, louvain, BASS, and PRECAST. 
The values of the best performances are in bold type.} \label{tab:sim_results_zeta_kappa_ari}
\end{table}

In terms of spatial domain clustering performance, BayesClint achieved the highest ARI in six of the sixteen scenarios (bottom of Table \ref{tab:sim_results_zeta_kappa_ari}), with BASS achieving the highest ARI in the other scenarios. However, as for cell type clustering, BayesClint's performance is highly similar to BASS's performance in scenarios where BASS is best. Across all evaluated methods, the ``regular'' cell type composition regime consistently yielded higher ARI values compared to the ``irregular'' regime.
This indicates that the particular pattern of the cell type composition does affect spatial domain clustering performance for all methods, but BayesClint is nonetheless able to infer the spatial domains from various cell type compositions quite well. Furthermore, as compared to cell type clustering, whereas BayesClint and BASS both maintained high fidelity in domain recovery, the remaining methods exhibited a significant decline in performance. This disparity underscores the clear advantage of multi-scale frameworks that explicitly model the dependency between single-cell identities and regional tissue structures.

For either active gene selection (Table \ref{tab:sim_results_active_acc}) or differentiating gene selection (Table \ref{tab:sim_results_deg_acc}), BayesClint achieves much higher accuracies than the other methods. For each scenario, BayesClint achieves a higher accuracy in differentiating gene selection than active gene selection. This is likely due to the differentiating genes being a subset of all the active genes. In terms of the AUC, the feature selection performance is more comparable between BayesClint and the competing methods (Web Appendix Section \ref{sec:additional_ft_select_results}). The reason for the vast discrepancy between the accuracy and the AUC for the gene selection methods is that, with the 0.05 threshold for the false discovery rate, the gene selection methods have a high sensitivity but a very low specificity as compared to BayesClint.

\begin{table}[]
\centering
\begin{tabular}{llllllll}
\hline
$N$ & $\theta$ & $P$ & $P_d$ & BayesClint & SPC & modelGeneVar & correlateGenes \\ 
\hline
1 & irr & 200 & 40 & \textbf{84.8} (1.3) & 67.3 (1.7) & 62.3 (0.2) & 77.7 (1.0) \\ 
1 & irr & 200 & 80 & \textbf{89.1} (1.3) & 45.2 (0.8) & 62.6 (0.4) & 78.8 (1.3) \\ 
1 & irr & 1000 & 40 & \textbf{86.5} (1.3) & 61.0 (2.1) & 60.4 (0.1) & 71.2 (1.3) \\ 
1 & irr & 1000 & 80 & \textbf{86.1} (1.8) & 71.0 (1.7) & 60.8 (0.1) & 73.5 (1.1) \\ 
\hdashline
1 & reg & 200 & 40 & \textbf{85.6} (1.3) & 67.4 (1.6) & 63.0 (0.3) & 75.8 (0.9) \\ 
1 & reg & 200 & 80 & \textbf{89.9} (1.6) & 45.4 (1.0) & 62.1 (0.2) & 78.8 (1.4) \\ 
1 & reg & 1000 & 40 & \textbf{87.9} (1.2) & 60.0 (2.1) & 60.3 (0.0) & 71.2 (1.3) \\ 
1 & reg & 1000 & 80 & \textbf{86.5} (1.5) & 69.2 (1.9) & 60.8 (0.1) & 75.0 (1.2) \\ 
\hdashline
3 & irr & 200 & 40 & \textbf{87.7} (1.8) & 68.1 (2.4) & 62.2 (0.4) & 40.3 (0.1) \\ 
3 & irr & 200 & 80 & \textbf{79.3} (2.5) & 40.4 (0.1) & 61.4 (0.3) & 40.1 (0.0) \\ 
3 & irr & 1000 & 40 & \textbf{96.9} (0.3) & 61.5 (2.2) & 59.9 (0.1) & 40.3 (0.0) \\ 
3 & irr & 1000 & 80 & \textbf{96.9} (0.5) & 67.5 (2.3) & 60.4 (0.1) & 40.3 (0.0) \\ 
\hdashline
3 & reg & 200 & 40 & \textbf{88.8} (1.6) & 71.7 (2.1) & 62.1 (0.3) & 40.3 (0.1) \\ 
3 & reg & 200 & 80 & \textbf{81.2} (2.6) & 41.0 (0.4) & 61.2 (0.2) & 40.2 (0.0) \\ 
3 & reg & 1000 & 40 & \textbf{96.5} (0.7) & 63.0 (2.3) & 60.1 (0.1) & 40.3 (0.0) \\ 
3 & reg & 1000 & 80 & \textbf{97.1} (0.2) & 65.6 (2.3) & 60.6 (0.1) & 40.3 (0.0) \\
\hline
\end{tabular}
\caption{Results of the selection of active genes. The average accuracy (multiplied by 100) across 50 Monte Carlo iterations is presented for each of the three competing methods under 16 scenarios determined by 4 factors. The standard error of the average is in parentheses. $N$ is the number of simulated tissue samples, $\theta$ refers to the cell type compositions (irregular or regular, abbreviated as `irr' or `reg', respectively), $P$ is the number of genes in total, and $P_d$ is the number of differentially expressed genes. Our proposed algorithm is denoted as BayesClint, whereas the competing methods are modelGeneVar and correlateGenes. 
The values of the best performances are in bold type.} \label{tab:sim_results_active_acc}
\end{table}

\begin{table}[]
\centering
\begin{tabular}{llllllll}
\hline
$N$ & $\theta$ & $P$ & $P_d$ & BayesClint & SPC & SPARK-X & BASS+DESeq2 \\ 
\hline
1 & irr & 200 & 40 & \textbf{94.6} (1.2) & 56.3 (1.9) & 6.4 (0.2) & 31.6 (2.3) \\ 
1 & irr & 200 & 80 & \textbf{89.9} (1.3) & 45.2 (0.8) & 11.3 (0.3) & 36.8 (2.0) \\ 
1 & irr & 1000 & 40 & \textbf{98} (1.3) & 54.4 (5.1) & 1.5 (0.1) & 8.9 (2.4) \\ 
1 & irr & 1000 & 80 & \textbf{94.1} (2.3) & 66.8 (3.1) & 2.9 (0.1) & 12.5 (2.7) \\ 
\hdashline
1 & reg & 200 & 40 & \textbf{94.3} (0.9) & 55.8 (1.5) & 1.3 (0.1) & 33.0 (2.4) \\ 
1 & reg & 200 & 80 & \textbf{91} (1.5) & 45.4 (1.0) & 2.5 (0.2) & 34.1 (1.7) \\ 
1 & reg & 1000 & 40 & \textbf{98.1} (1) & 55.5 (5.5) & 0.2 (0.0) & 4.5 (0.8) \\ 
1 & reg & 1000 & 80 & \textbf{97.9} (0.8) & 64.1 (3.7) & 0.5 (0.0) & 8.5 (1.6) \\ 
\hdashline
3 & irr & 200 & 40 & \textbf{89} (2) & 52.6 (2.4) & 9.1 (0.4) & 49.0 (2.9) \\ 
3 & irr & 200 & 80 & \textbf{80.5} (2.5) & 40.4 (0.1) & 15.3 (0.5) & 51.4 (1.2) \\ 
3 & irr & 1000 & 40 & \textbf{99.6} (0) & 53.0 (5.3) & 2.1 (0.1) & 11.2 (2.0) \\ 
3 & irr & 1000 & 80 & \textbf{98.2} (0.8) & 48.9 (3.6) & 3.8 (0.1) & 15.7 (1.2) \\ 
\hdashline
3 & reg & 200 & 40 & \textbf{88.6} (1.7) & 56.0 (1.9) & 0.7 (0.1) & 48.5 (2.7) \\ 
3 & reg & 200 & 80 & \textbf{82.5} (2.6) & 41.0 (0.4) & 1.6 (0.2) & 51.4 (1.2) \\ 
3 & reg & 1000 & 40 & \textbf{99.6} (0) & 54.4 (5.2) & 0.2 (0.0) & 12.3 (2.5) \\ 
3 & reg & 1000 & 80 & \textbf{96} (1.4) & 44.6 (3.7) & 0.4 (0.0) & 16.7 (1.8) \\
\hline
\end{tabular}
\caption{Results of the selection of differentially expressed genes. The average accuracy (multiplied by 100) across 50 Monte Carlo iterations is presented for each of the three competing methods under 16 scenarios determined by 4 factors. The standard error of the average is in parentheses. $N$ is the number of simulated tissue samples, $\theta$ refers to the cell type compositions (irregular or regular, abbreviated as `irr' or `reg', respectively), $P$ is the number of genes in total, and $P_d$ is the number of differentially expressed genes. Our proposed algorithm is denoted as BayesClint, whereas the competing methods are SPARK-X and BASS+DESeq2. 
The values of the best performances are in bold type.} \label{tab:sim_results_deg_acc}
\end{table}

This simulation study demonstrates that BayesClint is competitive with cutting-edge algorithms for cell type or spatial domain clustering in the worst case, and performs better than all algorithms in many cases. Additionally, BayesClint simultaneously performs very well on both kinds of gene selection.

\section{Real Data Analysis}
\label{sec:real_data_analysis}

We evaluate the performance of BayesClint with two STARmap datasets that focus on different regions of the mouse brain \citep{wang_three-dimensional_2018}. These datasets are provided at the cellular resolution and serve as benchmarks due to their established ground-truth cell types and spatial domains that allow us to demonstrate the multi-scale clustering capability of BayesClint. These datasets also offer complementary modeling challenges. The first dataset comprises a single tissue sample with nearly 1,000 gene expression measurements. This provides a high-dimensional setting that allows us to demonstrate the advantages of conducting dimension reduction and feature selection jointly with clustering, as done in BayesClint, in contrast with competing approaches. The second dataset includes three tissue samples but a more targeted panel of fewer than 200 genes. Thus, this dataset will demonstrate the necessity of multi-sample clustering, and allow us to evaluate our model’s capacity for joint inference and data integration across heterogeneous replicates. Together, these applications demonstrate the versatility of BayesClint across varying genomic scales and experimental designs.

\subsection{Evaluation and visualization of clustering results}
\label{sec:eval_and_viz}

For both STARmap datasets, we compare the performance of BayesClint against that of competing clustering methods in terms of spatial domain clustering, cell type clustering, and estimation of the cell type compositions of the spatial domains. For the clustering results, we evaluate the performance by calculating the ARI between each method's estimated partition of spatial domains or cell types, respectively, and the ground truth. 

Recall that the BayesClint algorithm assigns numerical cluster labels $\zeta_{mi}, \kappa_{mi}$ to each cell to partition them into cell types and spatial domains, respectively. To correspond the numerical cluster labels to the ground-truth cell type and spatial domain names, we determine the assignment of cell type and spatial domain names to the cluster indices $\zeta_{mi}, \kappa_{mi}$, respectively, that lead to the most alignment between the estimated labels and the ground truth. Then, we can use these cell type and spatial domain names in the visualization of results and for comparing the cell type compositions estimated according to Section \ref{sec:estimation_cluster_labels}, $\hat{\theta}_{ck}$, with the ground truth, $\theta_{ck}$.

We visualize the spatial domain clustering by plotting the spatial locations of the cells and coloring them according to their estimated spatial domain. We visualize the cell type clustering by scaling the log-normalized data, conducting PCA, and then using uniform manifold approximation and projection (UMAP) \citep{mcinnes_umap_2018} on the top 30 principal components to acquire 2-dimensional embeddings, which we then plot. We visualize the cell type composition $\theta_{ck}$ of each spatial domain via stacked barplots, and compare the estimated compositions with the ground truth in terms of the root mean squared error: RMSE = $\sqrt{\frac{1}{C K} \sum_{c = 1}^C \sum_{k = 1}^K (\hat{\theta}_{ck} - \theta_{ck})^2}$.

\subsection{Application to the mouse visual cortex STARmap data}
\label{sec:real_dat_mvc}

We first examined a publicly available STARmap dataset from a study of the mouse visual cortex (MVC) \citep{wang_three-dimensional_2018}. The MVC is an important cortical region of study, because it serves as a useful model to explore how sensory inputs are transformed into goal-directed perceptions and actions \citep{glickfeld_mouse_2014}. This dataset contained 1020 genes across 1207 cells, corresponding to 15 cell types and 7 layers, as manually annotated in \cite{wang_three-dimensional_2018}. We filtered out genes for which 90\% of cells or more had a zero read count \citep{li_bayesian_2021}. We then retained only the cells that had at least a 100 count for all genes combined \citep{ma_spatially_2022}; all of the cells met this criteria, so no cells were removed. After this quality control procedure, we had 886 gene measurements for the 1207 cells.
Next, we conducted library size normalization by dividing each gene count by the total gene count for the corresponding cell, multiplying it by 10,000, adding a pseudocount of 1, and transforming it to a natural log scale \citep{satija_spatial_2015}. Then, we centered and scaled each normalized gene count to have a mean of 0 and a standard deviation of 1. Based on the scree plot of the resulting correlation matrix of the data, we choose $r = 9$ as the maximum number of active components (Web Figure \ref{fig:mvc_screeplot}). 

In Web Appendix Section \ref{sec:CorK_mvc}, we apply the heuristic described in Section \ref{sec:estimation_cluster_labels} for selecting the number of cell types and spatial domains to the MVC dataset. While the heuristic identifies $C = 11$ cell types and $K = 4$ spatial domains, the ground-truth annotations for this dataset are established as 15 cell types and 7 spatial domains. To ensure a direct and fair comparison with competing methods and to validate the model's accuracy against known benchmarks, we fix $C$ and $K$ to these ground-truth values for the primary analysis.

We examined the convergence of our algorithm on this dataset. After running 4 MCMC chains, each with 6,500 burn-in and 15,000 MCMC iterations, we examined the pairwise correlation coefficients between the \textrm{PPI}s of all 886 genes. These indicated good concordance between the four chains, with all pairwise Pearson correlation coefficients falling in the range [0.938, 0.998] (see Web Figure \ref{fig:mvc_corrplot_4_chains}). After checking the convergence, we continue the analysis with one of the chains, containing 15,000 MCMC iterations.

\subsubsection{BayesClint achieves the highest spatial domain clustering performance on the MVC dataset and reveals a set of highly differentiating genes}
\label{sec:mvc_clust_ft_select_results}

\begin{figure}
    \centering
    \subfloat[MVC, Spatial Domains\label{fig:mvc_spatial_domains_ARIs}]{\includegraphics[width=0.5\textwidth]{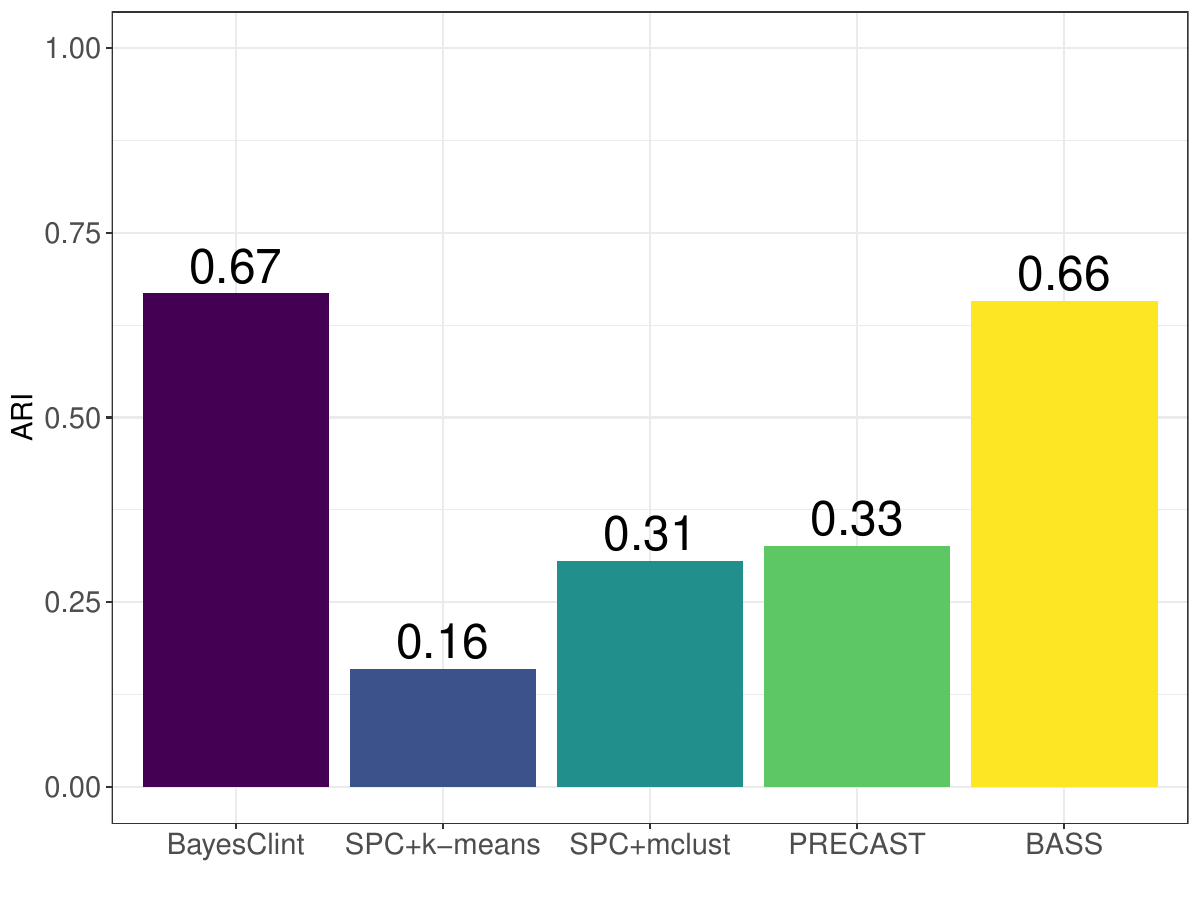}}
    \subfloat[MVC, Cell Types\label{fig:mvc_cell_types_ARIs}]{\includegraphics[width=0.5\textwidth]{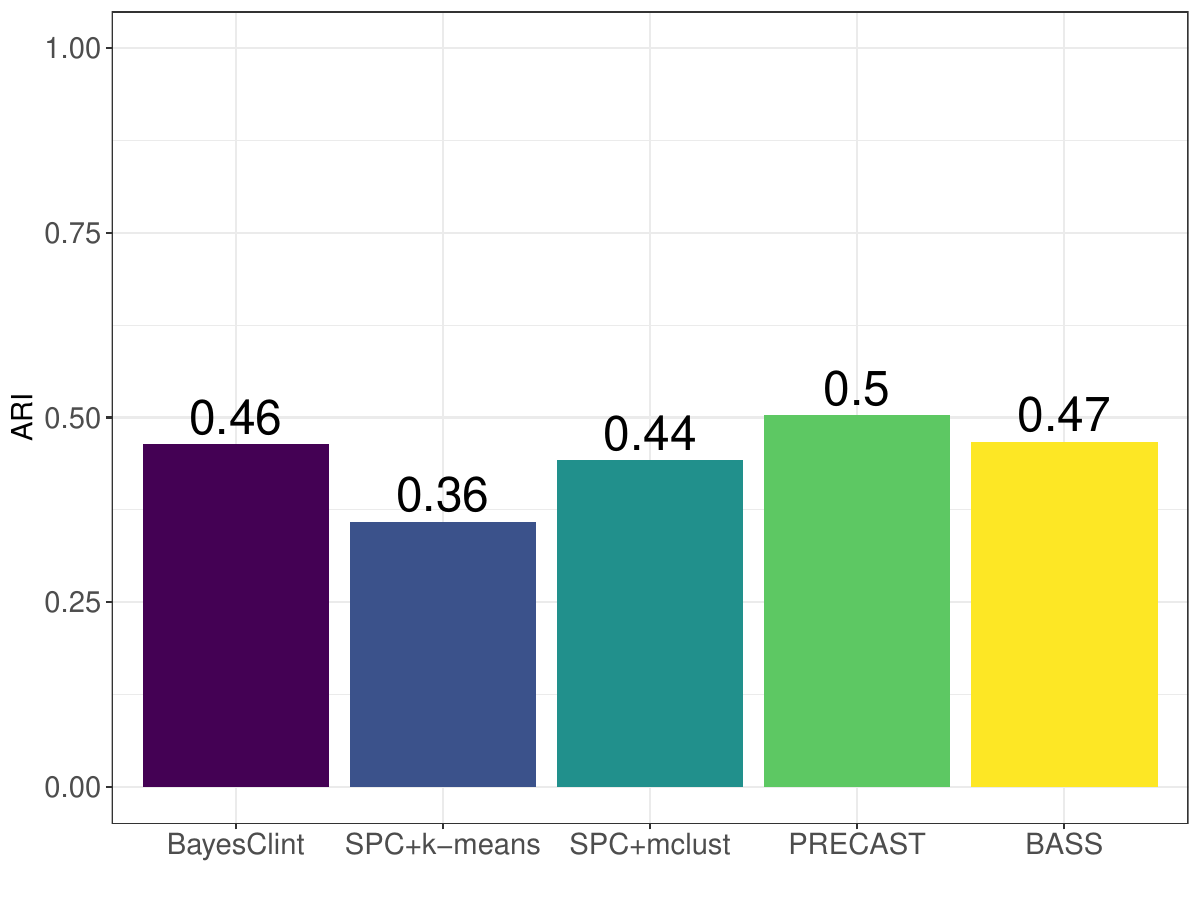}}

    \centering
    \subfloat[mPFC, Spatial Domains\label{fig:mpfc_spatial_domains_ARIs}]{\includegraphics[width=0.5\textwidth]{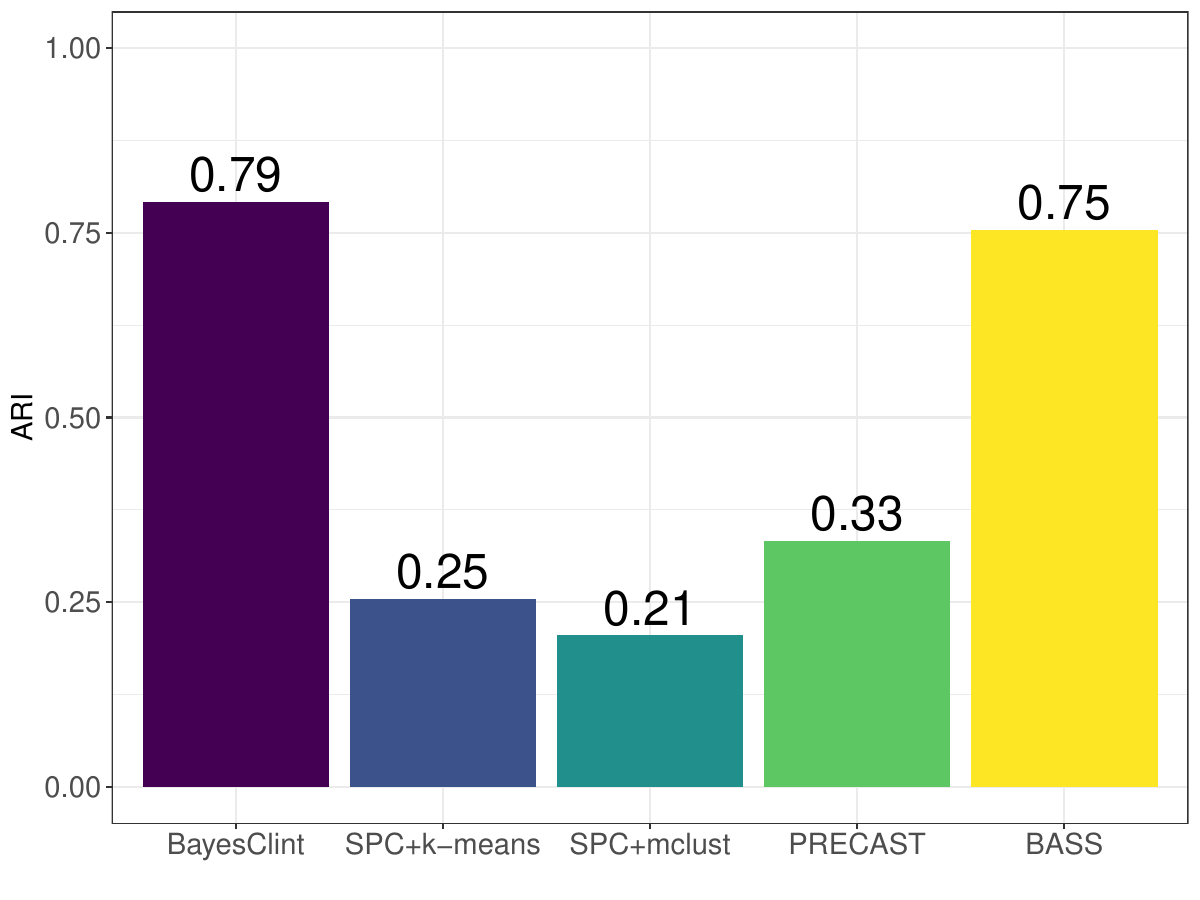}}
    \subfloat[mPFC, Cell Types\label{fig:mpfc_cell_types_ARIs}]{\includegraphics[width=0.5\textwidth]{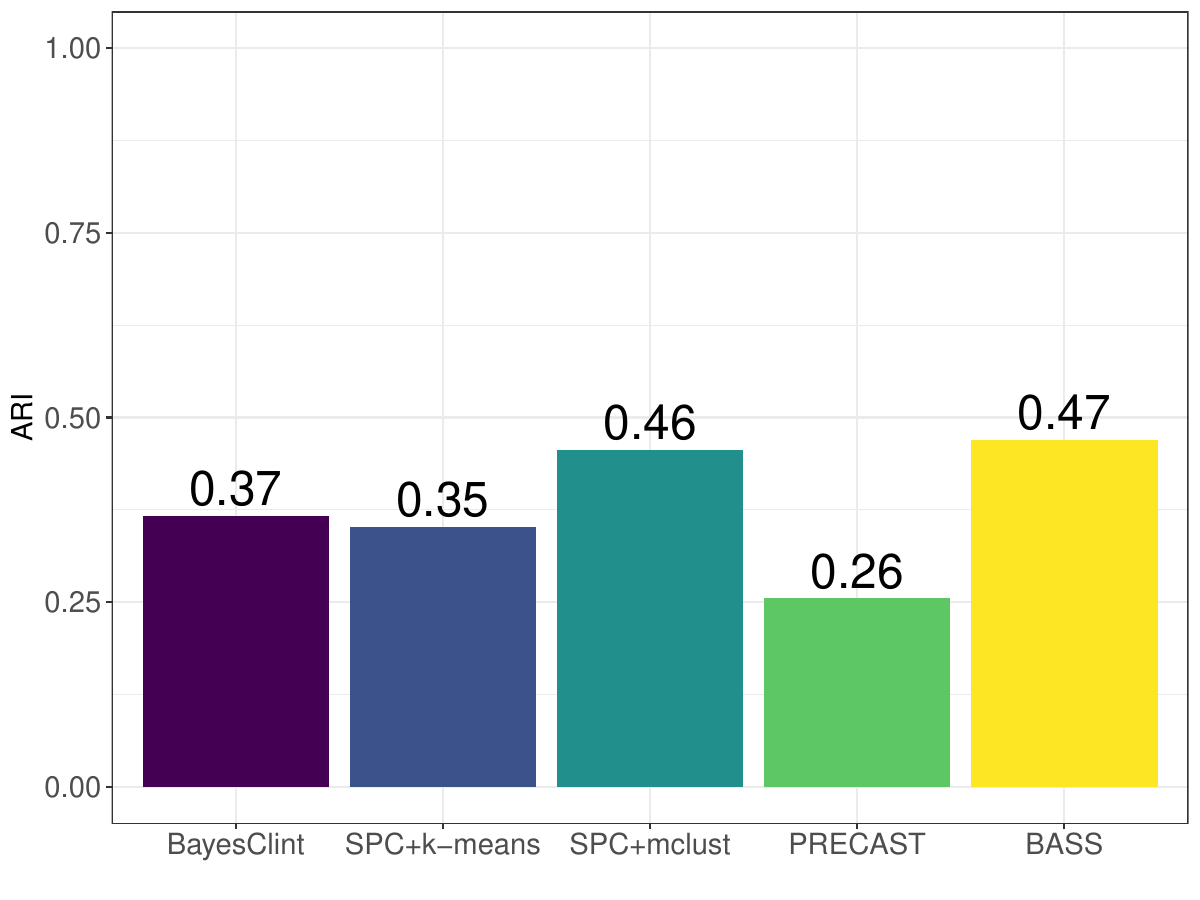}}
    
    \caption{ARIs for the five clustering methods on the MVC dataset, for a) spatial domain clustering and b) cell type clustering. ARIs for the five clustering methods on the mPFC dataset, for c) spatial domain clustering and d) cell type clustering.}
    \label{fig:mvc_mpfc_ARIs}
\end{figure}

We evaluate and visualize the results of BayesClint and competing algorithms on this dataset according to Section \ref{sec:eval_and_viz}. Web Figure \ref{fig:mvc_ground_truths} shows the ground truths for the three kinds of estimation. For spatial domain clustering, BayesClint and BASS yielded comparable ARI of 0.67 and 0.66, respectively, significantly outperforming the other three competing algorithms (Figure \ref{fig:mvc_spatial_domains_ARIs} and Web Figures \ref{fig:mvc_spatial_domains_clustering_results_bayesclint} and \ref{fig:mvc_spatial_domains_clustering_results}). Both methods failed to assign cells to the HPC spatial domain, suggesting that the HPC and adjacent `cc' domains are statistically indistinguishable within this clustering framework. For cell type clustering, PRECAST achieved the highest ARI (0.5), with BayesClint demonstrating competitive performance at 0.46 (Figure \ref{fig:mvc_cell_types_ARIs} and Web Figure \ref{fig:mvc_cell_types_clustering_results}). Lastly, the estimation accuracy of the cell type compositions between BayesClint and BASS (Web Figure \ref{fig:mvc_cell_type_composition_results}) was nearly equivalent, yielding marginal differences in the RMSE at 0.0916 and 0.0895, respectively.

A key advantage of BayesClint over existing frameworks, such as BASS, is its integrated feature selection mechanism. While BayesClint identified all 886 genes as active in the MVC dataset (Web Figure \ref{fig:mvc_bayesclint_active}), our two-stage selection process utilizing the ACAT procedure detailed in Section \ref{sec:select_active_differentiating} facilitates a more granular interpretation. Specifically, we identified a parsimonious subset of 86 genes that drive the primary variation across estimated cell types (Web Figure \ref{fig:mvc_bayesclint_differentiating}) in the top-ranked component (i.e., the component with the minimum combined $p$-value). This sparse representation reveals critical biological signals that are otherwise masked in high-dimensional space. We observed significant upregulation within the Reln, Oligo, and Smc cell types, as shown in the top ten rows of Web Figure \ref{fig:mvc_bayesclint_differentiating}. Such localized expression patterns are clearly discernible in the reduced gene set but become obscured when analyzing the full gene panel, as seen by comparing Web Figures \ref{fig:mvc_bayesclint_active} and \ref{fig:mvc_bayesclint_differentiating}.

\subsection{Application to the mouse medial prefrontal cortex STARmap data}
\label{sec:real_dat_mpfc}

To illustrate BayesClint's performance in integrating multiple tissue samples, we examined a publicly available STARmap dataset that consists of three tissue samples obtained from the medial prefrontal cortex (mPFC) of the brain from different mice \citep{wang_three-dimensional_2018}. The mPFC is a crucial cortical region due to its role in high-level cognitive functions like decision-making, memory, attention, and emotion \citep{carlen_what_2017}. This dataset originally contained 166 genes and 1,049, 1,053, and 1,088 cells for each of the tissue samples, respectively. Among the 166 genes, 112 of them are putative cell type markers and 48 of them are activity-regulated genes. The cells were manually annotated into 15 cell types and 4 spatial domains by \cite{wang_three-dimensional_2018} and \cite{li_bass_2022}, respectively. We used the same quality control procedure as described in Section \ref{sec:real_dat_mvc} on each tissue sample individually, and retained only the genes for which less than 90\% of cells had a zero read count (for each tissue sample) and only the cells which had at least a 100 count for all genes combined. This resulted in a reduction of the number of genes to 114 across tissue samples, but no cells were removed. We then performed library size normalization and scaling as described in Section \ref{sec:real_dat_mvc}. In particular, we centered and scaled each normalized gene count to have a mean of 0 and a standard deviation of 1 within each tissue sample, for each of the three samples. Based on the scree plot of the resulting correlation matrix of the data, we choose $r = 9$ as the maximum number of active components (Web Figure \ref{fig:mpfc_screeplot}). Since three tissue samples may have batch-specific systemic variation, we perform batch-effect removal across the three samples via Seurat v3 \citep{stuart_comprehensive_2019}. This batch effect-corrected dataset is used for BayesClint and all competing methods except for DESeq2 (this method requires the raw gene counts, and will treat the tissue sample index as a blocking factor). 

In Web Appendix Section \ref{sec:CorK_mpfc}, we apply the heuristic proposed in Section \ref{sec:estimation_cluster_labels} for selecting the number of cell types and spatial domains to the mPFC dataset. Using the heuristic, we selected $C = 11$ cell types and $K = 4$ spatial domains. However, because the number of cell types and the number of spatial domains in the ground truth are known (15 cell types and 4 spatial domains, respectively), we fix the number of cell types $C$ and the number of spatial domains $K$ in the BayesClint model to those values.

We examined the convergence of our algorithm on this dataset. After running 4 MCMC chains, each with 6,500 burn-in and 15,000 MCMC iterations, we examined the pairwise correlation coefficients between the \textrm{PPI}s of all 114 genes. These indicated good concordance between the four chains, with all pairwise Pearson correlation coefficients falling in the range [0.894, 0.972] (see Web Figure \ref{fig:mpfc_corrplot_4_chains}). After checking the convergence, we continue the analysis with one of the chains, containing 15,000 MCMC iterations.

\subsubsection{BayesClint achieves the highest spatial domain clustering performance on the mPFC dataset and reveals a set of highly differentiating genes}
\label{sec:mpfc_clust_ft_select_results}

We evaluate and visualize the results of BayesClint and competing algorithms on this dataset according to Section \ref{sec:eval_and_viz}. Figure \ref{fig:mpfc_ground_truths} shows the ground truths for the three kinds of estimation. For spatial domain clustering, BayesClint and BASS both have high ARIs (0.79 and 0.75, respectively), with BayesClint having the highest one, and both methods are significantly better than the other three competing algorithms (Figure \ref{fig:mpfc_spatial_domains_ARIs} and Web Figures \ref{fig:mpfc_spatial_domains_clustering_results_bayesclint} and \ref{fig:mpfc_spatial_domains_clustering_results}). For cell type clustering, BASS had the highest ARI (0.47), but BayesClint is not far behind with an ARI of 0.37 (Figure \ref{fig:mpfc_cell_types_ARIs} and Web Figure \ref{fig:mpfc_cell_types_clustering_results}). Lastly, the estimation accuracy of the cell type compositions between BayesClint and BASS (Web Figure \ref{fig:mpfc_cell_type_composition_results}) is comparable, with RMSEs of 0.0634 and 0.0685, respectively.

As mentioned earlier, a distinct advantage of BayesClint over existing methods is its integrated gene selection framework. Applied to the mPFC dataset, BayesClint identified 112 active genes (Web Figure \ref{fig:mpfc_bayesclint_active}). By employing the ACAT procedure (Section \ref{sec:select_active_differentiating}), we further isolated a subset of 79 genes associated with the top-ranked component (i.e., the component with the minimum combined $p$-value). These genes represent the most potent drivers of cell type differentiation within the estimated spatial architecture (Web Figure \ref{fig:mpfc_bayesclint_differentiating}). Notably, this statistically selected subset includes well-characterized markers such as Gad1, Mog, Egr2, and Slc17a7, all of which exhibit pronounced differential expression patterns across the identified cell type clusters.

\section{Discussion}
\label{sec:discussion}

We have presented a Bayesian hierarchical modeling framework, BayesClint, which builds upon cutting-edge SRT clustering methods in the literature and overcomes their limitations. Not only can BayesClint perform dimension reduction across multiple tissue samples and jointly perform cell type and spatial domain clustering, but it can also perform nested feature selection of active and differentiating genes. BayesClint does all of these tasks simultaneously through a Bayesian factor analysis model, where we posit a multi-scale mixture model for the factors, and enforce sparsity on the factor loadings matrix to select features. 

We have validated our algorithm through a comprehensive simulation study and application to two STARmap datasets of the mouse brain. In particular, we have shown that BayesClint is comparable with the cutting-edge clustering method BASS in the simulation study, has significantly higher accuracies than the gene selection methods for both active and differentiating gene selection in the simulation study, and is comparable with BASS in terms of spatial domain clustering in the STARmap datasets. A limitation of our method is that we fix the number of cell types and spatial domains a priori; we propose a heuristic for selecting the number of cell types and spatial domains (Section \ref{sec:estimation_cluster_labels}). There are a couple of other ways to overcome this limitation, e.g., by proposing a model selection criterion, or estimating the number of clusters by a Dirichlet process mixture model \citep{muller_bayesian_2015,li_bayesian_2017}. Another limitation of our method is the long computation time typical of MCMC algorithms, particularly when there is a high count of clusters ($C$ or $K$), genes, or cells. An effective strategy for decreasing this runtime is to run shorter MCMC chains in parallel. 

Several extensions of our model are worth investigating. Many current SRT datasets not only contain gene expression and spatial information, but also paired histology or pathology images \citep{hu_spagcn_2021}, or scRNA-seq reference data measured on the same tissue with cell type-specific gene expression information \citep{ma_accurate_2024}. BayesClint could be extended to incorporate such information to improve clustering and gene selection performance. In addition, many multi-omics experiments are beginning to incorporate temporal or spatial information \citep{velten_identifying_2022}, serving as motivation for extending BayesClint to multi-omics data.

\section{Data Availability} 
\label{sec:data_availability}

MVC RData was obtained from \citep{li_interpretable_2024}. mPFC RData was obtained from \citep{li_bass_2022}.

\section{Computer Specifications}

The main simulation study with the 16 scenarios was run with the high performance computing cluster at the Minnesota Supercomputing Institute. The sensitivity analyses and real data analyses were run on a MacBook Pro with 32 GB memory, 2.4 BHz 8-Core Intel Core i9 CPU, Radeon Pro 555X 4GB Intel UHD GPU, 250GB hard drive, and macOS Sequoia 15.7.4. The R version 4.4.1 (2024-06-14) -- ``Race for Your Life'' was used on the MacBook Pro.

\section{Software}
\label{sec:software}

We have created an R package called BayesClint for implementing the methods. It also contains a function used to generate the simulated data, the STARmap mPFC dataset analyzed in this manuscript, and a subsetted version of the STARmap mPFC dataset to serve as a quick example for the method. Its source R and C++ codes are available in the GitHub repository https://github.com/AlvinSheng/BayesClint.

\section{Supplementary Material}
\label{sec:supplementary_materials}

Supplementary material is available online at
\url{arXiv.org}.

\section*{Acknowledgments and Funding}

Thierry Chekouo was supported by a National Institutes of Health (NIH)  grant: 1R35GM150537. Sandra E. Safo was also supported by NIH NIGMS grant \#R35GM142695. Thierry Chekouo thanks Medtronic Inc. for their support in the form of a faculty fellowship. The views expressed in this manuscript are those of the authors and not the funders.

{\it Conflict of Interest}: None declared.

\bibliographystyle{unsrtnat}
\bibliography{references}

\end{document}


\title{BayesClint: Bayesian multi-scale clustering and multi-sample integration with feature selection for spatial transcriptomics data Supplementary Materials}
\date{}

\author{Alvin Sheng$^1$, Sandra E. Safo$^{1}$ and Thierry Chekouo$^{1,2}$ }

\maketitle  

$^1$Division of Biostatistics and Health Data Science,
University of Minnesota,
2221 University Avenue SE,
55414,
MN,
USA

$^2$E-mail address for correspondence: tchekouo@umn.edu 

\begin{abstract}
{These supplementary materials contain all of the Web Appendix Sections, Web Tables, and Web Figures referenced in the main article. If this appendix refers to a Section, Figure, or Table from the main article, it will be clearly specified with the phrase `the main article.'} 
\end{abstract}

\section{MCMC Implementation}
\label{sec:mcmc_impl}

We implement an MCMC algorithm for posterior inference that employs a partially collapsed Gibbs sampler \citep{van_dyk_partially_2008} to sample the binary selection indicators $\gamma_{lj}$ and the loadings matrix $\bA$. We use Gibbs steps in combination with Metropolis-Hastings steps to sample the other parameters. 

We present in the below subsections the conditional distributions and candidate distributions used to update the parameters during our MCMC algorithm. The following subsections are arranged in the order that each parameter or set of parameters is sampled. We update all parameters for a number of iterations (e.g., 65,000 MCMC iterations) until the samples have converged to a stationary distribution. 

\begin{figure}
    \centering
    \includegraphics[width=\linewidth]{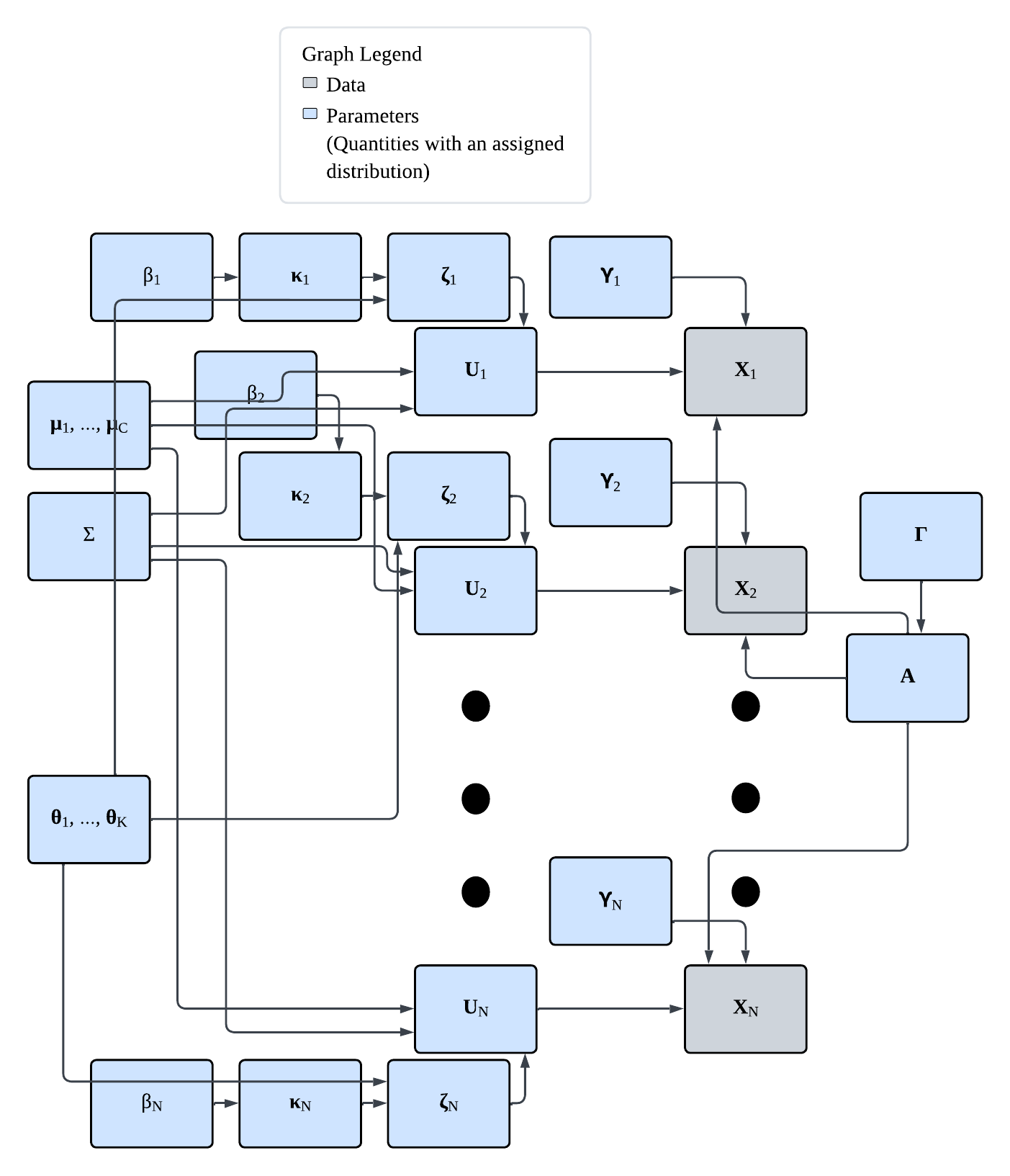}
    \caption{This figure represents the proposed Bayesian hierarchical model as a directed acyclic graph (DAG; \citealp{reich_bayesian_2019}). In this DAG, the nodes represent the observations and parameters as gray and blue boxes, respectively, and the directed edges denote the conditional dependencies described in Section 2 of the main article. If the conditional distribution for the observations or a given set of parameters depends on a second set of parameters, then there will be an arrow(s) from the second set of parameters to the observations or the given set of parameters. Hyperparameters that are fixed a priori instead of being assigned a prior distribution are omitted in the above DAG.}
    \label{fig:BayesClint_DAG}
\end{figure}

See Figure \ref{fig:BayesClint_DAG} for a graphical representation of the BayesClint model as computed by our MCMC implementation. See Section \ref{sec:list_of_notation} for a list of notation used in the main article and supplementary materials.

\subsection{Concatenated Factor Model}
\label{sec:concat_factor_model}

To sample the feature selection indicators $\gamma_{lj}$ (discussed in Section \ref{sec:gamma_eta_sampling}) and other parameters, we first concatenate the matrices in the factor model across tissue samples $m = 1, \ldots, N$ as follows. 

Let $\bX = (\bX_1^\intercal, \ldots, \bX_N^\intercal)^\intercal$ and $\bE = (\bE_1^\intercal, \ldots, \bE_N^\intercal)^\intercal$ be integrated matrices of dimension $n \times P$, where $n = \sum_{m = 1}^N n_m$. Let $\bU = (\bU_1^\intercal, \ldots, \bU_N^\intercal)^\intercal$ be integrated matrices of dimension $n \times r$. 
Then, we have that 

\begin{align} \label{eq:cat_factor_model}
    \bX = \bU \bA + \bE.
\end{align}

The integrated residual matrix $\bE$ would have the distribution

\begin{align}
    \textrm{vec}(\bE^\intercal) \sim \mathcal{N}(\bzero, \bUpsilon),
\end{align}

\noindent where $\bUpsilon$ is a block diagonal matrix with $\Upsilon_m, m = 1, \ldots, N$ on the diagonals.

\subsection{Sampling the binary selection indicators}
\label{sec:gamma_eta_sampling}

\subsubsection{Incorporating component selection indicators $\rho_l$}

We first discuss a more general situation where we have component selection as well as the selection of specific loadings. This approach has been used in \cite{chekouo_bayesian_2022}. In this approach, we first select the important components with $\rho_l, l = 1, \ldots, r$ (corresponding to the rows of $\bA$) and then given the component, we select the important genes $\gamma_{lj}, j = 1, \ldots, P$ (corresponding to the columns of $\bA$). The prior distribution for the component indicator, $\rho_l$, is the Bernoulli distribution with parameter $q$. In turn, $q$ has the hyperprior Beta$(a,b)$. By default, $a$ and $b$ are set to 1.

In the $l^{th}$ component, the prior distribution of $\gamma_{lj}$ is dependent on the indicator $\rho_l$ and is defined as

\begin{align}
    \gamma_{lj} \mid \rho_l \sim (1 - \rho_l) \delta_0 + \rho_l \textrm{Bernoulli}(q_\gamma),
\end{align}

\noindent where $\textrm{Bernoulli}(q_\gamma)$ is the Bernoulli distribution with probability $q_\gamma$ and $\delta_0$ is the mass distribution concentrated at zero. To infer a sparse matrix $\bA$, i.e., whether the $a_{lj}$ entry in $\bA$ is important or not, we impose the following sparsity-inducing prior on the matrix elements given the indicators $\gamma_{lj}$ and $\rho_l$:

\begin{align}
    a_{lj} \mid \rho_l, \gamma_{lj} \sim (1 - \rho_l \gamma_{lj}) \delta_0 + \rho_l \gamma_{lj} \mathcal{N}(0, 1), 
\end{align}

\noindent where $\delta_{0}$ is the mass distribution concentrated at $0$. 
We set $q_\gamma$ to 0.05, the prior probability to select features for a specific component, as we expect that a small number of features are active in every component $l$.

\subsubsection{Metropolis-Hastings step in the general case, with component selection}

Equation \ref{eq:cat_factor_model} can also be written as 

\begin{align} \label{vec_factor_model}
    (\bX)_{\cdot j} = \bU_{\cdot \bgamma_j} \ba_{\bgamma_j} + (\bE)_{\cdot j}, j = 1, \ldots, P,
\end{align}

\noindent where $\bgamma_j$ is the $r$-length vector $(\gamma_{1j}, \ldots, \gamma_{rj})^\intercal$, $\ba_{\bgamma_j}$ is the $j^{th}$-column sub-vector of $\bA$ with $\gamma_{lj} = 1$, and $\bU_{\cdot \bgamma_j}$ is a sub-matrix of $\bU$ with columns $l$ corresponding to $\gamma_{lj} = 1$. The error term $(\bE)_{\cdot j}$ has the distribution $\mathcal{N}(\bzero, \mathcal{U}_j)$, where $\mathcal{U}_j$ is a block diagonal matrix with $\upsilon^2_{1j} I_{n_1}, \ldots, \upsilon^2_{Nj} I_{n_N}$ on the diagonals. Then, we transform Equation \ref{vec_factor_model} by pre-multiplying $\mathcal{U}_j^{-1/2}$:

\begin{align} \label{trans_vec_factor_model}
    \mathcal{U}_j^{-1/2} (\bX)_{\cdot j} = \mathcal{U}_j^{-1/2} \left(\bU_{\cdot \bgamma_j}\right) \ba_{\bgamma_j} + \mathcal{U}_j^{-1/2} (\bE)_{\cdot j}, j = 1, \ldots, P.
\end{align}

\noindent Let $(\bX)_{\cdot j}^* = \mathcal{U}_j^{-1/2} (\bX)_{\cdot j}$. Then, after integrating $\ba_{\bgamma_j}$, we have the transformed probability distribution function

\begin{align} \label{collapsed_factor_likel}
    p((\bX)_{\cdot j}^* \mid \bU_{\cdot \bgamma_j}, \bgamma_{j}, \mathcal{U}_j) = \nonumber \\
    \mathcal{N}((\bX)_{\cdot j}^*; \bzero, \mathcal{V}_j),
\end{align}

\noindent where $\mathcal{V}_j = \mathcal{U}_j^{-1/2} \left(\bU_{\cdot \bgamma_j} \right) \left(\bU_{\cdot \bgamma_j} \right)^\intercal \mathcal{U}_j^{-1/2} + I_n$. We have then,

\begin{align} \label{eq:gamma_rho_likelprior}
p(\brho, \bGamma \mid \bX, \bU, \bUpsilon) &\propto \prod_{j=1}^{P} p((\bX)^*_{\cdot j} \mid \bU_{\cdot \bgamma_j}, \bgamma_{j}, \mathcal{U}_j) p(\bgamma_{j} \mid \brho) p(\brho) \nonumber \\
&\propto \prod_{j=1}^{P} \mathcal{N}((\bX)_{\cdot j}^*; \bzero, \mathcal{V}_j) \prod_{l=1}^{r} p(\gamma_{lj} \mid \rho_l) p(\rho_l) \nonumber \\
&\quad = \prod_{j=1}^{P} G_j(\brho, \bgamma_{j}) \prod_{l=1}^{r} p(\rho_l),
\end{align}

\noindent where $\bGamma = (\gamma_{lj})$ and $G_j(\brho, \bgamma_{j}) = \mathcal{N}((\bX)_{\cdot j}^*; \bzero, \mathcal{V}_j) \prod_{l=1}^{r} p(\gamma_{lj} \mid \rho_l)$. To sample $(\brho, \bGamma)$ from Equation \ref{eq:gamma_rho_likelprior}, we adopt a Metropolis-Hastings step. The proposal distribution will be defined as

\begin{align} \label{eq:gamma_rho_proposal_distr}
    Q(\rho_l', \bgamma_{l \cdot}' \mid \rho_l, \bgamma_{l \cdot}) = \rho_l \delta_0(\rho_l') \delta_{\bzero}(\bgamma_{l \cdot}') + (1 - \rho_l) \delta_1(\rho_l') p(\bgamma_{l \cdot}' \mid \bX, \bU, \bUpsilon, \rho_l' = 1)
\end{align}

\noindent where

\begin{align}
    p(\bgamma_{l \cdot}' \mid \bX, \bU, \bUpsilon, \rho_l' = 1) = \prod_{j=1}^{P} \mathcal{P}_{lj}^{\gamma_{lj}'} (1 - \mathcal{P}_{lj})^{1 - \gamma_{lj}'}, \quad \mathcal{P}_{lj} = \frac{G_j(\brho^1, \bgamma_{j}^1)}{G_j(\brho^1, \bgamma_{j}^1) + G_j(\brho^1, \bgamma_{j}^0)},
\end{align}

\noindent $\brho^1 = (\rho_1, \dots, \rho_l = 1, \dots, \rho_r)$, $\bgamma_{j}^1 = (\gamma_{1j}, \dots, \gamma_{lj} = 1 \dots, \gamma_{rj})$ and $\bgamma^0_{j} = (\gamma_{1j}, \dots, \gamma_{lj} = 0 \dots, \gamma_{rj}).$

From proposal (\ref{eq:gamma_rho_proposal_distr}), if the current component is not active, that is, if $\rho_l = 0$, we propose to activate this group by setting $\rho_l' = 1$, and sample from the conditional distribution, $p(\bgamma_{l \cdot}' \mid \bX, \bU, \bUpsilon, \rho_l' = 1)$, of the indicators for variable selection $\bgamma_{l \cdot}$. We then determine whether to accept this proposal or not by the Metropolis-Hastings acceptance-rejection rule. Conversely, if the current component is active, that is, $\rho_l = 1$, we propose to make it inactive by setting $\rho_l' = 0$ and setting all the indicators $\bgamma_{l \cdot}'$ with this component to zero. Again, we determine whether to accept this proposal or not by the acceptance-rejection rule. 

\subsubsection{Metropolis step in the case of no component selection}

In the case of no component selection, we still use the conditional distribution in Equation \ref{eq:gamma_rho_likelprior}, but with the prior probability of component selection $q$ fixed to 1. Effectively, this also fixes $\rho_l, l = 1, \ldots, r$ to 1, indicating that all components are selected. Then, the proposal distribution for each binary selection indicator $\gamma_{lj}$ then simply becomes a symmetric, deterministic flip: $Q(\gamma_{lj}' | \gamma_{lj}) \sim \gamma_{lj} \delta_0(\gamma_{lj}') + (1 - \gamma_{lj}) \delta_1(\gamma_{lj}')$. We determine whether to accept this proposal or not by the Metropolis acceptance-rejection rule.

\subsection{Sampling $q$, prior probability of component selection}

In the case where we do have component selection indicators $\rho_l$, the full conditional distribution of $q$, the prior probability of component selection, is proportional to

\begin{align}
    p(q \mid \brho) \propto q^{(\sum_{l=1}^r \rho_l + a) - 1} (1 - q)^{(r - \sum_{l=1}^r \rho_l + b) - 1}
\end{align}

\noindent This is the kernel of a $\text{Beta}(\sum_{l=1}^r \rho_l + a, r - \sum_{l=1}^r \rho_l + b)$ distribution.

\subsection{Sampling $\upsilon^2_{mj}$, variance of the independent errors $\bepsilon_{m \cdot j}$}

We easily show that the full conditional distributions of $\upsilon^2_{mj}$’s are obtained as

\begin{align}
    p(\upsilon_{mj}^2 \mid \cdot) = \textrm{Inverse-Gamma}\left(\upsilon_{mj}^2; a_0 + \frac{n_m}{2}, b_0 + \frac{1}{2} \sum_{i=1}^{n_m} (x_{mij} - \ba_{\cdot j}^\intercal \bu_{mi})^2 \right) \nonumber \\
    = \textrm{Inverse-Gamma}\left(\upsilon_{mj}^2; a_0 + \frac{n_m}{2}, b_0 + \frac{1}{2} \{(\bX_m)_{\cdot j} - \bU_{m} \ba_{\cdot j} \}^\intercal \{(\bX_m)_{\cdot j} - \bU_{m} \ba_{\cdot j} \} \right).
\end{align}

\subsection{Sampling $\bA$}

Full conditional distributions of $\ba_{\bgamma_j}$ are obtained as 

\begin{align}
    p(\ba_{\bgamma_j} \mid \cdot) &= \mathcal{N}(\ba_{\bgamma_j} ; V_a M_a, V_a) \nonumber \\
    V_a &= \left( \left(\bU_{\cdot \bgamma_j} \right)^\intercal \mathcal{U}_j^{-1} \left(\bU_{\cdot \bgamma_j} \right) + I_{n_{\gamma_j}} \right)^{-1} \nonumber \\
    M_a &= \left(\bU_{\cdot \bgamma_j} \right)^\intercal \mathcal{U}_j^{-1} (\bX)_{\cdot j} 
\end{align}

\noindent where $n_{\gamma_j}$ is the number of elements in $\ba_{\bgamma_j}$. We also have $a_{lj} = 0$ if $\gamma_{lj} = 0$.

\subsection{Sampling $\bu_{mi}$}

The full conditional distribution of $\bu_{mi}$ is obtained as 

\begin{align}
    p(\bu_{mi} \mid \cdot) &= \mathcal{N}(\bu_{mi} ; V_u M_u, V_u) \nonumber \\
    V_u &= [\bA \{\textrm{diag}(\upsilon^{-2}_{m1}, \ldots, \upsilon^{-2}_{mP})\} \bA^{\intercal} + \Sigma^{-1}]^{-1} \nonumber \\
    M_u &= \bA \{\textrm{diag}(\upsilon^{-2}_{m1}, \ldots, \upsilon^{-2}_{mP})\} \bx_{mi} + \Sigma^{-1} \bmu_{\zeta_{mi}}.
\end{align}

\subsection{Sampling cell type labels $\zeta_{mi}$}

The full conditional distribution of the cell type label takes the form of a categorical distribution, where the probability of being assigned cell type $c$ for the $i^{th}$ cell on tissue sample $m$ is given by

\begin{align*}
p\left(\zeta_{mi} = c \mid \cdot \right) 
&\propto p\left( \bu_{mi} \mid \zeta_{mi} = c, \bmu_{c}, \Sigma \right) 
p\left( \zeta_{mi} = c \mid \kappa_{mi} = k, \theta_{ck} \right) \\
&\propto \exp \left\{ -\frac{1}{2} \left( \bu_{mi} - \bmu_{c} \right)^\intercal 
\Sigma^{-1} \left( \bu_{mi} - \bmu_{c} \right) \right\} 
\cdot \theta_{ck}.
\end{align*}

\subsection{Sampling $\bmu_{c}$}

The full conditional distribution of $\bmu_{c}$ is given as

\begin{align}
    p(\bmu_{c} \mid \cdot) &= \mathcal{N}(\bmu_{c}; V_\mu M_\mu, V_\mu) \nonumber \\
V_\mu &= \left( \sum_{m=1}^{N} | T_{m}(c) | \Sigma^{-1} + (\Sigma_0)^{-1} \right)^{-1} \nonumber \\
M_\mu &= \sum_{m=1}^{N} \sum_{i \in T_{m}(c)} \Sigma^{-1} \bu_{mi} + (\Sigma_0)^{-1} \bd,
\end{align}

\noindent where $T_m(c)$ is an index set of cells with cell type label $c$ on section $m$, and $|\cdot|$ is the cardinality of the corresponding index set.

\subsection{Sampling $\Sigma$}

The full conditional distribution of the variance-covariance matrix $\Sigma$ for the factors $\bu_{mi}$ takes the following parametric form:

\begin{align}
    &p(\Sigma \mid \cdot) \\
    &\propto \prod_{m=1}^{N} \prod_{i=1}^{n_m} \mathcal{N}(\bu_{mi}; \bmu_{\zeta_{mi}}, \Sigma)
\times \text{Inverse-Wishart}(\Sigma; df_0, \Omega_0) \nonumber \\
&\propto \lvert \Sigma \rvert^{-\frac{\sum_{m=1}^N n_m}{2}} \exp \left( -\frac{1}{2} \sum_{m=1}^{N} \sum_{i=1}^{n_m} (\bu_{mi} - \bmu_{\zeta_{mi}})^\intercal \Sigma^{-1} (\bu_{mi} - \bmu_{\zeta_{mi}}) \right) \nonumber \\
&\quad
\times \lvert \Sigma \rvert^{-\frac{r + df_0 + 1}{2}} \exp \left( -\frac{1}{2} \text{trace}(\Omega_0 \Sigma^{-1}) \right) \nonumber \\
&\propto \lvert \Sigma \rvert^{-\frac{(\sum_{m=1}^N n_m + r + df_0 + 1)}{2}} \nonumber \\
&\quad \times \exp \left( -\frac{1}{2} \text{trace} \left[ \left( \sum_{m=1}^{N} \sum_{i=1}^{n_m} (\bu_{mi} - \bmu_{\zeta_{mi}}) (\bu_{mi} - \bmu_{\zeta_{mi}})^\intercal + \Omega_0 \right) \Sigma^{-1} \right] \right),
\end{align}

\noindent which is the kernel of an inverse Wishart distribution. Therefore, we have the full conditional distribution

\begin{align}
    p(\Sigma \mid \cdot) = \text{Inverse-Wishart} &(\sum_{m=1}^N n_m + df_0, \sum_{m=1}^{N} \sum_{i=1}^{n_m} (\bu_{mi} - \bmu_{\zeta_{mi}}) (\bu_{mi} - \bmu_{\zeta_{mi}})^\intercal + \Omega_0 ).    
\end{align}

\subsection{Sampling cell type compositions $\btheta_k$}

The full conditional distribution of $\btheta_k$ is obtained as

\begin{align}
    f(\btheta_k \mid \bzeta, \bkappa) \propto f(\bzeta \mid \bkappa, \btheta_k) f(\btheta_k) \propto \prod_{c = 1}^C \theta_{c, k}^{\sum_{m=1}^N |T_m(c,k)| + c_0 - 1}, 
\end{align}

\noindent where $T_m(c,k)$ is the index set of cells with cell type label $c$ and spatial domain label $k$ in the $m^{th}$ tissue sample and $|\cdot|$ is the cardinality of the corresponding index set. Therefore, the full conditional distribution of $\btheta_k$ takes the form of a Dirichlet distribution, that is,

\begin{align}
    \btheta_k \mid \bzeta, \bkappa \sim \text{Dir}\left( \sum_l |T_m(1,k)| + c_0, \dots, \sum_l |T_m(C,k)| + c_0 \right).
\end{align}

\subsection{Sampling spatial domain labels $\kappa_{mi}$}

The full conditional distribution of the spatial domain label takes the form of a categorical distribution, where the probability of being assigned spatial domain $k$ for the $i^{th}$ cell on tissue sample $m$ is given by

\begin{align}
    p(\kappa_{mi} = k \mid \cdot) \propto p(\zeta_{mi} = c \mid \kappa_{mi} = k, \theta_{ck}) p(\kappa_{mi} = k \mid \beta_m) \nonumber \\
    = \theta_{ck} \exp \left\{ \beta_m \sum_{i' \in \mathcal{N}_{mi}} \bI(\kappa_{mi'} = k) \right\},
\end{align}

\noindent where $\mathcal{N}_{mi}$ is the set of indices $i'$ such that cell $i'$ is adjacent to cell $i$ in section $m$. \\

\subsection{Sampling the spatial smoothing parameter $\beta_m$}
\label{sec:betam_sampling}

The full conditional distribution of $\beta_m, m = 1, \ldots, N$ takes the following form:

\begin{align}
\label{eq:full_condit_beta}
    f(\beta_m \mid \bkappa_m, B_m) &\propto p(\bkappa_m \mid B_m, \beta_m) p(\beta_m) \nonumber \\ &\frac{1}{d(B_m, \beta_m)} \exp \left\{\beta_m \sum_{i \sim i'} \bI(\kappa_{mi} = \kappa_{mi'}) \right\} \bI(0 \leq \beta_m \leq \beta_{max}).
\end{align}

\noindent The probability mass function for the Potts model and the prior for $\beta_m$ as shown in Equation \ref{eq:full_condit_beta} were discussed in Section 2.2 of the main article. The Metropolis update for $\beta_m$ uses a uniform candidate distribution $Q(\beta_m' \mid \beta_m) = \frac{1}{0.2} \bI(\beta_m - 0.1 \leq \beta_m' \leq \beta_m + 0.1)$. Simulated annealing is also used for this Metropolis step during the burn-in phase to encourage more exploration of the parameter space at the beginning of burn-in \citep{reich_bayesian_2019}. Then, the acceptance probability for the proposed value $\beta_m'$ is given by 

\begin{align} \label{eq:beta_acceptance_prob}
    R = \text{min} \left[ 1, \exp \left\{ \log(d(B_m, \beta_m)) - \log(d(B_m, \beta_m')) + (\beta_m' - \beta_m) \sum_{i \sim i'} \bI(\kappa_{mi} = \kappa_{mi'}) \right\} \bI(0 \leq \beta_m' \leq \beta_{max}) \right].
\end{align}

\subsubsection{Evaluating the normalizing constant for the Potts model}
\label{sec:eval_d}

The Metropolis update step involves the major computational challenge of evaluating the log of the normalizing constant, $\log\{d(B_m, \beta_m)\}$. Therefore, we use the thermodynamic integration approach to estimate $\log\{d(B_m, \beta_m)\}$ \citep{reich_spatial_2014,green_hidden_2002,smith_estimation_2006}. This approach is based on the identity

\begin{align}
    \frac{\partial}{\partial \beta_m} \log\{d(B_m, \beta_m)\} = E \left\{ \sum_{i \sim i'} \bI(\kappa_{mi} = \kappa_{mi'}) \bigg| B_m, \beta_m \right\}, \label{eq:deriv_expec_id_beta}
\end{align}

\noindent where the expectation is with respect to the Potts model for $\kappa_{m1}, \ldots, \kappa_{mn_m}$ with parameters $B_m, \beta_m$. the identity is a property of exponential families; see Equation 3.4.4 in Theorem 3.4.2 of \cite{casella_statistical_2002}. Because $\log\{ d(B_m, \beta_m) \}$ is an antiderivative of the expectation in terms of $\beta_m$, we have

\begin{align} \label{eq:beta_integ_approx}
    \int_0^{\beta_m} E \left\{ \sum_{i \sim i'} \bI(\kappa_{mi} = \kappa_{mi'}) \bigg| B_m, \beta \right\} d\beta = \log\{d(B_m, \beta_m)\} - \log\{d(B_m, 0)\}.
\end{align}

Approximating the integral of the expectation in (\ref{eq:beta_integ_approx}) will provide us with the quantities needed in the Metropolis update for $\beta_m$; the $\log\{d(B_m, 0)\}$ term above will be cancelled out in the process of computing the difference $\log\{d(B_m, \beta_m)\} - \log\{d(B_m, \beta_m')\}$ in Equation \ref{eq:beta_acceptance_prob}. To this end, we examine a grid of $\beta$ values $\{0, 0.01, 0.02, \ldots, \beta_{max} \}$. For each $\beta$ value, we generate 1000 realizations of the Potts model with parameters $B_m, \beta$ and calculate the average of $\sum_{i \sim i'} \bI(\kappa_{mi} = \kappa_{mi'})$ across realizations, denoted $\hat{E}\left\{ \sum_{i \sim i'} \bI(\kappa_{mi} = \kappa_{mi'}) \bigg| B_m, \beta \right\}$. We generate the 1000 realizations of the Potts model (after discarding 200 realizations as burn in) via the Swendsen-Wang algorithm \citep{swendsen_nonuniversal_1987} as implemented in the R package PottsUtils (version 0.3.3.1) \citep{feng_mri_2012}. After obtaining these averages for the grid of $\beta$ values, we fit a 10th-order polynomial spline to this curve of $\hat{E}\left\{ \sum_{i \sim i'} \bI(\kappa_{mi} = \kappa_{mi'}) \bigg| B_m, \beta \right\}$ values and integrate the polynomial. This integrated polynomial then has $\beta$ as input and an estimate of $\log\{d(B_m, \beta)\} - \log\{d(B_m, 0)\}$ as output, which is the quantity needed to compute the acceptance probability in (\ref{eq:beta_acceptance_prob}).

\section{Additional simulation study results}

\begin{table}
\caption{Overall features of BayesClint and cutting-edge SRT clustering algorithms.}
\label{tab:bayesclint_competing_comparison}
\begin{center}
\begin{tabular}{lcccc}
\hline
Method & Dimension Reduction & Feature Selection & Multi-Sample & Multi-Scale Clustering \\
\hline
BayesClint & \ding{51} & \ding{51} & \ding{51} & \ding{51} \\
BayesCafe & \ding{51} & \ding{51} &  & \\
PRECAST & \ding{51} &  & \ding{51} & \\
BASS & & & \ding{51} & \ding{51} \\
\hline
\end{tabular}
\end{center}
\end{table}


\begin{figure}
    \centering
    \includegraphics[width=0.45\linewidth]{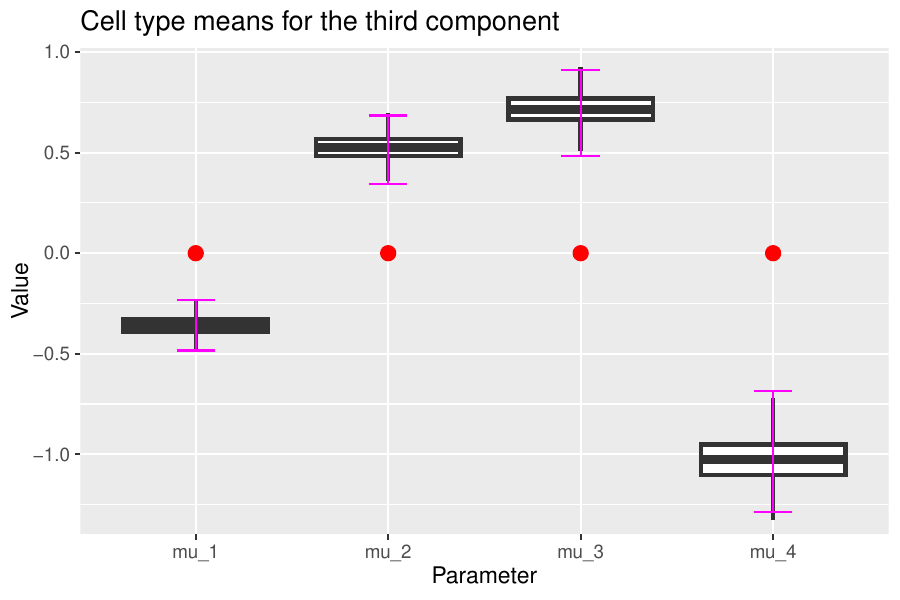}
    \includegraphics[width=0.45\linewidth]{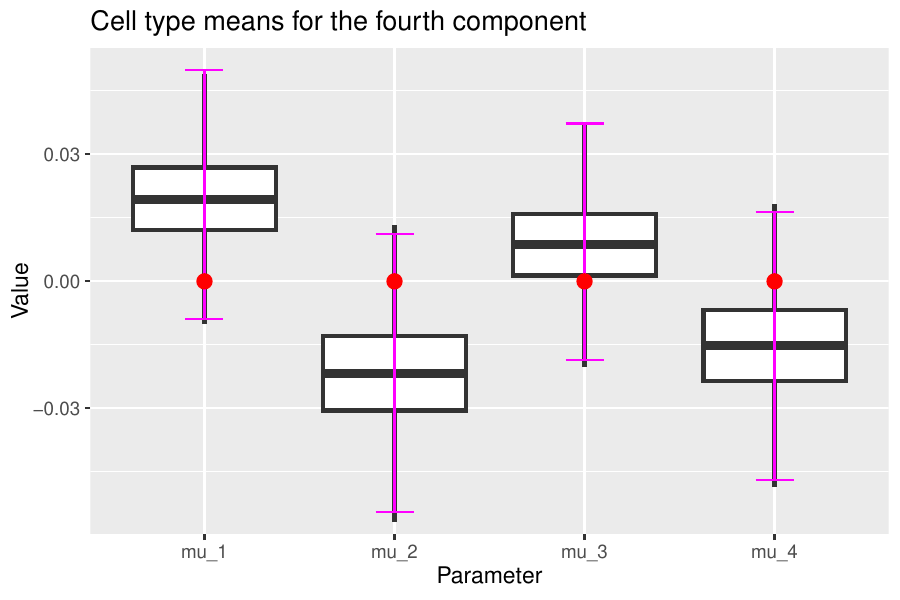}
    \caption{An example of a differentiating component (Left) and a non-differentiating component (Right) from the simulation study. Boxplots representing the posterior distribution of the four cell type means are shown for each component, along with the 95\% credible intervals.}
    \label{fig:diff_vs_nondiff_component}
\end{figure}


\begin{figure}
    \centering
    \includegraphics[width=0.45\linewidth]{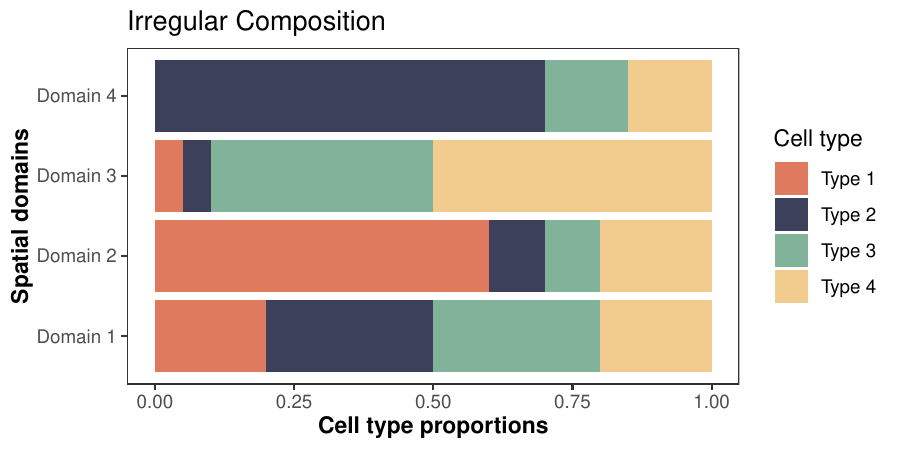}
    \includegraphics[width=0.45\linewidth]{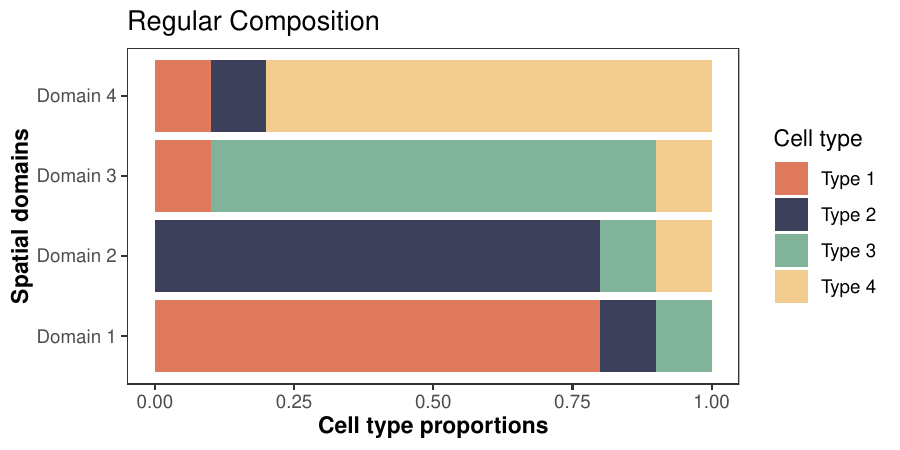}
    \caption{The irregular cell type composition (Left) and regular cell type composition (Right) used in the simulation study.}
    \label{fig:irreg_vs_reg}
\end{figure}

\begin{figure}
    \centering
    \includegraphics[width=\linewidth]{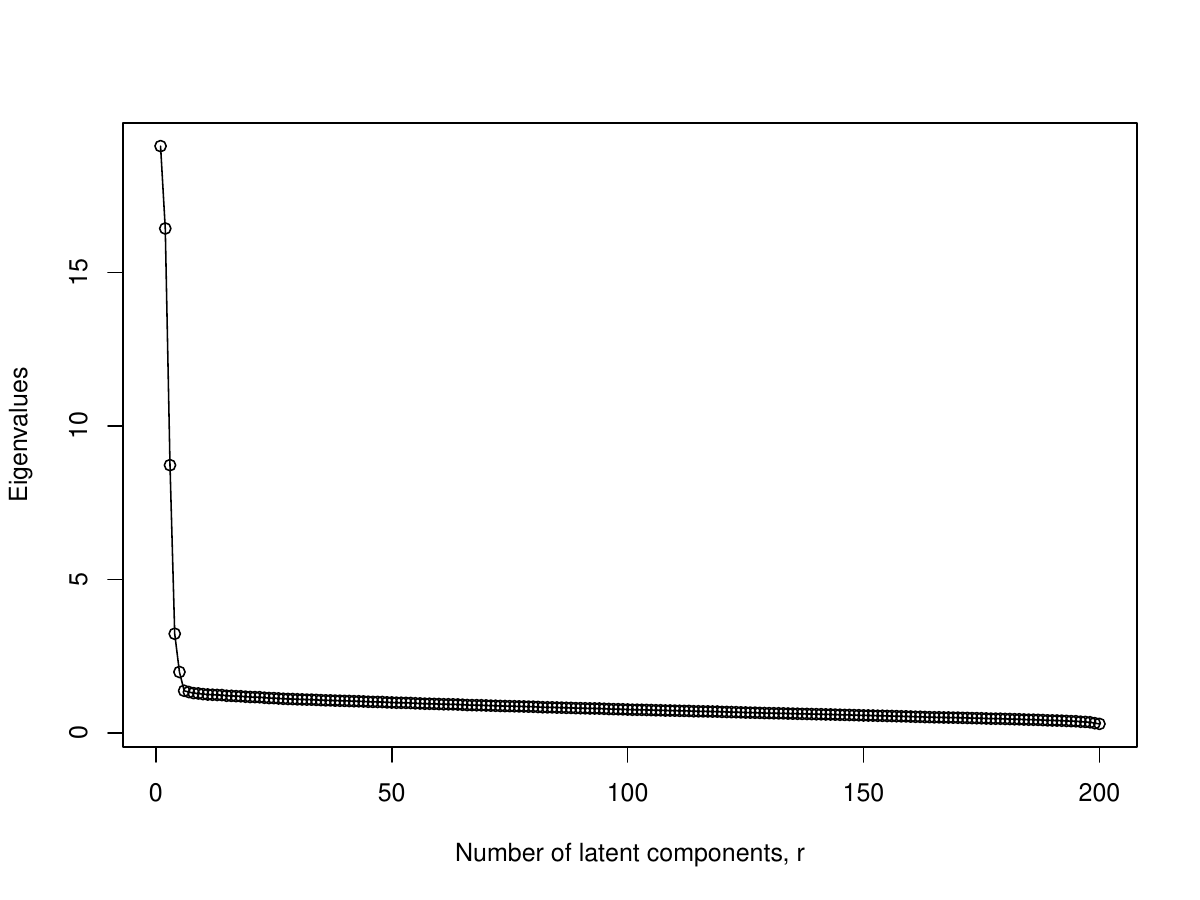}
    \caption{Scree plot for an example simulation study scenario with $N = 3$, irregular $\theta$ composition, $P = 200$, and $P_d = 40$. According to the elbow method, the number of latent components $r$ should be set to 4.}
    \label{fig:sim_study_screeplot}
\end{figure}

\begin{figure}
    \centering
    \includegraphics[width=\linewidth]{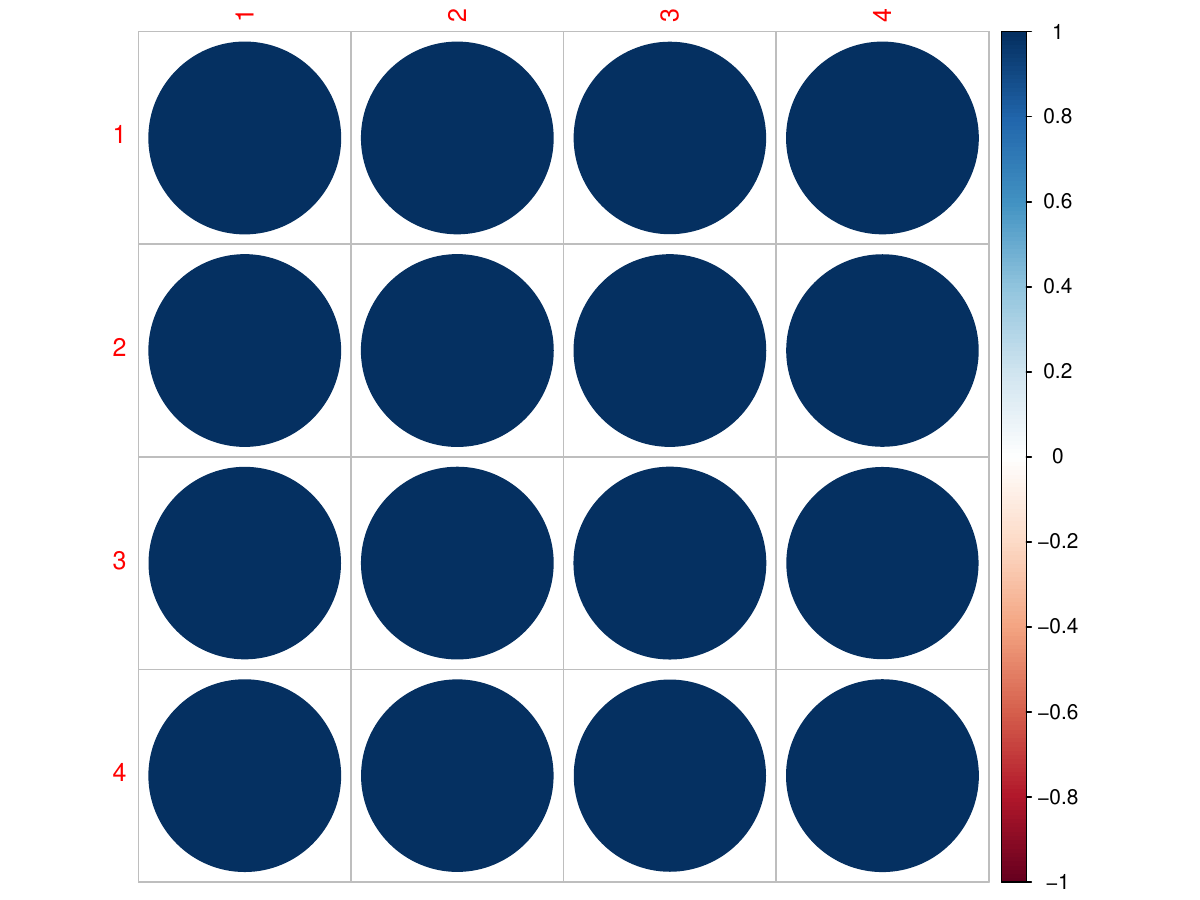}
    \caption{Pairwise correlation coefficients across the four MCMC chains, between the PPI values of all $P$ genes, for an example simulation study scenario with $N = 3$, irregular $\theta$ composition, $P = 200$, and $P_d = 40$.}
    \label{fig:sim_study_corrplot_4_chains}
\end{figure}

\subsection{Additional details on competing algorithms}
\label{sec:additional_competing_details}

\underline{Uniform data pre-processing for all methods}

\begin{itemize}
    \item For BayesClint and all methods, we do log-normalization and centering and scaling (except for PRECAST, which explicitly only requires centering, and DESeq2, which explicitly requires the raw counts).
    \item We also scale datasets separately by tissue sample, both for analysis and visualization.
\end{itemize}

\noindent \underline{Competing clustering algorithms}

\begin{itemize}
    \item Probabilistic embedding, clustering, and alignment for integrating spatial transcriptomics data (PRECAST) by \cite{liu_probabilistic_2023}.
    \begin{itemize}
        \item Multi-sample clustering algorithm for array-based ST technologies. 
        \item Cannot distinguish between cell type and spatial domain clustering. Will use same partition for both cell types and spatial domains when measuring clustering performance.
    \end{itemize}
    \item Bayesian Analytics for Spatial Segmentation (BASS) by \cite{li_bass_2022}.
    \begin{itemize}
        \item Multi-scale clustering algorithm for single-cell resolution ST data.
        \item 6500 burn-in and 6500 MCMC iterations, as for BayesClint.
        \item Will be combined with DESeq2 for differentiating gene selection
    \end{itemize}
    \item Sparse Principal Components Analysis (SPC) by \cite{witten_penalized_2009}.
    \begin{itemize}
        \item We do 5-fold cross-validation to select the column regularizing term sum of absolute values of elements of the right singular matrix.
        \item We consider a gene as selected if at least one of the $r = 4$ components has a non-zero element, similarly to BayesClint.
        \item Will be combined with $k$-means \citep{hartigan_algorithm_1979} and mclust for clustering (SPC + $k$-means, SPC + mclust). SPC improves performance over $k$-means and mclust used by themselves. Like PRECAST, $k$-means and mclust are not able to distinguish between cell type and spatial domain clustering. However, these methods cannot account for the spatial dependence among neighboring cells.
        \item SPC will be used by itself for gene selection. Will only get accuracies, not area under the receiver operating characteristic curve.
        \item Cannot distinguish between active and differentiating genes. Will use same selection for both kinds of genes when measuring selection accuracy.
        \item For multi-sample SPC implementation (for either clustering in combination with $k$-means or mclust, or gene selection by itself), expression matrices will be concatenated prior to analysis.
    \end{itemize}
\end{itemize}

\noindent \underline{Competing gene selection methods}

\begin{itemize}    
    \item modelGeneVar from scran (active gene selection) \citep{lun_step-by-step_2016}
    \item correlatePairs from scran (active gene selection) \citep{lun_step-by-step_2016}
    \item Spark-X (differentiating gene selection) \citep{zhu_spark-x_2021}
    \begin{itemize}
        \item We use Spark-X as opposed to Spark, since that is recommended for cases with more than 3000 cells
        \item We are considering this method as differentiating gene selection rather than active gene selection, because it may capture the spatial variation caused by the differing cell type compositions across spatial domains.
    \end{itemize}
    \item BASS+DESeq2 (differentiating gene selection) \citep{love_moderated_2014}
    \begin{itemize}
        \item The aforementioned gene selection methods all use $p$-values to select the genes, and use Benjamini-Hochberg by default to adjust them (except SPARK-X, which uses Benjamini-Yekutieli). We will convert SPARK-X $p$-values to Benjamini-Hochberg $p$-values (via stats::p.adjust), to be consistent with the other methods. Benjamini-Yekutieli is more conservative, as it is valid for all dependency structures. Benjamini-Hochberg and BFDR assume independence among hypothesis tests.
        \item To adjust for multiple tissue samples in scenarios with $N=3$, we include the tissue sample index as a blocking factor just for BASS+DESeq2, since it requires the raw, batch-uncorrected counts. The other three methods use data that are assumed to be batch corrected, so they don't use the tissue sample index as the blocking factor. This is the same approach used in the real data analyses with more than one tissue sample. 
    \end{itemize}
\end{itemize}

\subsection{Additional Feature Selection Results}
\label{sec:additional_ft_select_results}

The feature selection performance is measured by calculating the area under the curve (AUC) of the receiver operating characteristic, a commonly used metric for evaluating binary classifiers. The AUC considers both the true positive rate and false positive rate at various thresholds. The AUC ranges from 0 to 1, where a higher value indicates more accurate results. In the case of BayesClint, the $\textrm{PPI}_j$ and $\textrm{PPI}_{d,j}$ will be used to calculate the AUC for active or differentiating gene selection, respectively. For the other gene selection methods, the unadjusted $p$-value will be used to calculate the AUC \citep{habibzadeh_use_2024}.

For the active gene selection performance, our method BayesClint achieved a high performance across all scenarios (Table \ref{tab:sim_results_active_auc}). For scenarios with three tissue samples and 1000 genes, our active gene selection is almost perfect (with an AUC of 0.99 or greater). This indicates that BayesClint is able to integrate information across multiple tissue samples to perform highly accurate feature selection among many genes. For the differentially expressed gene selection performance, our method BayesClint achieves an almost perfect performance (with an AUC of 0.99 or greater) for a majority of scenarios, and in fact achieves perfect performance when the number of tissue samples is three, the cell compositions are regular, the total number of genes is 1000, and the number of differentially expressed genes is 40 (Table \ref{tab:sim_results_deg_auc}). Even in cases where competing methods correlateGenes, SPARK-X, and BASS+DESeq2 outperform BayesClint in active or differentiating gene selection, BayesClint is still very competitive.

\begin{table}[] 
\centering
\begin{tabular}{lllllll}
\hline
$N$ & $\theta$ & $P$ & $P_d$ & BayesClint & modelGeneVar & correlateGenes \\ 
\hline
1 & irr & 200 & 40 & 85.8 (1.5) & 69.3 (0.5) & \textbf{93.6} (0.3) \\ 
1 & irr & 200 & 80 & 98.6 (0.3) & 86.8 (0.4) & \textbf{99.3} (0.1) \\ 
1 & irr & 1000 & 40 & \textbf{95.4} (0.8) & 55.9 (0.4) & 87.1 (0.7) \\ 
1 & irr & 1000 & 80 & \textbf{92.5} (1.6) & 59.3 (0.3) & 88.7 (0.6) \\ 
\hdashline
1 & reg & 200 & 40 & 87.8 (1.5) & 70.2 (0.4) & \textbf{93.2} (0.3) \\ 
1 & reg & 200 & 80 & 98.6 (0.5) & 88.5 (0.4) & \textbf{99.3} (0.1) \\ 
1 & reg & 1000 & 40 & \textbf{96.1} (0.6) & 55.6 (0.4) & 86.8 (0.7) \\ 
1 & reg & 1000 & 80 & \textbf{91.9} (1.6) & 59.7 (0.3) & 88.9 (0.7) \\
\hdashline
3 & irr & 200 & 40 & 91.6 (1.7) & 69.5 (0.5) & \textbf{96.1} (0.1) \\ 
3 & irr & 200 & 80 & \textbf{95.3} (1.4) & 86.0 (0.4) & 94.8 (0.2) \\ 
3 & irr & 1000 & 40 & \textbf{99.7} (0.1) & 55.4 (0.4) & 98.9 (0.0) \\ 
3 & irr & 1000 & 80 & \textbf{99.3} (0.6) & 59.6 (0.3) & 98.6 (0.0) \\ 
\hdashline
3 & reg & 200 & 40 & 92.5 (1.5) & 68.7 (0.5) & \textbf{95.9} (0.2) \\ 
3 & reg & 200 & 80 & \textbf{95.6} (1.3) & 85.8 (0.4) & 94.9 (0.2) \\ 
3 & reg & 1000 & 40 & \textbf{99.1} (0.8) & 55.4 (0.3) & 98.7 (0.0) \\ 
3 & reg & 1000 & 80 & \textbf{99.8} (0.1) & 59.2 (0.3) & 98.6 (0.0) \\
\hline
\end{tabular}
\caption{Results of the selection of active genes. The average area under the curve (multiplied by 100) across 50 Monte Carlo iterations is presented for each of the three competing methods under 16 scenarios determined by 4 factors. The standard error of the average is in parentheses. $N$ is the number of simulated tissue samples, $\theta$ refers to the cell type compositions (irregular or regular, abbreviated as `irr' or `reg', respectively), $P$ is the number of genes in total, and $P_d$ is the number of differentially expressed genes. Our proposed algorithm is denoted as BayesClint, whereas the competing methods are modelGeneVar and correlateGenes. 
The values of the best performances are in bold type.} \label{tab:sim_results_active_auc}
\end{table}

\begin{table}[]
\centering
\begin{tabular}{lllllll}
\hline
$N$ & $\theta$ & $P$ & $P_d$ & BayesClint & SPARK-X & BASS+DESeq2 \\ 
\hline
1 & irr & 200 & 40 & 99.2 (0.3) & 92.1 (0.6) & \textbf{100} (0) \\ 
1 & irr & 200 & 80 & 98.9 (0.3) & 92.3 (0.5) & \textbf{99.8} (0.1) \\ 
1 & irr & 1000 & 40 & \textbf{99.7} (0.3) & 92.8 (0.5) & 99.1 (0.8) \\ 
1 & irr & 1000 & 80 & 98.8 (0.6) & 92.3 (0.4) & \textbf{99.8} (0.1) \\ 
\hdashline
1 & reg & 200 & 40 & 99.3 (0.2) & \textbf{100} (0) & \textbf{100} (0) \\ 
1 & reg & 200 & 80 & 98.8 (0.4) & 99.9 (0.0) & \textbf{99.9} (0) \\ 
1 & reg & 1000 & 40 & 99.4 (0.4) & \textbf{100} (0) & \textbf{100} (0) \\ 
1 & reg & 1000 & 80 & 99.7 (0.2) & \textbf{100} (0) & \textbf{100} (0) \\ 
\hdashline
3 & irr & 200 & 40 & 97.4 (0.7) & 90.0 (0.6) & \textbf{100} (0) \\ 
3 & irr & 200 & 80 & 95.8 (1.4) & 89.4 (0.6) & \textbf{100} (0) \\ 
3 & irr & 1000 & 40 & \textbf{100} (0) & 89.5 (0.6) & 99.4 (0.6) \\ 
3 & irr & 1000 & 80 & 99.6 (0.3) & 90.6 (0.5) & \textbf{100} (0) \\ 
\hdashline
3 & reg & 200 & 40 & 97.2 (0.6) & \textbf{100} (0) & \textbf{100} (0) \\ 
3 & reg & 200 & 80 & 96.0 (1.3) & \textbf{100} (0) & \textbf{100} (0) \\ 
3 & reg & 1000 & 40 & \textbf{100} (0) & \textbf{100} (0) & \textbf{100} (0) \\ 
3 & reg & 1000 & 80 & 98.6 (0.7) & \textbf{100} (0) & \textbf{100} (0) \\
\hline
\end{tabular}
\caption{Results of the selection of differentially expressed genes. The average area under the curve (multiplied by 100) across 50 Monte Carlo iterations is presented for each of the three competing methods under 16 scenarios determined by 4 factors. The standard error of the average is in parentheses. $N$ is the number of simulated tissue samples, $\theta$ refers to the cell type compositions (irregular or regular, abbreviated as `irr' or `reg', respectively), $P$ is the number of genes in total, and $P_d$ is the number of differentially expressed genes. Our proposed algorithm is denoted as BayesClint, whereas the competing methods are SPARK-X and BASS+DESeq2. 
The values of the best performances are in bold type.} \label{tab:sim_results_deg_auc}
\end{table}

\subsection{Sensitivity Analyses}
\label{sec:all_sensitans}

We assess the impact of the choice of hyperparameters $r$, $q_\gamma$, $C$, and $K$ on BayesClint's clustering and feature selection performance. We focus on the scenario with $N = 3$, irregular $\theta$ composition, $P = 200$, and $P_d = 40$ for these sensitivity analyses, and use 50 Monte Carlo iterations as in the main simulation study.

\subsubsection{Number of components $r$ sensitivity analysis}
\label{sec:r_sensitan}

\begin{figure}
    \centering
    \includegraphics[width=1\linewidth]{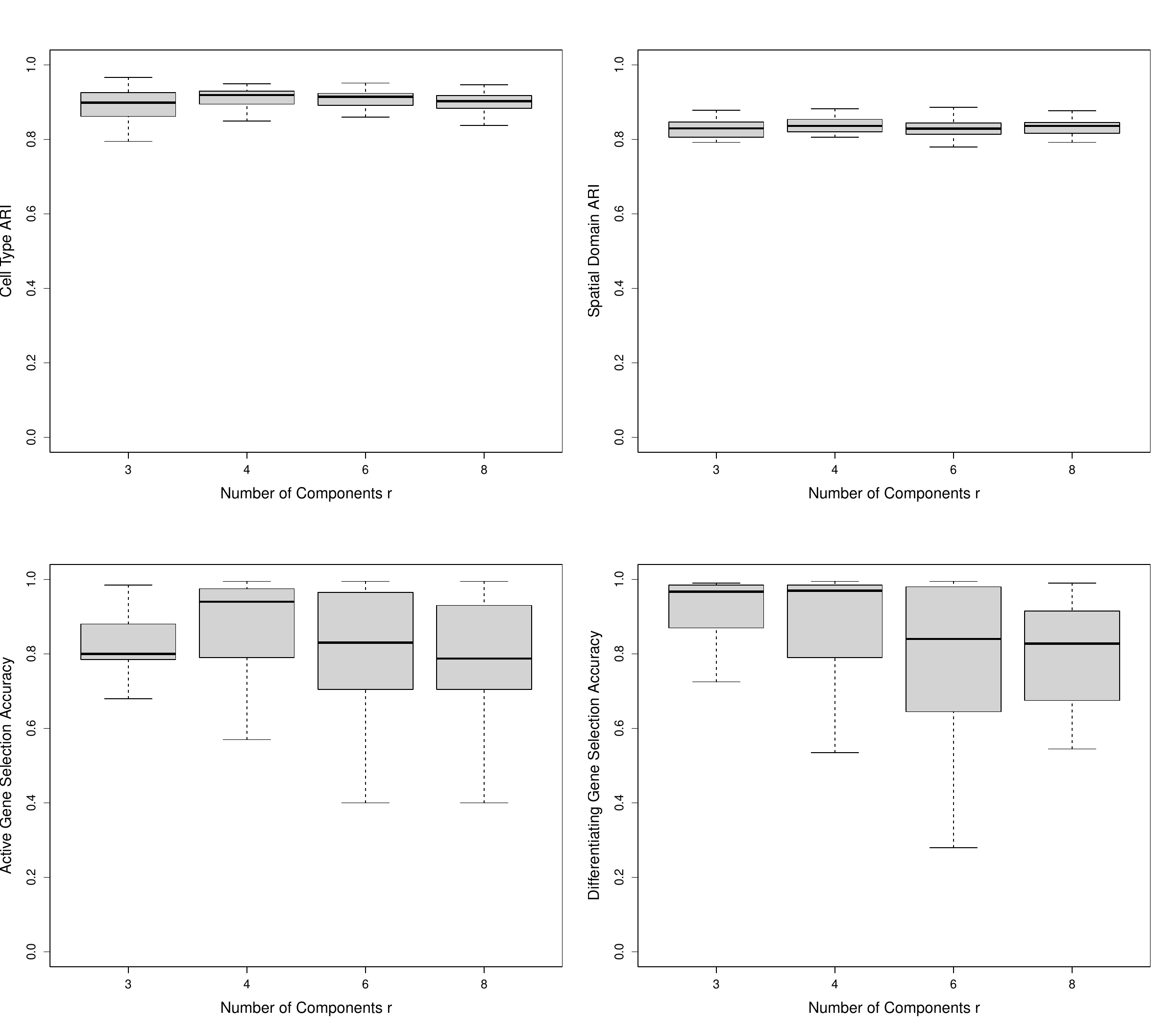}
    \caption{Number of components $r$ sensitivity analysis.}
    \label{fig:r_metrics_boxplots}
\end{figure}

We examine how the performance of BayesClint changes as we vary the number of components $r$ across the values $3, 4, 6,$ and $8$ (Figure \ref{fig:r_metrics_boxplots}). We observed a negligible variation of clustering performance, for either cell type or spatial domain clustering, across the $r$ values. However, we observe that for active gene selection, the $r$ value selected through the elbow method, 4, leads to a significantly better performance than the other $r$ values. For differentiating gene selection, both $r$ values 3 and 4 lead to a significantly better performance than the other $r$ values.

\subsubsection{Prior probability of selection $q_\gamma$ sensitivity analysis}
\label{sec:qv_sensitan}

\begin{figure}
    \centering
    \includegraphics[width=1\linewidth]{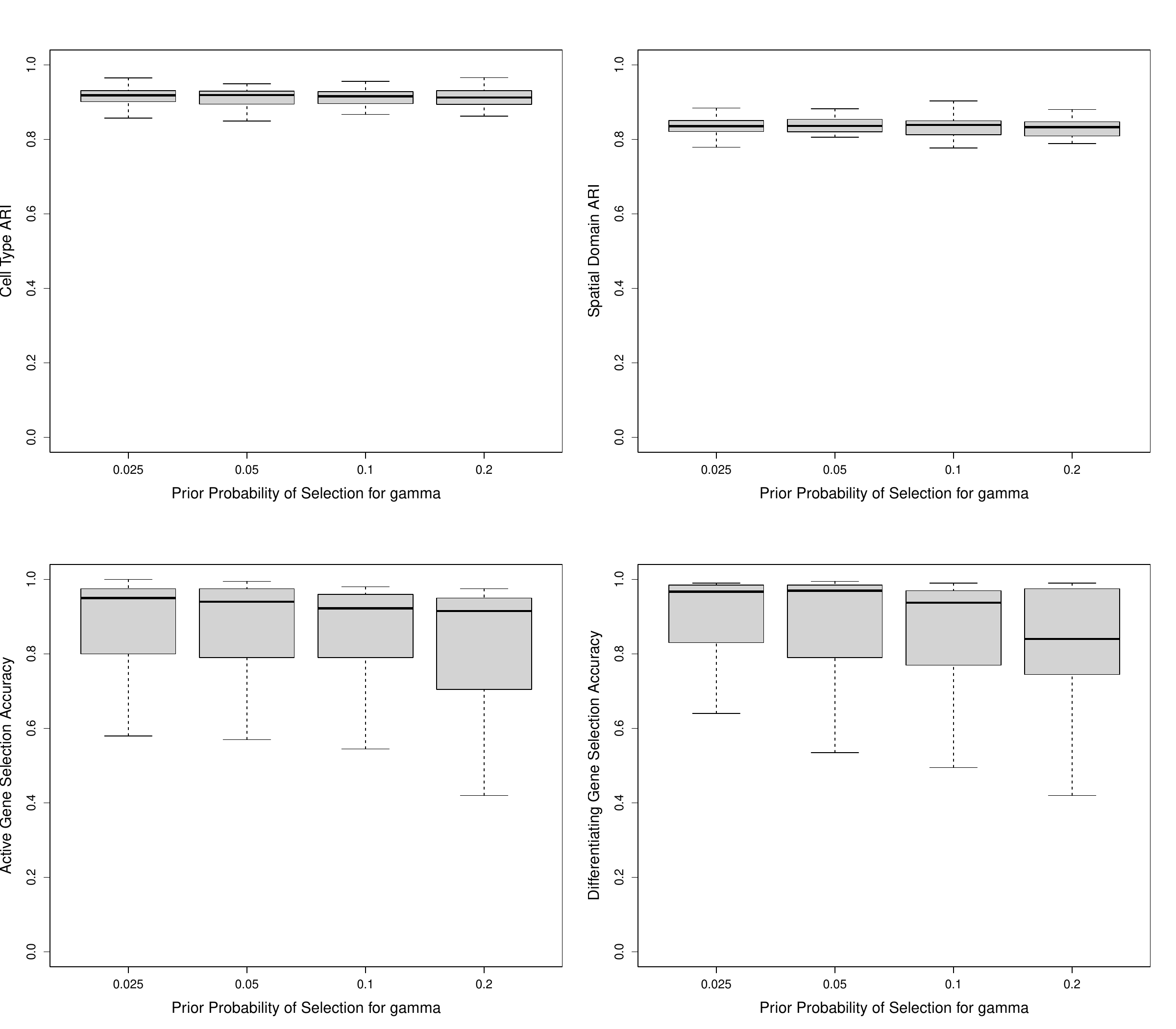}
    \caption{Prior probability of selection $q_\gamma$ sensitivity analysis.}
    \label{fig:qv_metrics_boxplots}
\end{figure}

We examine how the performance of BayesClint changes as we vary the prior probability of selection $q_\gamma$ across the values $0.025, 0.05, 0.1,$ and $0.2$ (Figure \ref{fig:qv_metrics_boxplots}). We observed a negligible variation of clustering performance, for either cell type or spatial domain clustering, across the $r$ values. We also observed a negligible variation of active gene selection performance. However, we observe that for differentiating gene selection, both $q_\gamma$ values 0.025 and 0.05 lead to a significantly better performance than the other $r$ values.

\subsubsection{Misspecification of the number of cell types or number of spatial domains sensitivity analysis}
\label{sec:CorK_sensitan}

We begin by examining a sensitivity analysis regarding the effects of misspecifying the numbers of cell types or spatial domains on clustering and feature selection performance. Following this evaluation, we introduce a heuristic for choosing the appropriate numbers of spatial domains and cell types, which involves incrementing the number of clusters and examining the relative cell proportions in the resulting clusters. In every case, we fit the BayesClint model with 6,500 burn-in and 6,500 MCMC iterations for 50 Monte Carlo datasets, as in the main simulation study. 

\begin{figure}
    \centering
    \includegraphics[width=1\linewidth]{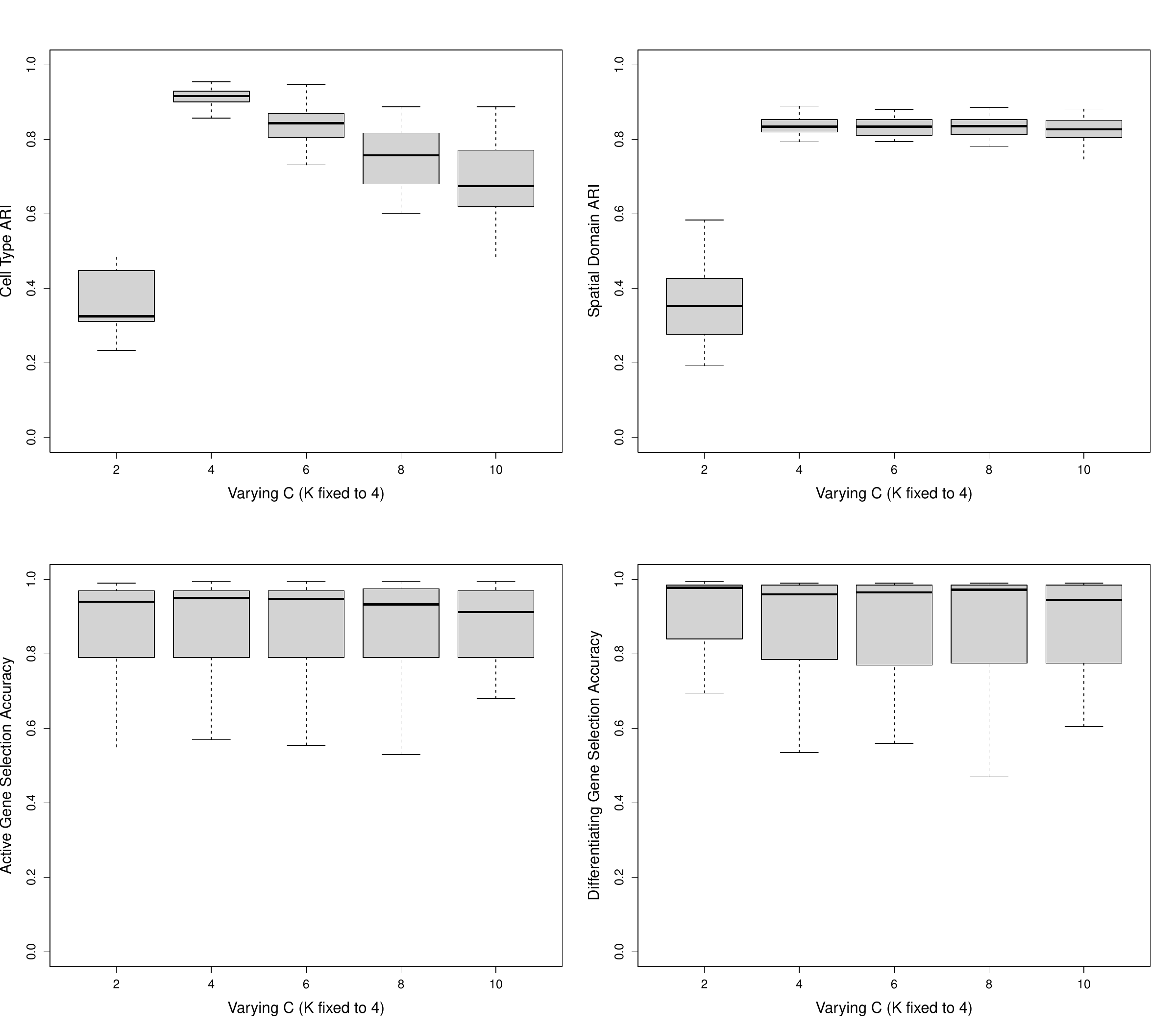}
    \caption{Misspecification of the number of cell types sensitivity analysis.}
    \label{fig:Cchange_metrics_boxplots}
\end{figure}

For the sensitivity analysis, we focus on one of the scenarios from the simulation study, where $N = 3$, irregular $\theta$ composition, $P = 200$, and $P_d = 40$. We first held the number of spatial domains at the truth ($K = 4$), while varying the number of cell types $C = 2, 4, 6, 8, 10$. The performance of BayesClint under these conditions are shown in Figure \ref{fig:Cchange_metrics_boxplots}. For cell type clustering, performance is the worst with only two clusters, the best with the true number of clusters (four), and gradually deteriorates with an increasing number of clusters. However, for spatial domain clustering, performance is worst with only two clusters, but then the variation of the performance becomes negligible for four or greater numbers of clusters.  
For active or differentiating gene selection, however, we observed a negligible variation of performance. This is expected, as changing the number of cell type clusters would primarily affect clustering, not feature selection.

\begin{figure}
    \centering
    \includegraphics[width=1\linewidth]{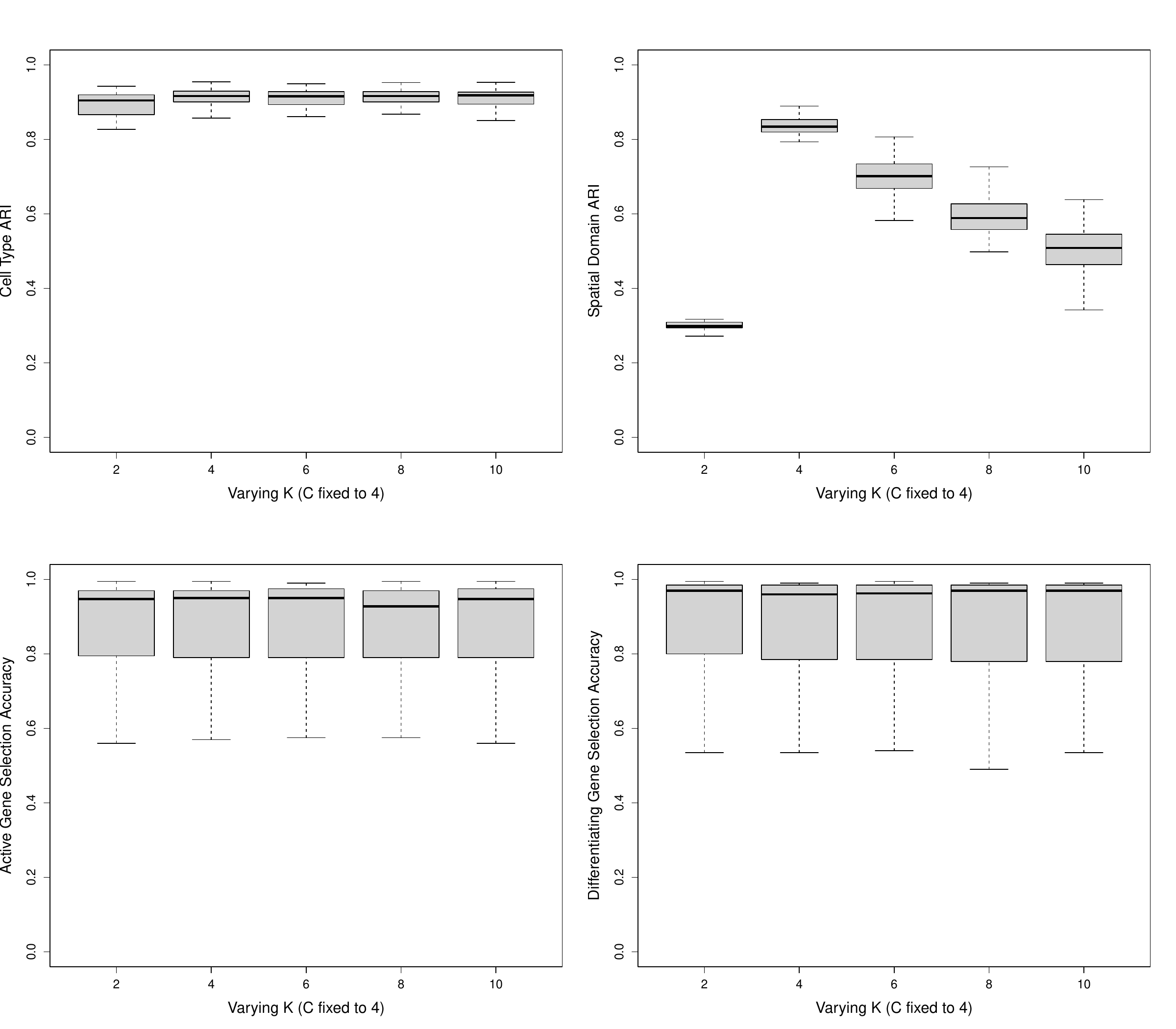}
    \caption{Misspecification of the number of spatial domains sensitivity analysis.}
    \label{fig:Kchange_metrics_boxplots}
\end{figure}

Next, we held the number of cell types at the truth ($C = 4$), while varying the number of spatial domains $K = 2, 4, 6, 8, 10$. The performance of BayesClint under these conditions are shown in Figure \ref{fig:Kchange_metrics_boxplots}. For cell type clustering, the variation in performance is negligible. However, for spatial domain clustering, performance is the worst with only two clusters, the best with the true number of clusters (four), and gradually deteriorates with an increasing number of clusters. As for the misspecification of cell types sensitivity analysis, for active or differentiating gene selection, we observed a negligible variation of performance.

In summary, BayesClint's clustering and feature selection performances are robust to overly large numbers of cell types or spatial domains to an extent. In particular, misspecifying the number of cell types or spatial domains mainly impacts the cell type or spatial domain ARI, respectively, in that clustering performance is worst for underspecified numbers of clusters, best for the correctly specified number of clusters, and gradually deteriorates with overspecified numbers of clusters.

As mentioned in Section 3.3 of the main article, analyzing SRT datasets generally involves using a pathologist's specialized knowledge to identify the numbers of cell types and spatial domains. If these numbers must be determined solely through the data, we suggest the following heuristic: one should begin by setting an initially high number of cell types and adjusting the number of spatial domains first, since both kinds of clustering are relatively more robust to an overspecified number of cell type clusters $C$, as opposed to an overspecified number of spatial domain clusters $K$. Then, one can increase the number of spatial domains until the smallest domain constitutes less than 6\% of the total cell count in the pre-processed dataset. Once the number of spatial domains is determined, one can then increase the number of cell types until the rarest cell type constitutes less than 2.5\% of the entire pre-processed dataset.

This heuristic was informed by a Monte Carlo experiment that also involves the example scenario with $N = 3$, irregular $\theta$ composition, $P = 200$, and $P_d = 40$. We first held the number of cell types constant at the truth (four), and incremented the number of spatial domains through the values $K = 2, 4, 6, 8, 10$ when fitting the BayesClint model for 50 Monte Carlo datasets. The average proportion of cells in the dataset 
that falls in the resulting spatial domain clusters is shown in Table \ref{tab:clust_props_sim1}. The moment that $K$ is incremented from the true number of clusters, four, to the next value, six, the average proportion of cells in the least prevalent cluster drops from 22.22\% to 5.00\%, passing the 6\% cut-off.

\begin{table}
\caption{Average proportion of cells (\%) assigned to each spatial domain cluster ID by BayesClint for the specified numbers of spatial domains $K = 2, 4, 6, 8, 10$ (with the number of cell types held at the truth, which is 4), for the simulation study scenario with $N = 3$, irregular $\theta$ composition, $P = 200$, and $P_d = 40$. The average (with the standard deviation in parentheses) was taken over 50 Monte Carlo iterations. The clusters are ordered from least prevalent (cluster ID 1) to most prevalent (cluster ID 10) before the averages and standard deviations are evaluated.}
\label{tab:clust_props_sim1}
\begin{center}
\begin{NiceTabular}{w{l}{4cm}|ccccc}
\hline
\rule[-5mm]{15mm}{0cm}
\diagbox{Spa. domain\\ cluster ID}{Specified $K$} & 2 & 4 & 6 & 8 & 10 \\[26pt]
\hline
1 & 29.09 (0.4) & 22.22 (0.2) & 5.00 (0.4) & 3.34 (0.3) & 1.87 (0.2) \\ 
2 & 70.91 (0.4) & 24.98 (0.1) & 9.76 (0.4) & 6.01 (0.3) & 3.61 (0.2) \\ 
3 & & 25.93 (0.1) & 15.77 (0.4) & 8.20 (0.3) & 5.12 (0.2) \\ 
4 & & 26.88 (0.2) & 19.66 (0.4) & 10.51 (0.3) & 7.12 (0.2) \\ 
5 & & & 23.20 (0.3) & 13.86 (0.3) & 8.64 (0.2) \\ 
6 & & & 26.61 (0.2) & 16.52 (0.3) & 10.26 (0.2) \\ 
7 & & & & 18.86 (0.4) & 12.33 (0.2) \\ 
8 & & & & 22.72 (0.3) & 14.30 (0.3) \\ 
9 & & & & & 16.87 (0.3) \\ 
10 & & & & & 19.87 (0.4) \\
\hline
\end{NiceTabular}
\end{center}
\end{table}

Next, we held the number of spatial domains constant at the truth (four), and incremented the number of cell types through the values $C = 2, 4, 6, 8, 10$ when fitting the BayesClint model for 50 Monte Carlo datasets. The average proportion of cells in the dataset 
that falls in the resulting cell type clusters is shown in Table \ref{tab:clust_props_sim2}. The moment that $C$ is incremented from the true number of clusters, four, to the next value, six, the average proportion of cells in the least prevalent cluster drops from 20.92\% to 2.02\%, passing the 2.5\% cut-off.

\begin{table}
\caption{Average proportion of cells (\%) assigned to each cell type cluster ID by BayesClint for the specified numbers of cell types $C = 2, 4, 6, 8, 10$ (with the number of spatial domains held at the truth, which is 4), for the simulation study scenario with $N = 3$, irregular $\theta$ composition, $P = 200$, and $P_d = 40$. The average (with the standard deviation in parentheses) was taken over 50 Monte Carlo iterations. The clusters are ordered from least prevalent (cluster ID 1) to most prevalent (cluster ID 10) before the averages and standard deviations are evaluated.}
\label{tab:clust_props_sim2}
\begin{center}
\begin{NiceTabular}{w{l}{4cm}|ccccc}
\hline
\rule[-5mm]{15mm}{0cm}
\diagbox{Cell type\\ cluster ID}{Specified $C$} & 2 & 4 & 6 & 8 & 10 \\[22pt]
\hline
1 & 34.45 (1.8) & 20.92 (0.5) & 2.02 (0.3) & 1.10 (0.2) & 0.43 (0.1) \\ 
2 & 65.55 (1.8) & 23.61 (0.2) & 6.09 (0.6) & 2.31 (0.3) & 1.12 (0.2) \\ 
3 & & 26.38 (0.1) & 19.02 (0.4) & 4.74 (0.5) & 2.28 (0.3) \\ 
4 & & 29.09 (0.6) & 21.81 (0.3) & 9.08 (0.4) & 3.69 (0.3) \\ 
5 & & & 24.12 (0.2) & 16.45 (0.5) & 6.34 (0.4) \\ 
6 & & & 26.94 (0.2) & 19.73 (0.4) & 9.06 (0.4) \\ 
7 & & & & 22.16 (0.3) & 15.28 (0.4) \\ 
8 & & & & 24.43 (0.3) & 17.97 (0.4) \\ 
9 & & & & & 20.73 (0.4) \\ 
10 & & & & & 23.10 (0.4) \\
\hline
\end{NiceTabular}
\end{center}
\end{table}

\subsubsection{Applying heuristic to the MVC dataset}
\label{sec:CorK_mvc}

Following the proposed heuristic, we begin by setting the number of cell types to a high number, $C = 19$, and fitting the BayesClint model for various numbers of spatial domains, $K = 2, 4, 6, 8, 10$. After determining the number of spatial domains according to the heuristic (using the 6\% cut-off), we then fix the number of spatial domains $K$ to that value, and then fit the BayesClint model for various numbers of cell types, $C = 11, 13, 15, 17, 19$. Finally, we would select the number of cell types according to the heuristic (using the 2.5\% cut-off). In each case, we would fit the BayesClint model with 15,000 MCMC iterations, after discarding 6,500 iterations as burn-in. According to Tables \ref{tab:clust_props_mvc_justK} and \ref{tab:clust_props_mvc_justC}, the heuristic would lead to the selection of $C = 11$ cell types and $K = 4$ spatial domains. The BayesClint model fitted with these values achieved a spatial domain clustering ARI of 0.54 and a cell type clustering ARI of 0.50.

\begin{table}
\caption{Proportion of cells (\%) assigned to each spatial domain cluster ID by BayesClint for the specified numbers of spatial domains $K = 2, 4, 6, 8, 10$ (with the number of cell types held at 19), for the MVC dataset. The clusters are ordered from least prevalent (cluster ID 1) to most prevalent (cluster ID 10).}
\label{tab:clust_props_mvc_justK}
\begin{center}
\begin{NiceTabular}{|w{l}{4cm}|ccccc|}
\hline
\rule[-5mm]{15mm}{0cm}
\diagbox{Spa. domain\\ cluster ID}{Specified $K$} & 2 & 4 & 6 & 8 & 10 \\[26pt]
\hline
1 & 49.38 & 15.08 & 3.40 & 0.00 & 0.00 \\ 
2 & 50.62 & 22.95 & 13.34 & 0.83 & 0.00 \\ 
3 & & 26.35 & 14.91 & 3.56 & 0.00 \\ 
4 & & 35.63 & 21.29 & 13.26 & 1.99 \\ 
5 & & & 22.62 & 15.49 & 4.56 \\ 
6 & & & 24.44 & 20.55 & 13.34 \\ 
7 & & &  & 21.38 & 13.92 \\ 
8 & & & & 24.94 & 20.96 \\ 
9 & & & & & 20.96 \\ 
10 & & & & & 24.28 \\
\hline
\end{NiceTabular}
\end{center}
\end{table}

\begin{table}
\caption{Proportion of cells (\%) assigned to each cell type cluster ID by BayesClint for the specified numbers of spatial domains $C = 11, 13, 15, 17, 19$ (with the number of spatial domains held at 4), for the MVC dataset. The clusters are ordered from least prevalent (cluster ID 1) to most prevalent (cluster ID 19).}
\label{tab:clust_props_mvc_justC}
\begin{center}
\begin{NiceTabular}{w{l}{4cm}|ccccc}
\hline
\rule[-5mm]{15mm}{0cm}
\diagbox{Cell type\\ cluster ID}{Specified $C$} & 11 & 13 & 15 & 17 & 19 \\[22pt]
\hline
1 & 2.82 & 1.33 & 0.00 & 0.17 & 0.00 \\ 
2 & 4.81 & 2.82 & 0.00 & 1.16 & 1.24 \\ 
3 & 6.30 & 2.90 & 1.49 & 1.24 & 1.41 \\ 
4 & 6.30 & 3.48 & 1.66 & 1.99 & 1.91 \\ 
5 & 6.46 & 4.72 & 2.73 & 3.65 & 1.99 \\ 
6 & 7.37 & 4.97 & 3.40 & 3.98 & 2.57 \\ 
7 & 8.78 & 5.63 & 3.73 & 3.98 & 2.82 \\ 
8 & 10.27 & 6.30 & 4.56 & 5.14 & 3.48 \\ 
9 & 11.18 & 6.55 & 5.63 & 5.55 & 3.65 \\ 
10 & 11.60 & 9.78 & 6.05 & 6.46 & 3.98 \\ 
11 & 24.11 & 12.51 & 6.79 & 6.63 & 4.39 \\ 
12 & & 15.41 & 7.71 & 6.63 & 6.05 \\ 
13 & & 23.61 & 7.79 & 6.96 & 6.21 \\ 
14 & & & 22.37 & 9.53 & 6.63 \\ 
15 & & & 26.10 & 9.53 & 7.37 \\ 
16 & & & & 10.94 & 9.53 \\ 
17 & & & & 16.49 & 9.61 \\ 
18 & & & & & 10.52 \\ 
19 & & & & & 16.65 \\
\hline
\end{NiceTabular}
\end{center}
\end{table}

\subsubsection{Applying heuristic to the mPFC dataset}
\label{sec:CorK_mpfc}

We apply the proposed heuristic in the same way as for the MVC dataset (Section \ref{sec:CorK_mvc}). According to Tables \ref{tab:clust_props_mpfc_justK} and \ref{tab:clust_props_mpfc_justC}, the heuristic would lead to the selection of $C = 11$ cell types and $K = 4$ spatial domains. The BayesClint model fitted with these values achieved a spatial domain clustering ARI of 0.76 and a cell type clustering ARI of 0.43.

\begin{table}
\caption{Proportion of cells (\%) assigned to each spatial domain cluster ID by BayesClint for the specified numbers of spatial domains $K = 2, 4, 6, 8, 10$ (with the number of cell types held at 19), for the mPFC dataset. The clusters are ordered from least prevalent (cluster ID 1) to most prevalent (cluster ID 10).}
\label{tab:clust_props_mpfc_justK}
\begin{center}
\begin{NiceTabular}{w{l}{4cm}|lllll}
\hline
\rule[-5mm]{15mm}{0cm}
\diagbox{Spa. domain\\ cluster ID}{Specified $K$} & 2 & 4 & 6 & 8 & 10 \\[26pt]
\hline
1 & 44.73 & 10.66 & 5.52 & 1.63 & 0.75 \\ 
2 & 55.27 & 14.26 & 10.97 & 10.00 & 2.13 \\ 
3 & & 32.23 & 11.35 & 10.56 & 2.16 \\ 
4 & & 42.85 & 15.77 & 10.82 & 3.48 \\ 
5 & & & 25.33 & 14.01 & 5.42 \\ 
6 & & & 31.07 & 14.58 & 8.81 \\ 
7 & & & & 15.74 & 8.87 \\ 
8 & & & & 22.66 & 13.32 \\ 
9 & & & & & 26.52 \\ 
10 & & & & & 28.53 \\
\hline
\end{NiceTabular}
\end{center}
\end{table}

\begin{table}
\caption{Proportion of cells (\%) assigned to each cell type cluster ID by BayesClint for the specified numbers of spatial domains $C = 11, 13, 15, 17, 19$ (with the number of spatial domains held at 4), for the mPFC dataset. The clusters are ordered from least prevalent (cluster ID 1) to most prevalent (cluster ID 19).}
\label{tab:clust_props_mpfc_justC}
\begin{center}
\begin{NiceTabular}{w{l}{4cm}|ccccc}
\hline
\rule[-5mm]{15mm}{0cm}
\diagbox{cell type\\ cluster ID}{Specified $C$} & 11 & 13 & 15 & 17 & 19 \\[22pt]
\hline
1 & 2.79 & 0.16 & 2.60 & 1.03 & 0.09 \\ 
2 & 5.11 & 2.73 & 2.63 & 2.07 & 1.22 \\ 
3 & 6.83 & 3.04 & 2.76 & 3.01 & 2.04 \\ 
4 & 6.90 & 3.82 & 2.95 & 3.04 & 2.23 \\ 
5 & 7.65 & 4.70 & 2.98 & 3.73 & 2.57 \\ 
6 & 8.56 & 5.55 & 3.70 & 3.98 & 3.23 \\ 
7 & 8.93 & 7.62 & 4.76 & 4.23 & 3.26 \\ 
8 & 9.62 & 9.22 & 5.80 & 4.33 & 3.73 \\ 
9 & 10.44 & 10.19 & 5.92 & 4.58 & 4.64 \\ 
10 & 12.32 & 10.53 & 8.53 & 4.70 & 4.67 \\ 
11 & 20.85 & 10.97 & 8.59 & 5.02 & 5.20 \\ 
12 & & 11.19 & 8.75 & 5.64 & 5.58 \\ 
13 & & 20.28 & 10.47 & 5.96 & 5.74 \\ 
14 & & & 11.32 & 7.87 & 5.99 \\ 
15 & & & 18.24 & 8.18 & 6.27 \\ 
16 & & & & 11.94 & 7.30 \\ 
17 & & & & 20.69 & 8.50 \\ 
18 & & & & & 10.56 \\ 
19 & & & & & 17.18 \\
\hline
\end{NiceTabular}
\end{center}
\end{table}




\begin{figure}
    \centering
    \includegraphics[width=\linewidth]{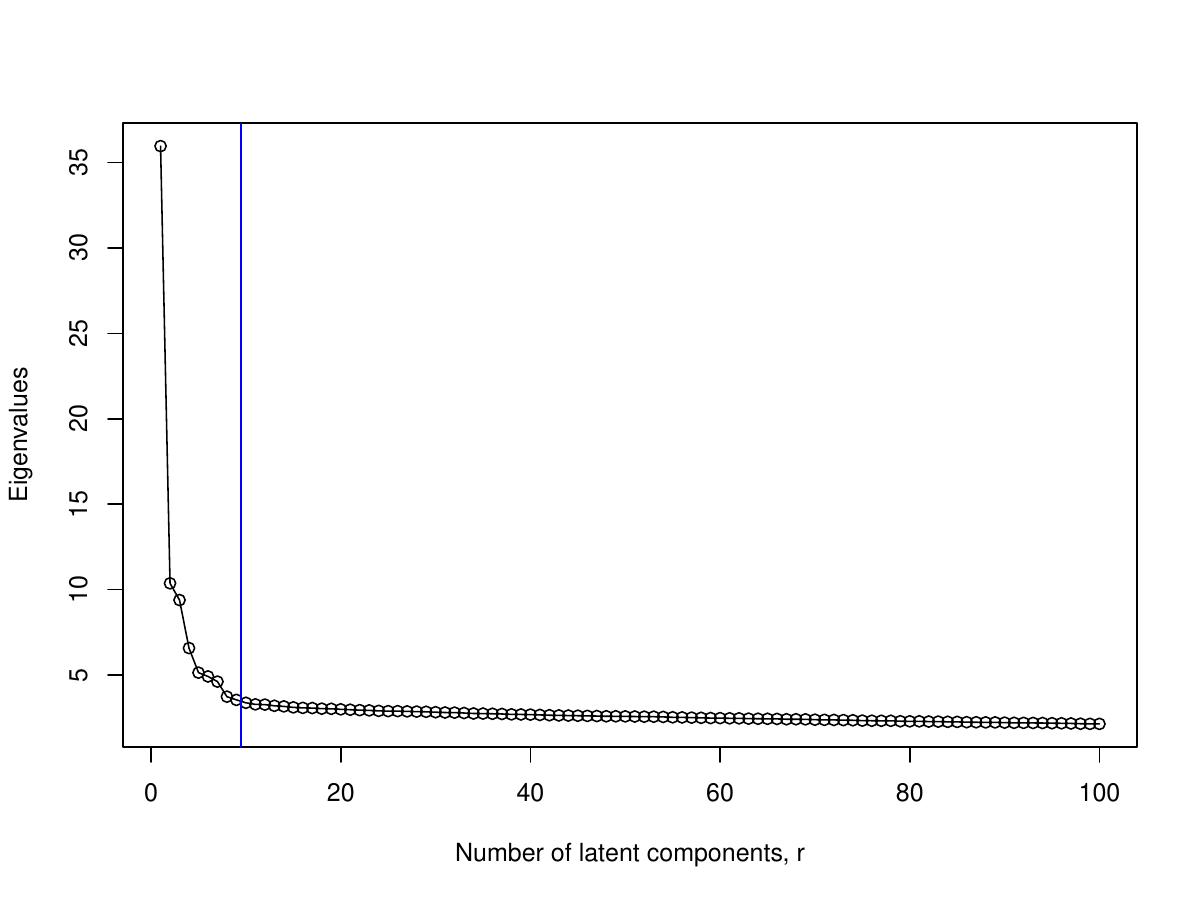}
    \caption{Scree plot for the MVC dataset, focusing on the first 100 eigenvalues. According to the elbow method (indicated by the blue line), we select the number of latent components $r$ to be 9.}
    \label{fig:mvc_screeplot}
\end{figure}

\begin{figure}
    \centering
    \includegraphics[width=1\linewidth]{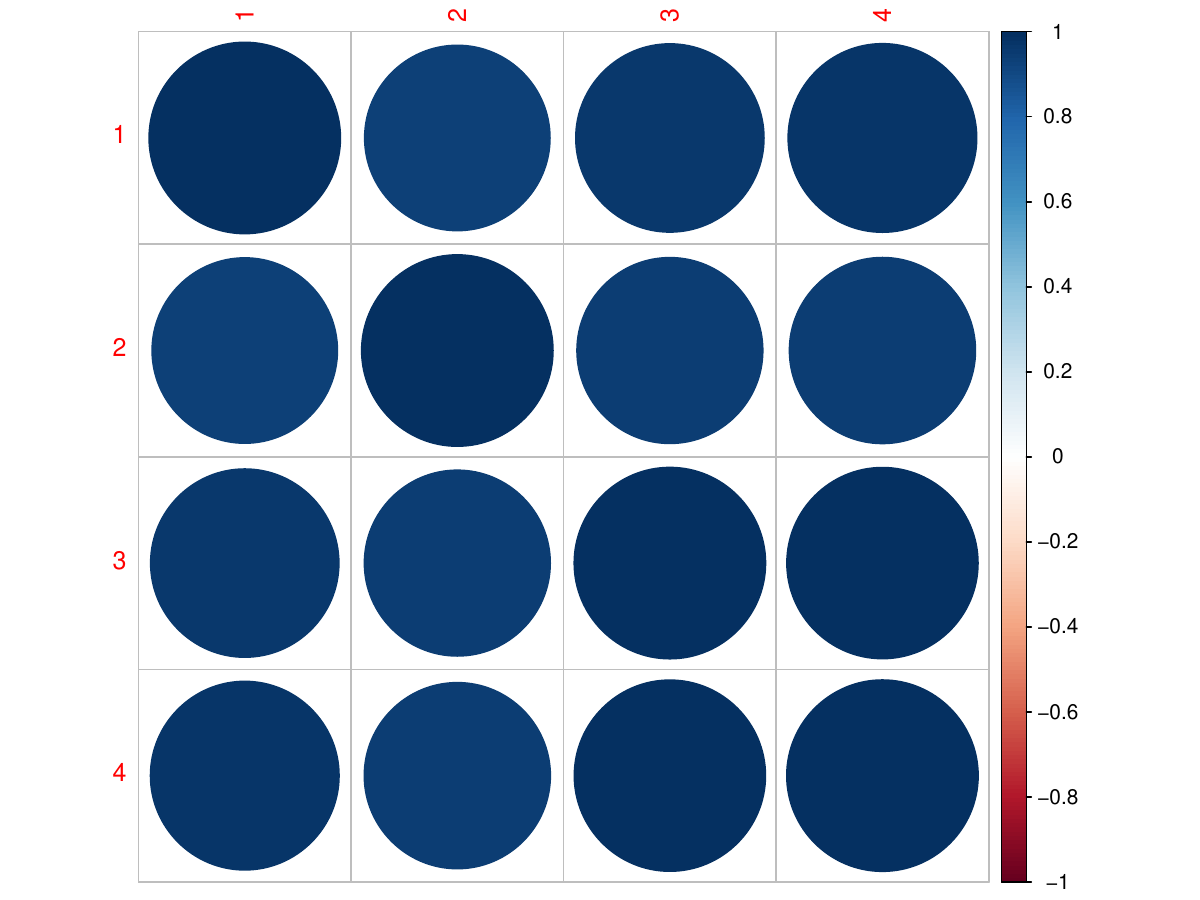}
    \caption{Pairwise correlation coefficients across the four MCMC chains, between the PPI values of all $P$ genes, for the MVC dataset.}
    \label{fig:mvc_corrplot_4_chains}
\end{figure}

\begin{figure}

    \centering

    \subfloat[Spatial Domain Clustering\label{fig:mvc_ground_truth_spatial_domains}]{\includegraphics[width=\textwidth]{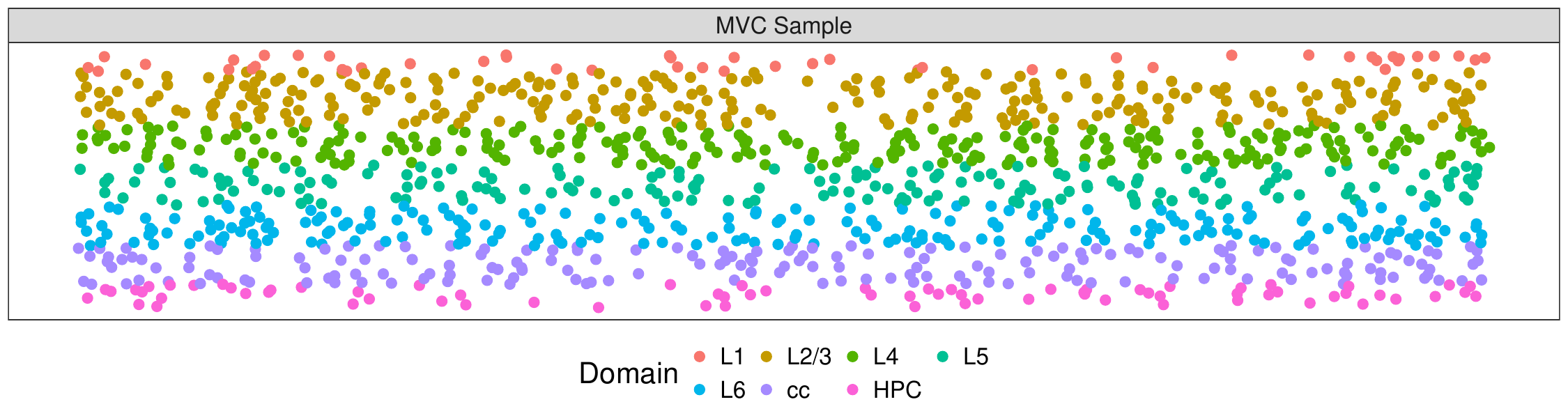}} 
    
    \subfloat[Cell Type Clustering\label{fig:mvc_ground_truth_cell_types}]{\includegraphics[width=0.8\textwidth]{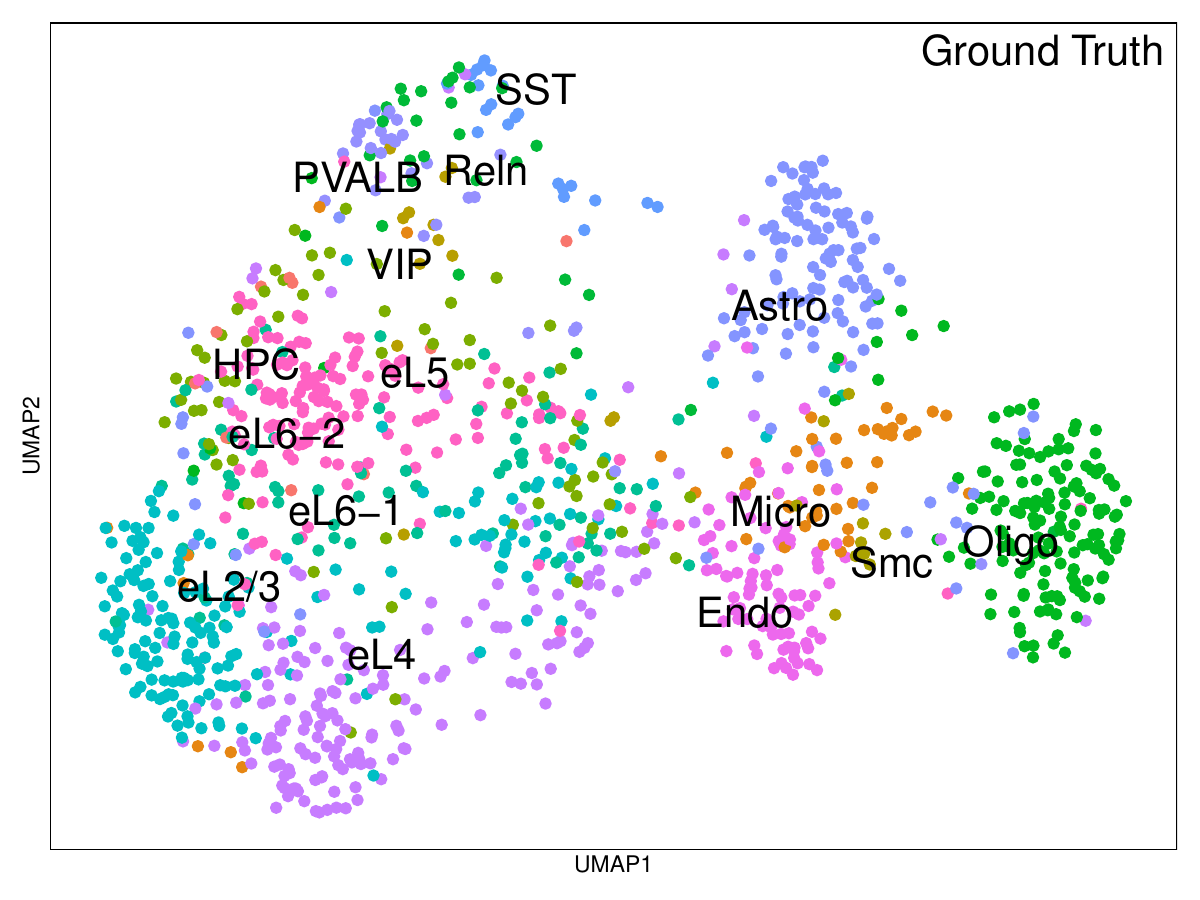}} 

    \subfloat[Cell Type Compositions\label{fig:mvc_ground_truth_cell_compositions}]{\includegraphics[width=\textwidth]{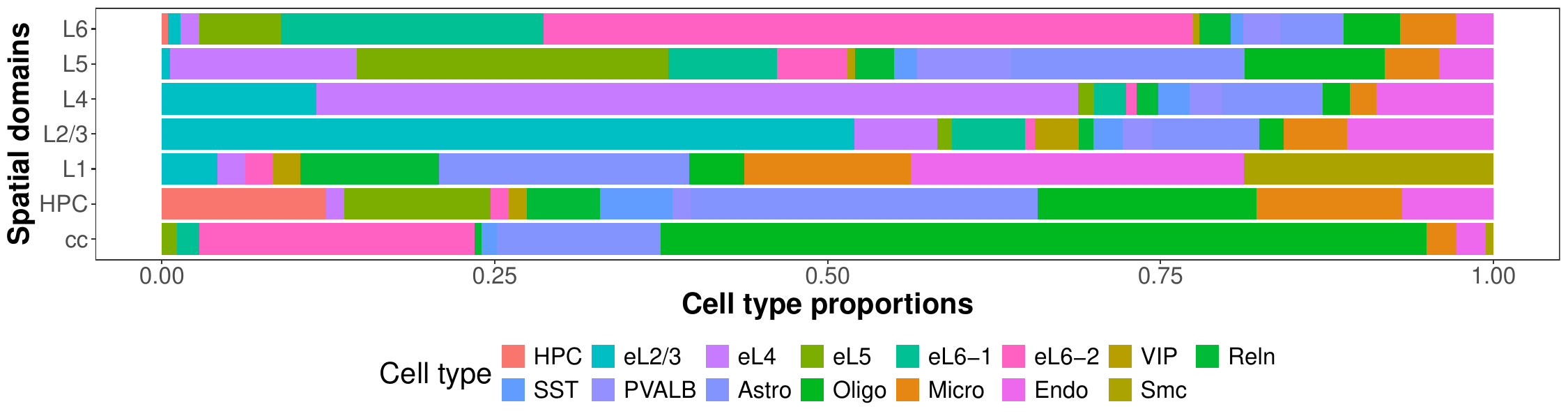}} 

    \caption{Ground truth clustering for the MVC dataset: a) spatial domain clustering, b) cell type clustering, and c) cell compositions.}
    \label{fig:mvc_ground_truths}
\end{figure}

\begin{figure}

    \centering
    \includegraphics[width=\textwidth]{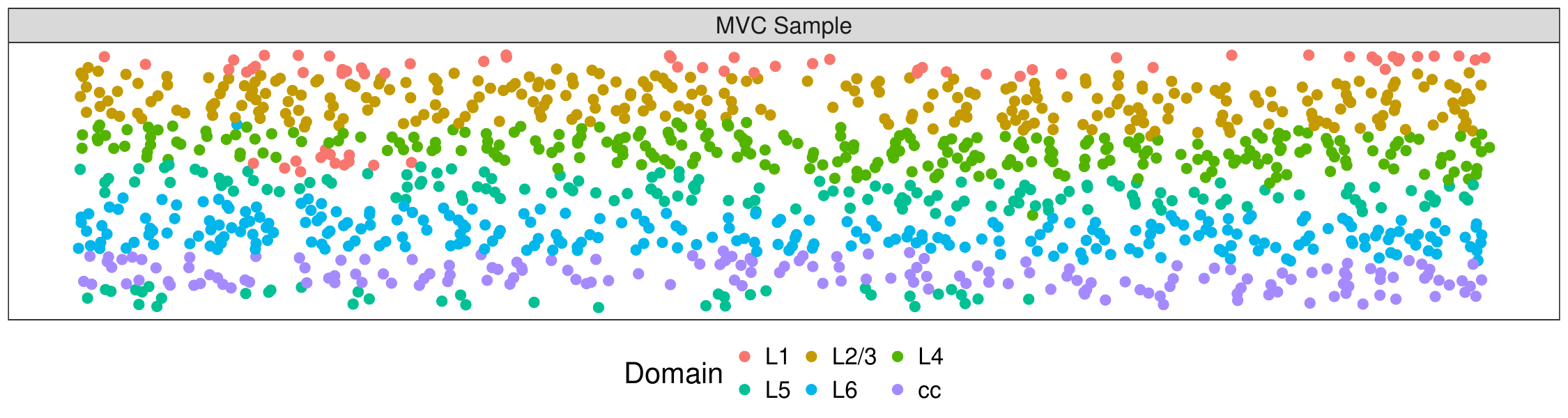}

    \caption{For the MVC dataset, spatial clustering results for BayesClint. This should be compared with the ground truth in Figure \ref{fig:mvc_ground_truth_spatial_domains}.}
    \label{fig:mvc_spatial_domains_clustering_results_bayesclint}
\end{figure}

\begin{figure}

    \subfloat[SPC+$k$-means\label{fig:mvc_spc_kmeans_spatial_domains}]{\includegraphics[width=\textwidth]{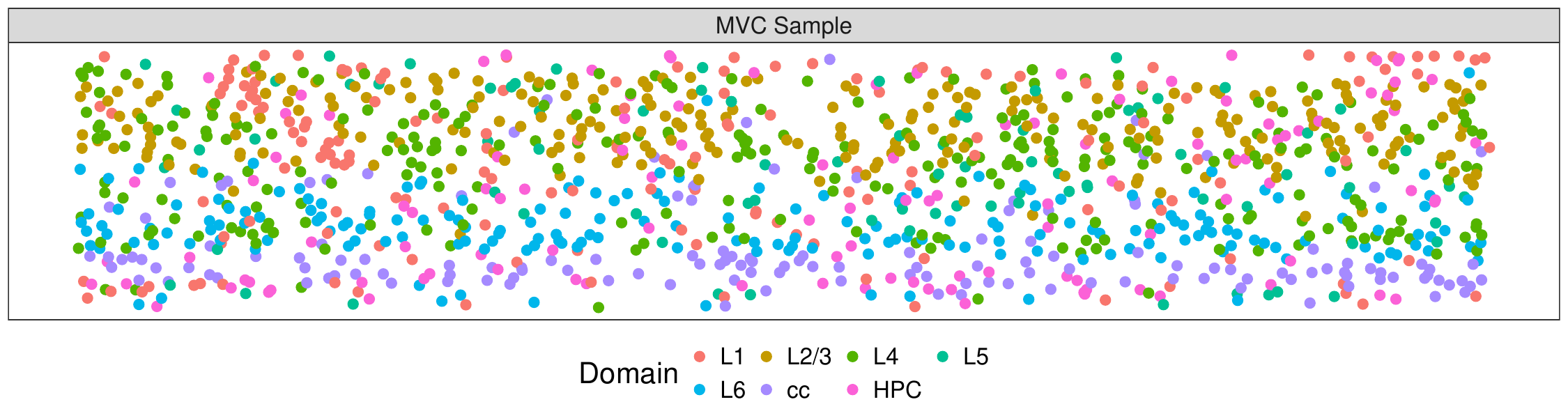}}

    \subfloat[SPC+mclust\label{fig:mvc_spc_mclust_spatial_domains}]{\includegraphics[width=\textwidth]{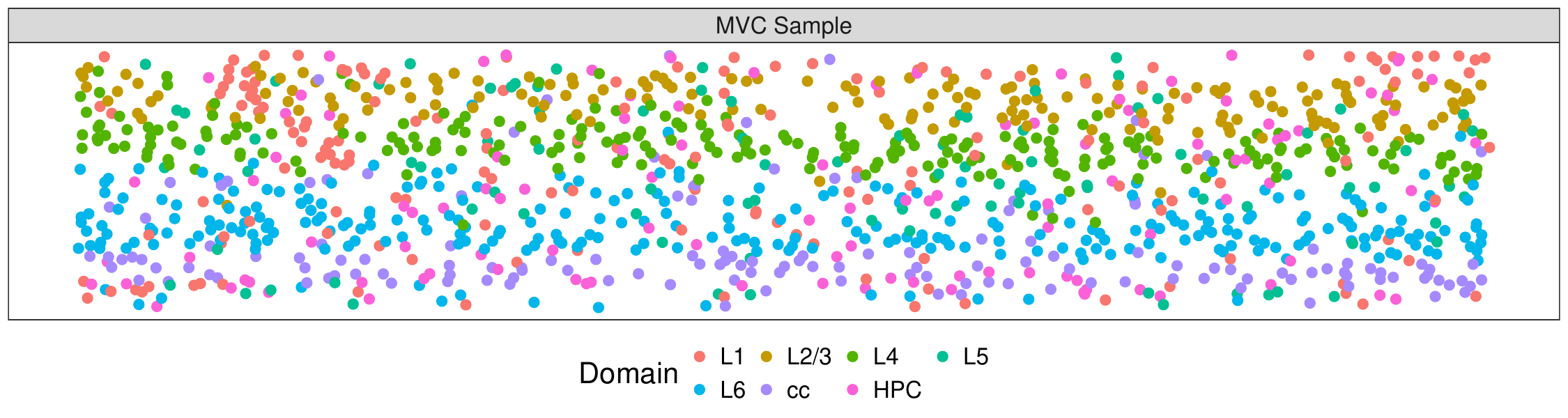}}

    \subfloat[PRECAST\label{fig:mvc_precast_spatial_domains}]{\includegraphics[width=\textwidth]{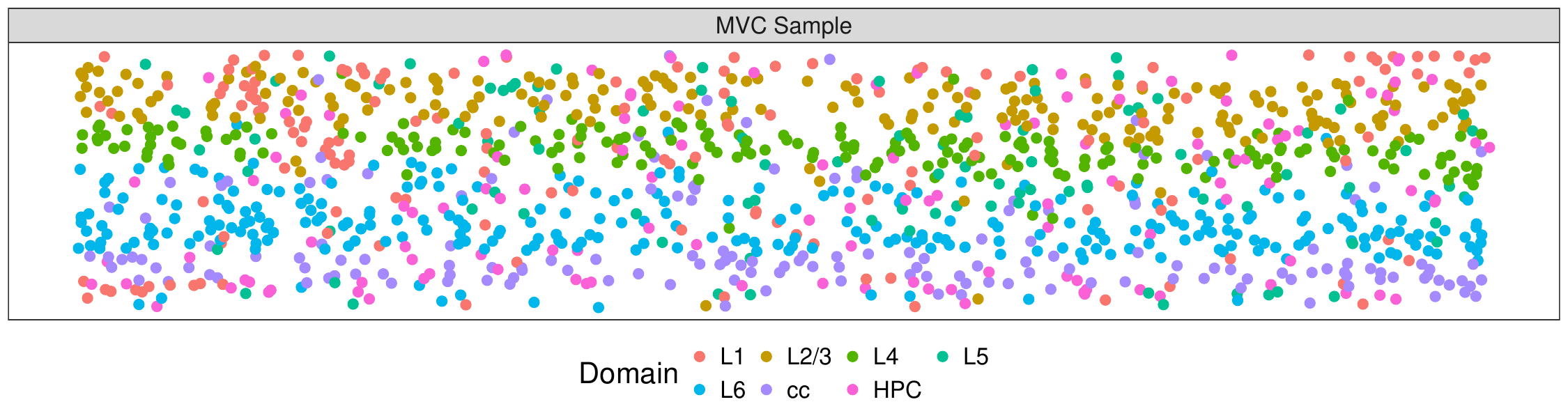}}

    \subfloat[BASS\label{fig:mvc_bass_spatial_domains}]{\includegraphics[width=\textwidth]{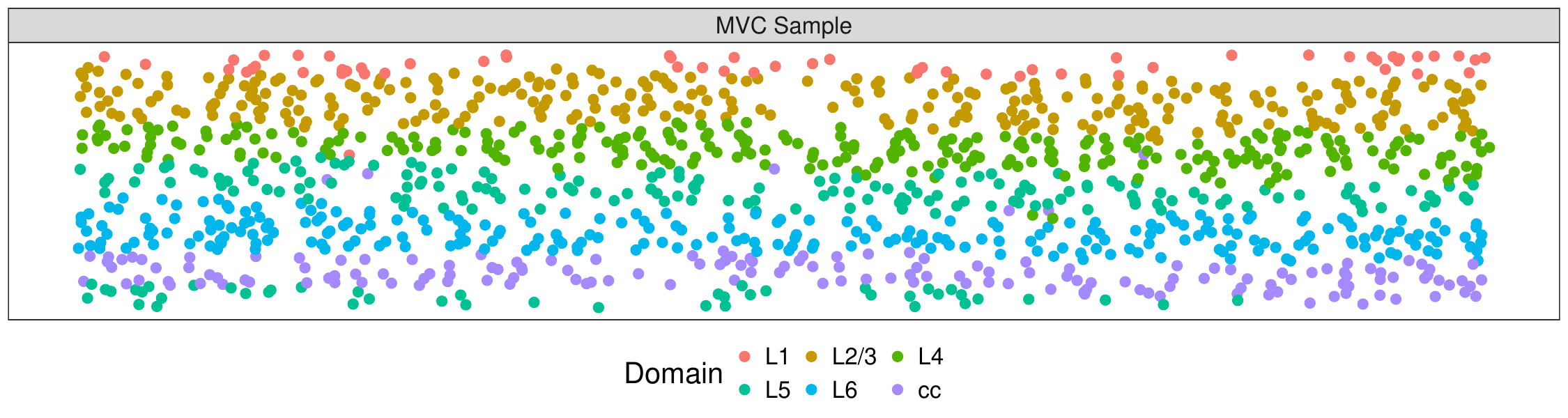}}

    \caption{For the MVC dataset, spatial clustering results for competing methods. This should be compared with the ground truth in Figure \ref{fig:mvc_ground_truth_spatial_domains}.}
    \label{fig:mvc_spatial_domains_clustering_results}
\end{figure}

\begin{figure}

    \centering
    \subfloat[BayesClint\label{fig:mvc_bayesclint_umap_cell_types}]{\includegraphics[width=0.5\textwidth]{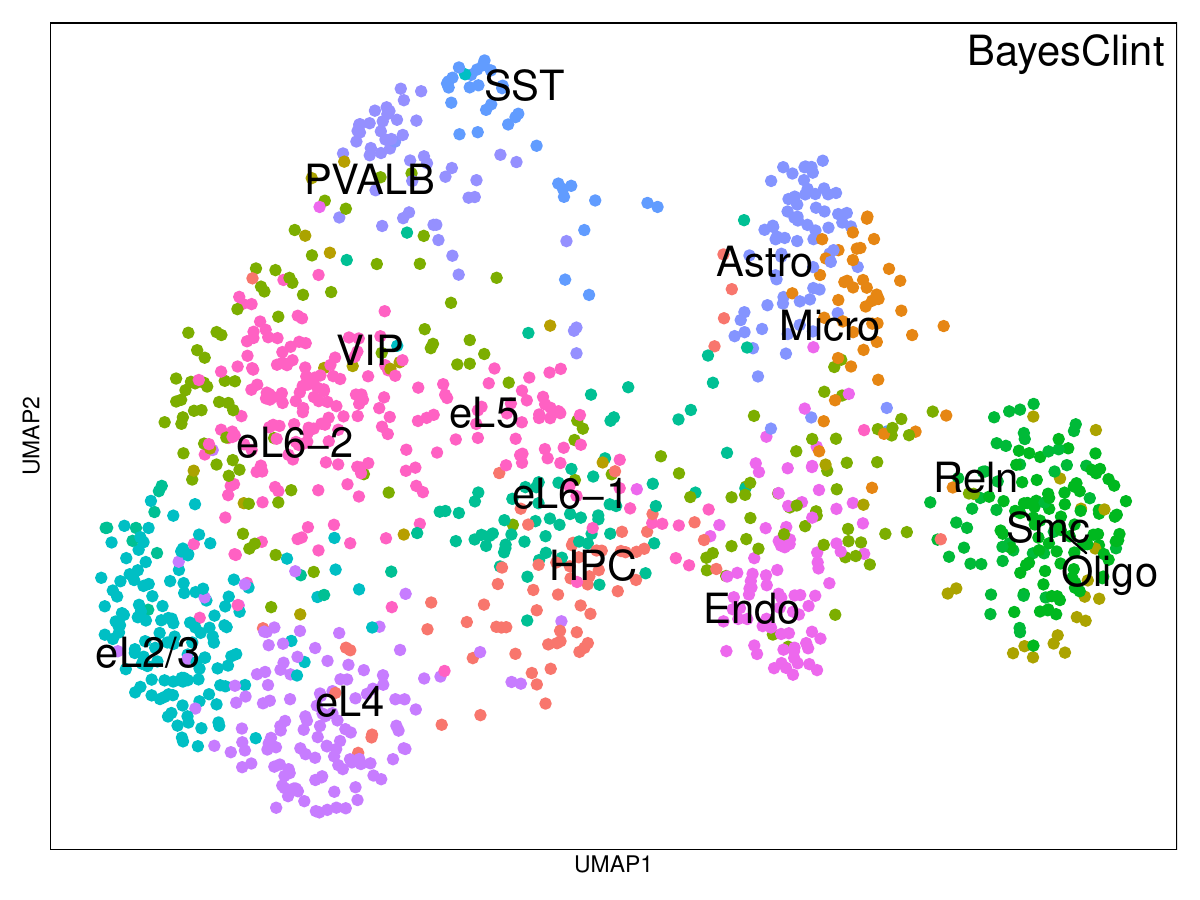}}
    \subfloat[SPC+$k$-means\label{fig:mvc_spc_kmeans_umap_cell_types}]{\includegraphics[width=0.5\textwidth]{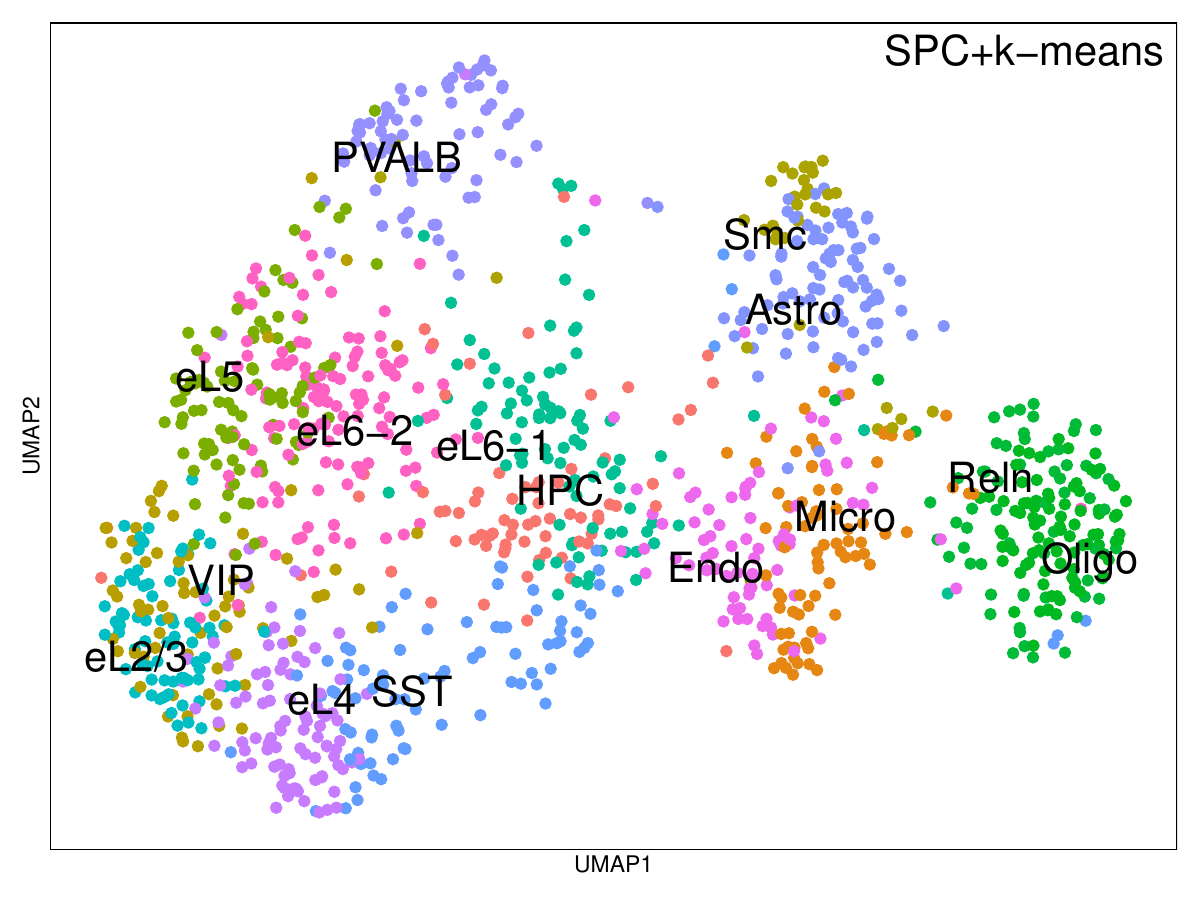}}

    \subfloat[SPC+mclust\label{fig:mvc_spc_mclust_umap_cell_types}]{\includegraphics[width=0.5\textwidth]{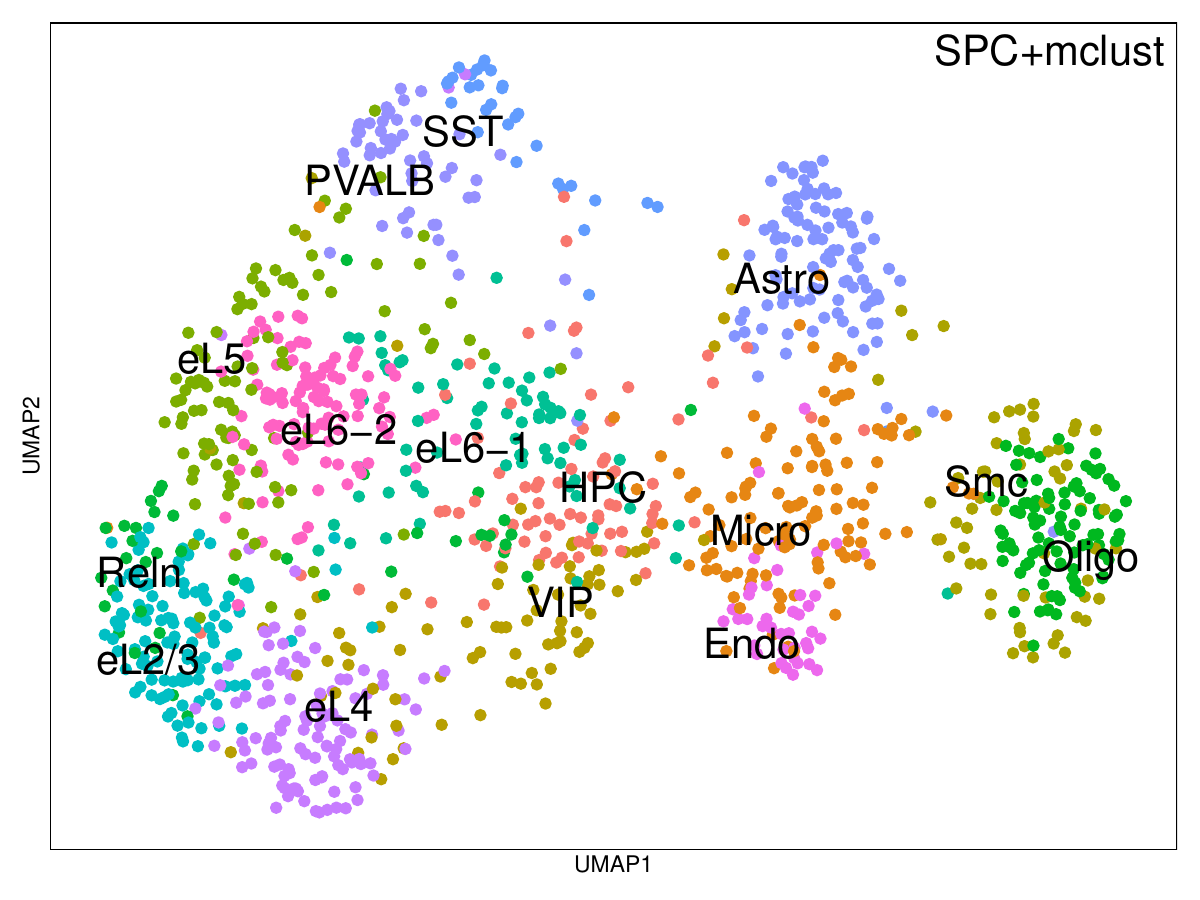}}
    \subfloat[PRECAST\label{fig:mvc_precast_umap_cell_types}]{\includegraphics[width=0.5\textwidth]{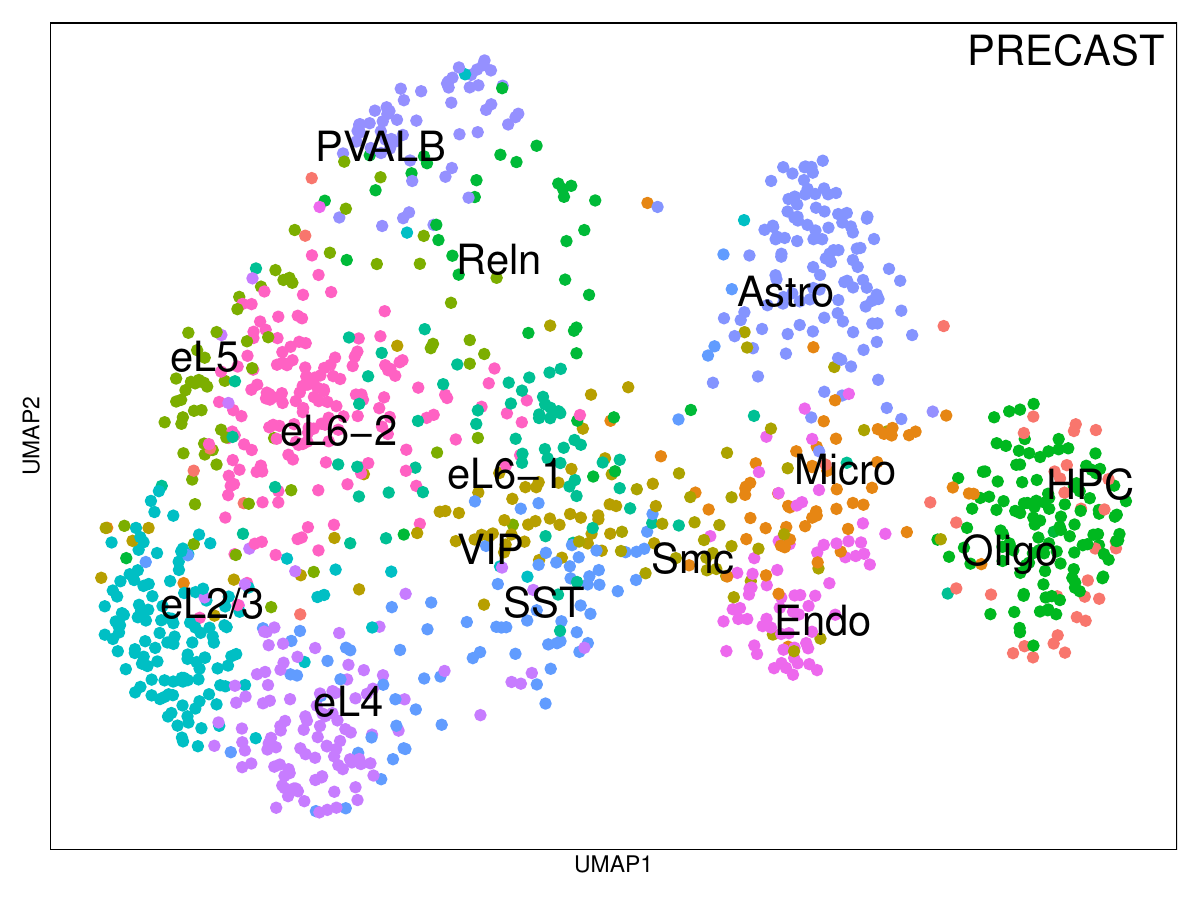}}

    \subfloat[BASS\label{fig:mvc_bass_umap_cell_types}]{\includegraphics[width=0.5\textwidth]{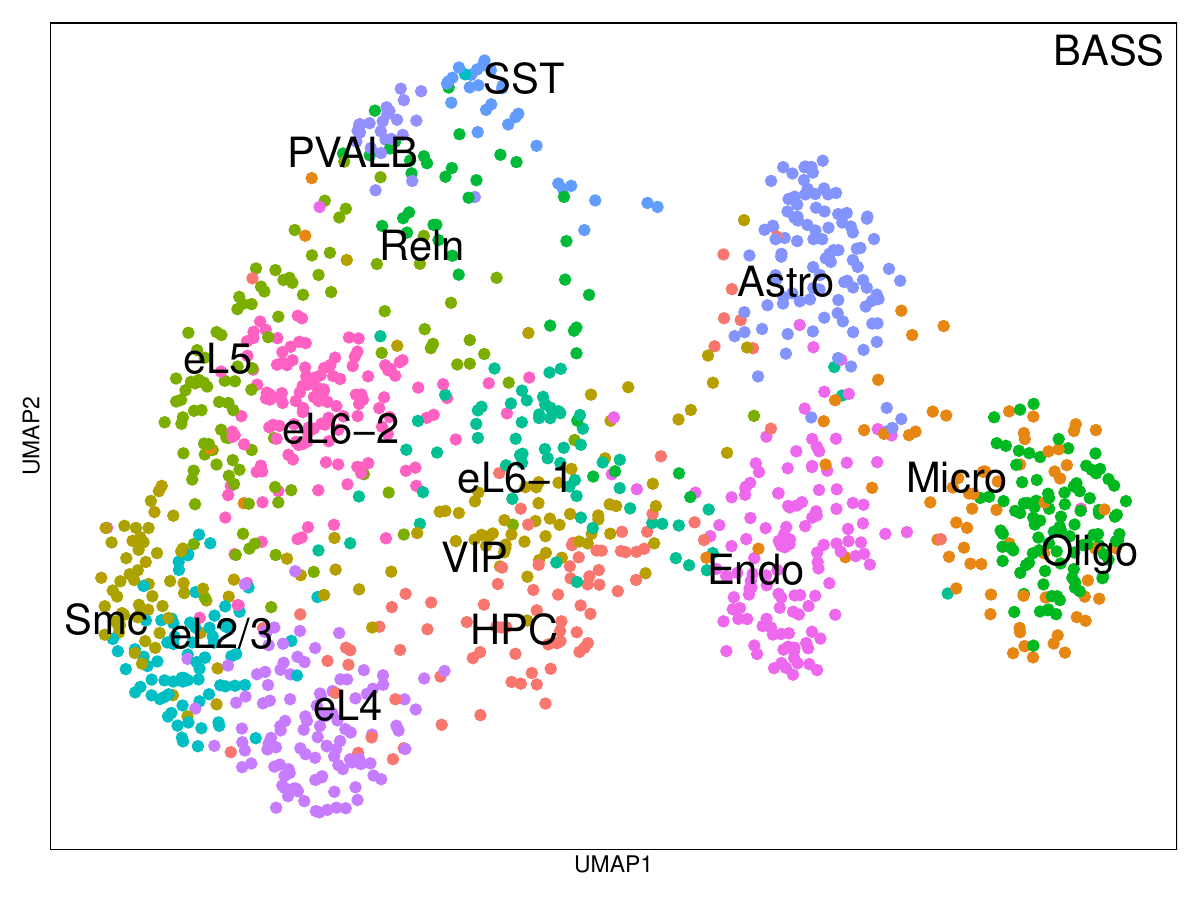}}

    \caption{For the MVC dataset, cell type clustering results. This should be compared with the ground truth in Figure \ref{fig:mvc_ground_truth_cell_types}.}
    \label{fig:mvc_cell_types_clustering_results}
\end{figure}

\begin{figure}

    \centering
    \subfloat[BayesClint\label{fig:mvc_bayesclint_cell_compositions}]{\includegraphics[width=\textwidth]{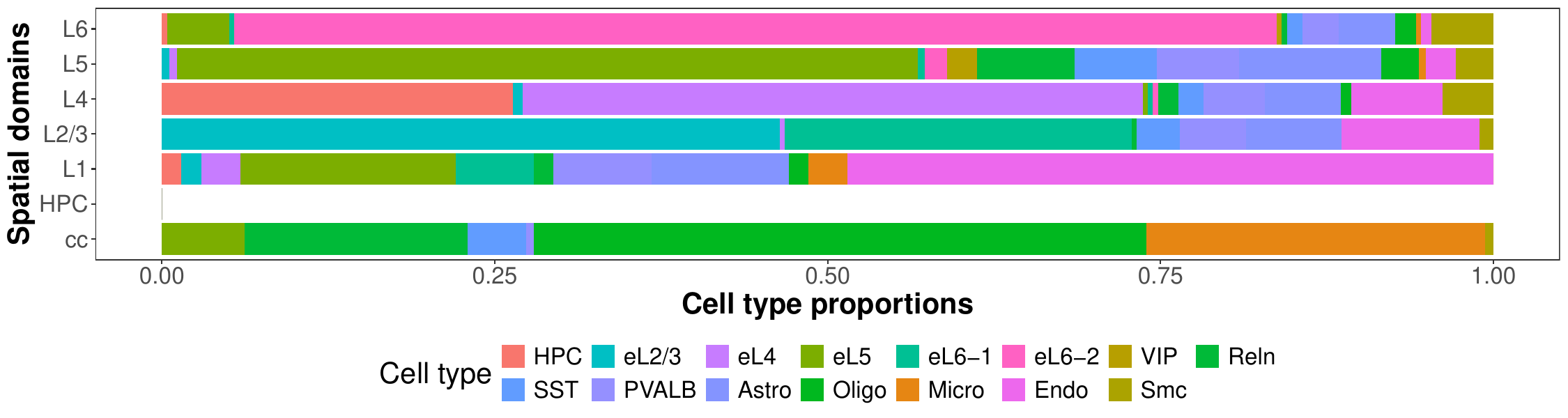}}

    \subfloat[BASS\label{fig:mvc_bass_cell_compositions}]{\includegraphics[width=\textwidth]{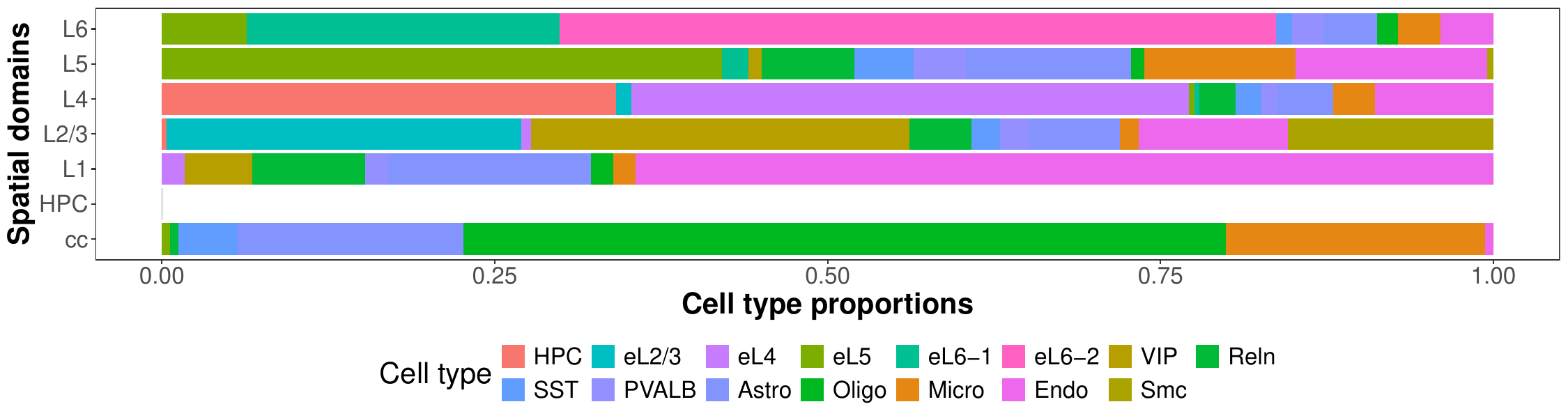}}

    \caption{For the MVC dataset, cell type composition results. This should be compared with the ground truth in Figure \ref{fig:mvc_ground_truth_cell_compositions}.}
    \label{fig:mvc_cell_type_composition_results}
\end{figure}

\begin{figure}
    \centering
    \includegraphics[width=0.75\linewidth]{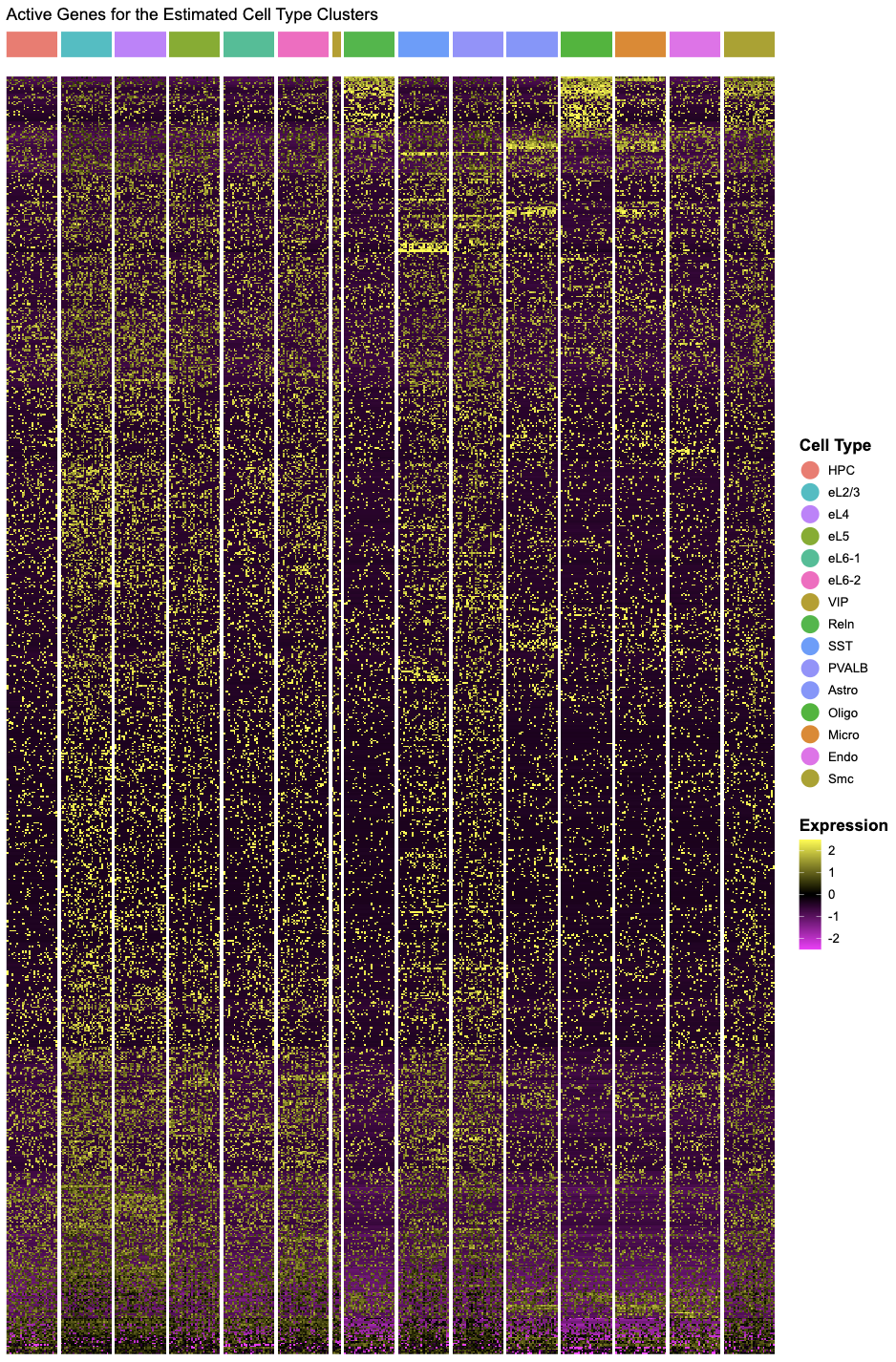}
    \caption{For the MVC dataset, gene expressions for the active genes selected by BayesClint. To create the heatmap, we first hierarchically cluster the genes' log-normalized expression values to group them. The row-wise ordering of the genes in the pair of heatmaps for MVC are made consistent for the set of active genes and the subset of differentiating genes. We cluster the cells according to the cell type clusters returned from BayesClint. We randomly downsample to 30 cells per cell type, for visualization purposes. Finally, we used doHeatmap() function from the PRECAST R package on the scaled values (to display the relative expression level centered at zero).}
    \label{fig:mvc_bayesclint_active}
\end{figure}

\begin{figure}
    \centering
    \includegraphics[width=0.75\linewidth]{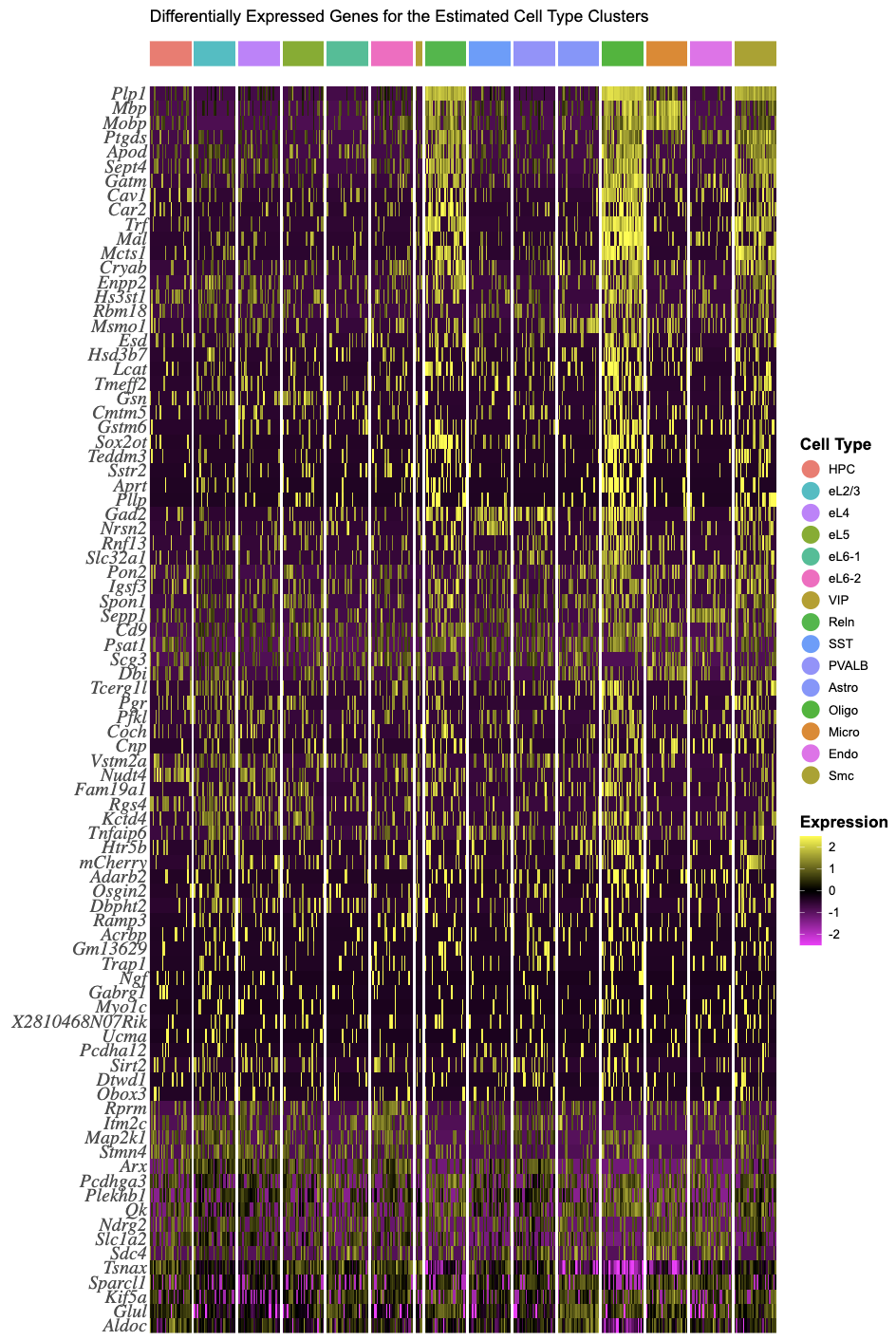}
    \caption{For the MVC dataset, gene expressions for the subset of 86 genes associated with the top-ranked component (i.e., the component with the minimum combined $p$-value), selected by BayesClint. To create the heatmap, we first hierarchically cluster the genes' log-normalized expression values to group them. The row-wise ordering of the genes in the pair of heatmaps for MVC are made consistent for the set of active genes and the subset of genes associated with the top-ranked component. We cluster the cells according to the cell type clusters returned from BayesClint. We randomly downsample to 30 cells per cell type, for visualization purposes. Finally, we used doHeatmap() function from the PRECAST R package on the scaled values (to display the relative expression level centered at zero).}
    \label{fig:mvc_bayesclint_differentiating}
\end{figure}


\begin{figure}
    \centering
    \includegraphics[width=\linewidth]{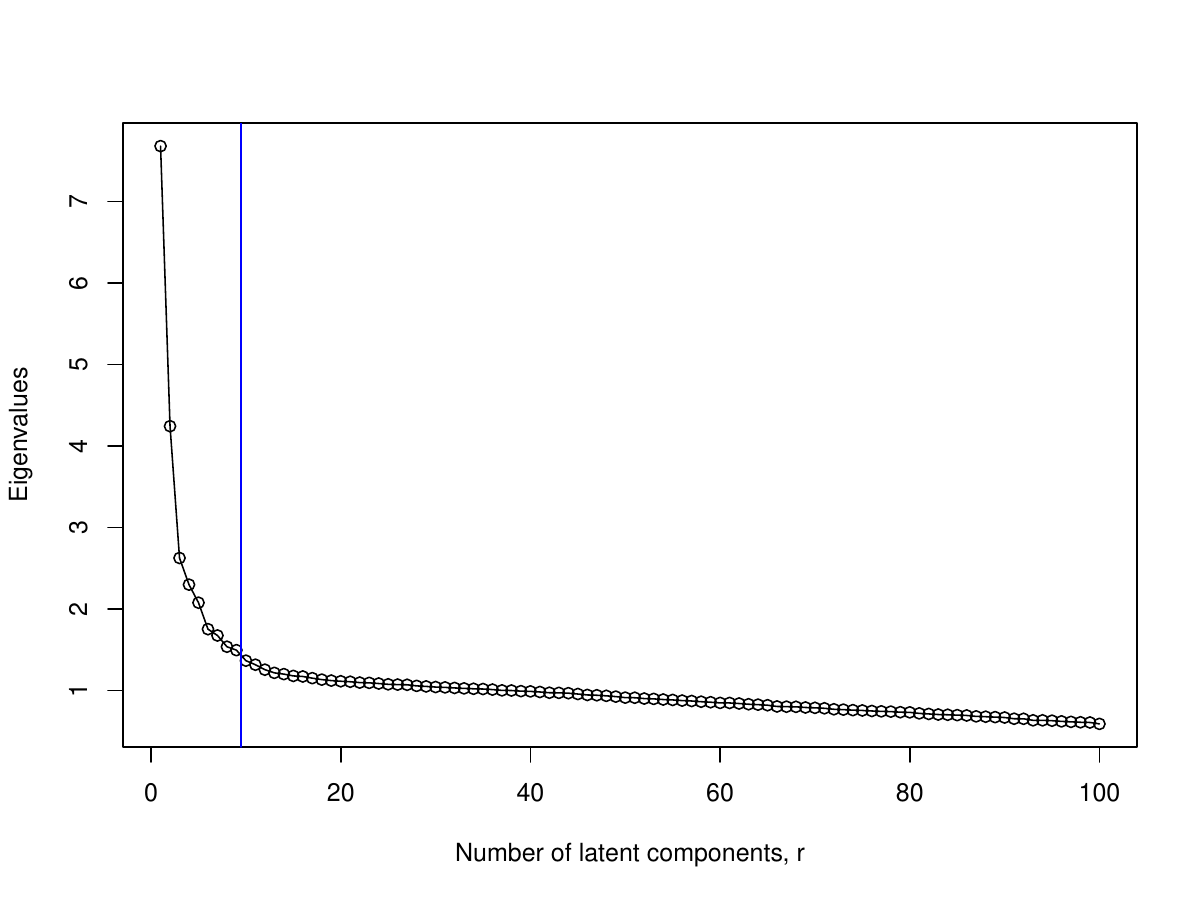}
    \caption{Scree plot for the mPFC dataset, focusing on the first 100 eigenvalues. According to the elbow method (indicated by the blue line), we select the number of latent components $r$ to be 9.}
    \label{fig:mpfc_screeplot}
\end{figure}

\begin{figure}
    \centering
    \includegraphics[width=1\linewidth]{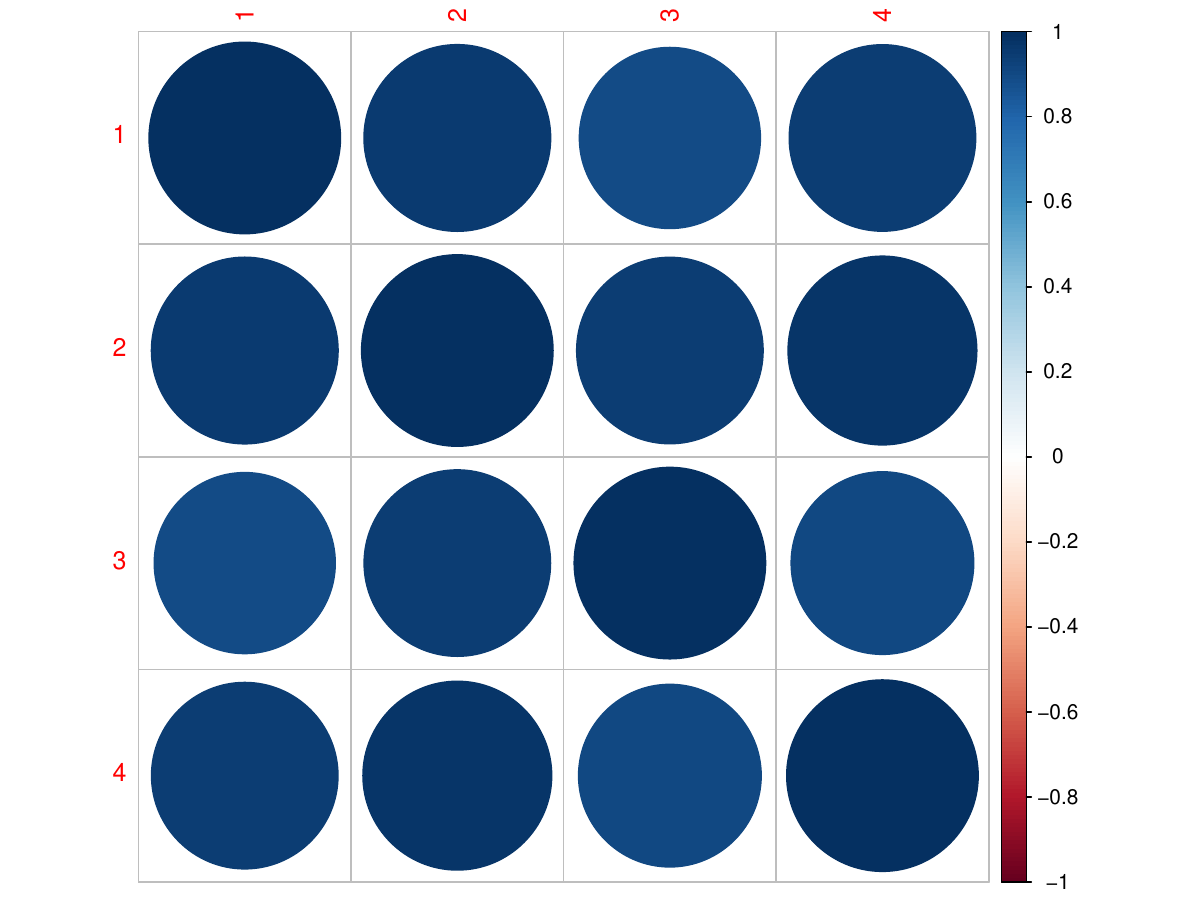}
    \caption{Pairwise correlation coefficients across the four MCMC chains, between the PPI values of all $P$ genes, for the mPFC dataset.}
    \label{fig:mpfc_corrplot_4_chains}
\end{figure}

\begin{figure}

    \centering

    \subfloat[Spatial Domain Clustering\label{fig:mpfc_ground_truth_spatial_domains}]{\includegraphics[width=\textwidth]{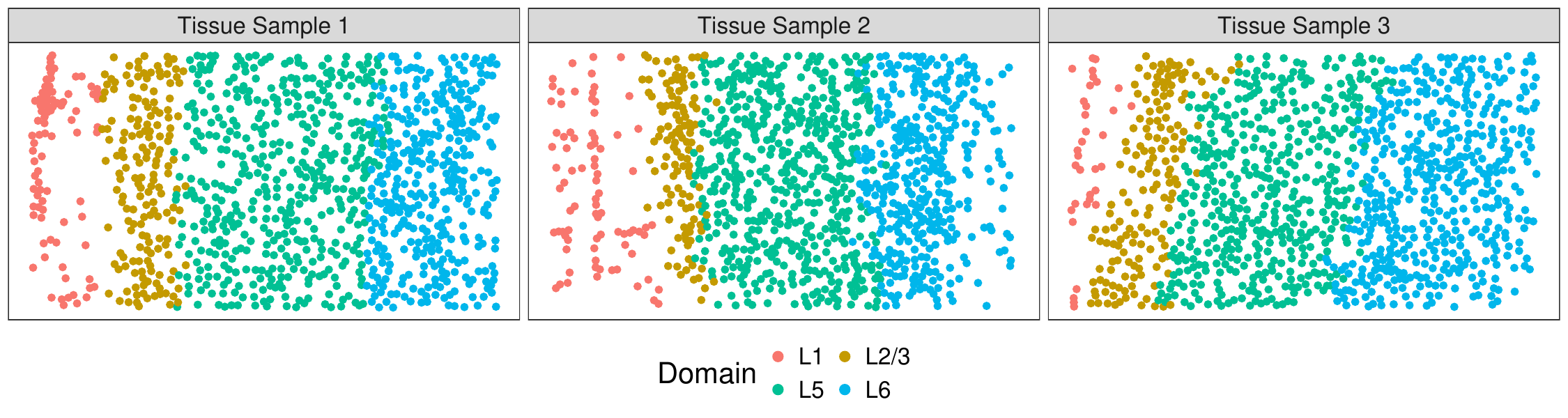}} 
    
    \subfloat[Cell Type Clustering\label{fig:mpfc_ground_truth_cell_types}]{\includegraphics[width=0.8\textwidth]{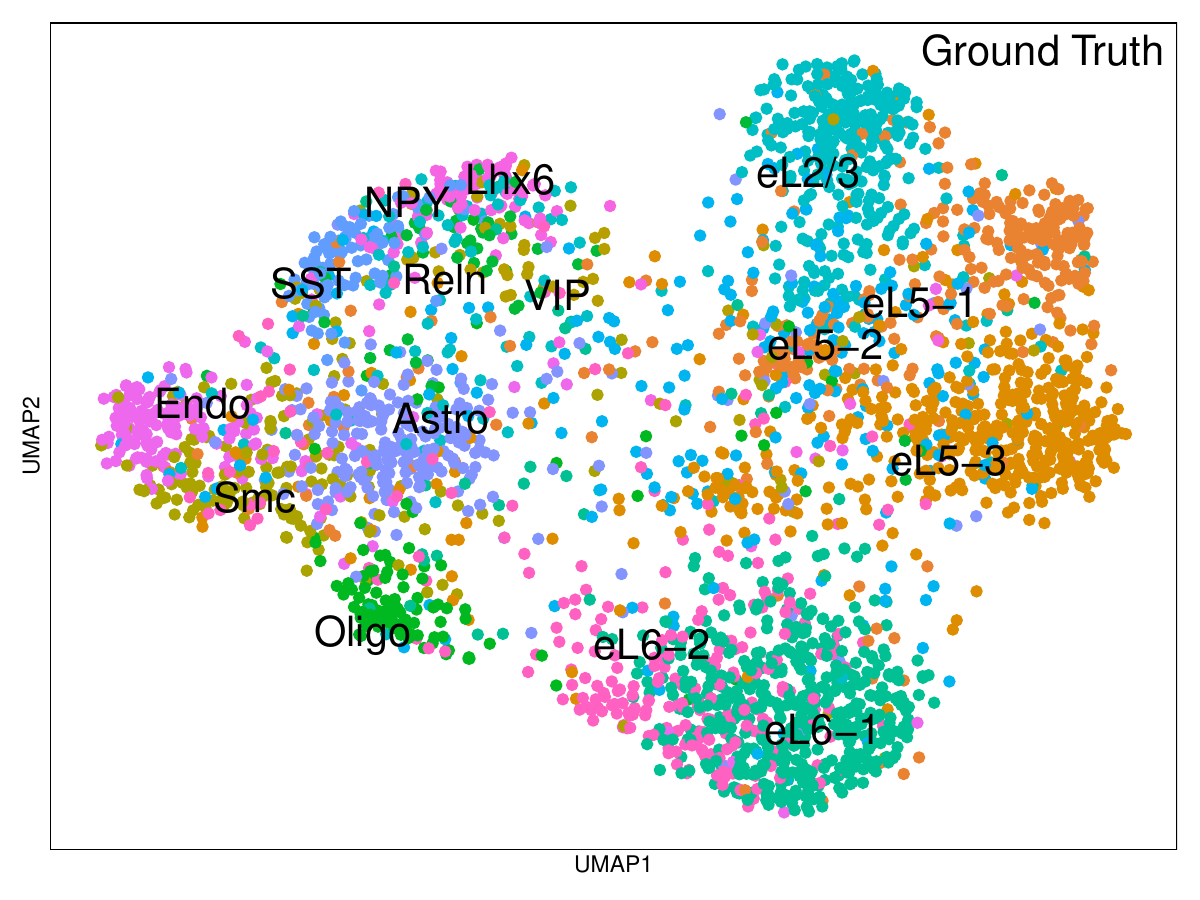}} 

    \subfloat[Cell Type Compositions\label{fig:mpfc_ground_truth_cell_compositions}]{\includegraphics[width=\textwidth]{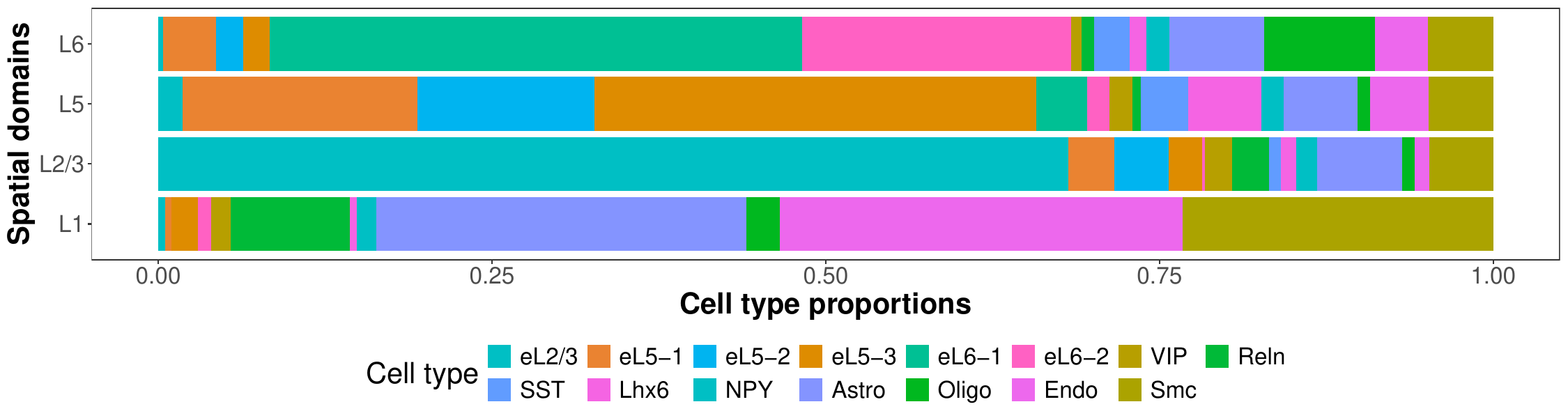}} 

    \caption{Ground truth clustering for the mPFC dataset: a) spatial domain clustering, b) cell type clustering, and c) cell compositions.}
    \label{fig:mpfc_ground_truths}
\end{figure}

\begin{figure}

    \centering
    \includegraphics[width=\textwidth]{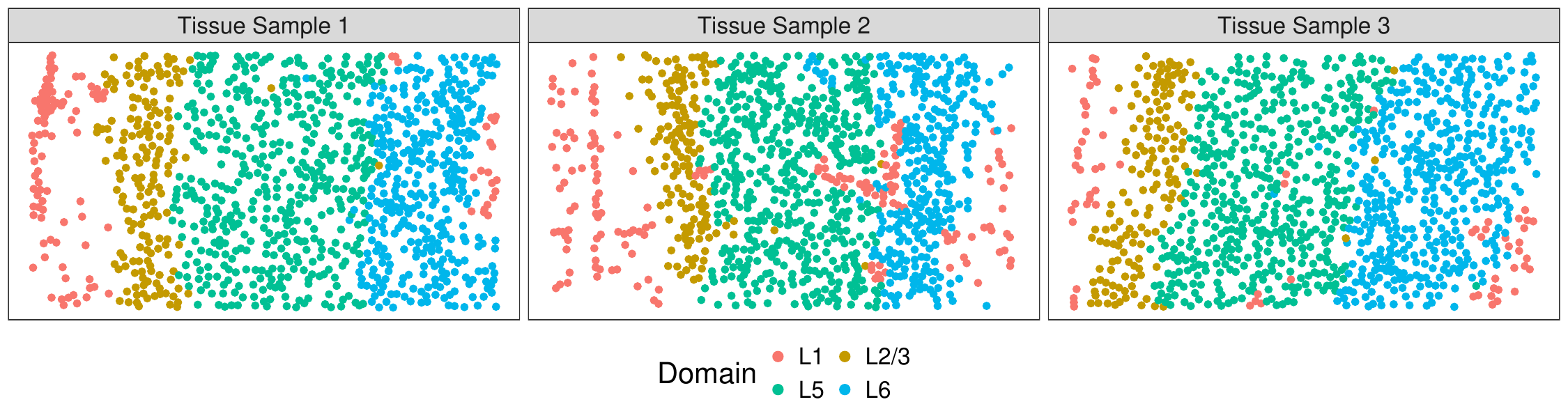}

    \caption{For the mPFC dataset, spatial clustering results for BayesClint. This should be compared with the ground truth in Figure \ref{fig:mpfc_ground_truth_spatial_domains}.}
    \label{fig:mpfc_spatial_domains_clustering_results_bayesclint}
\end{figure}

\begin{figure}

    \subfloat[SPC+$k$-means\label{fig:mpfc_spc_kmeans_spatial_domains}]{\includegraphics[width=\textwidth]{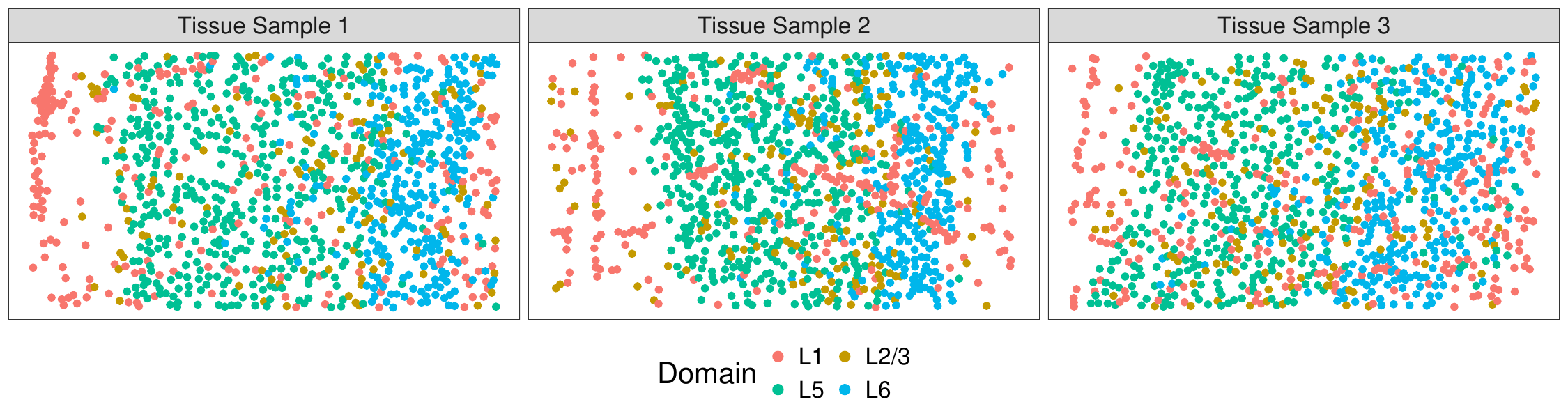}}

    \subfloat[SPC+mclust\label{fig:mpfc_spc_mclust_spatial_domains}]{\includegraphics[width=\textwidth]{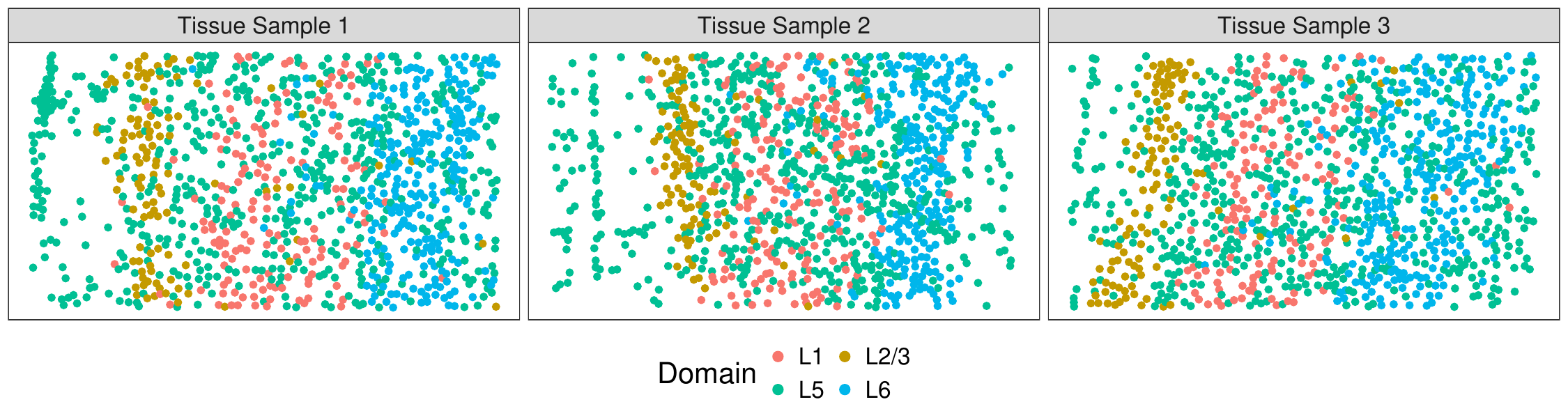}}

    \subfloat[PRECAST\label{fig:mpfc_precast_spatial_domains}]{\includegraphics[width=\textwidth]{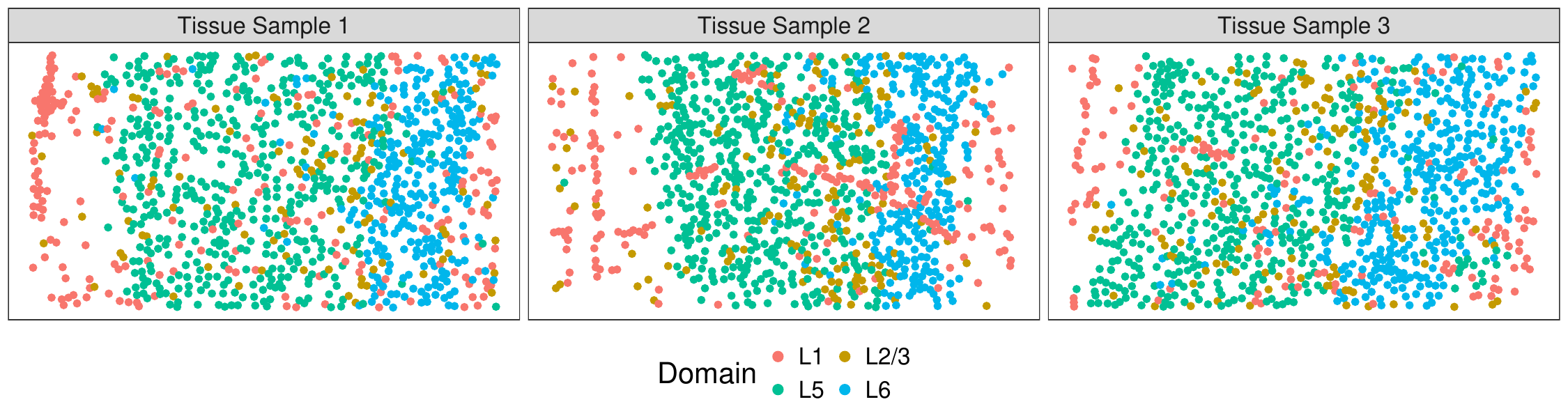}}

    \subfloat[BASS\label{fig:mpfc_bass_spatial_domains}]{\includegraphics[width=\textwidth]{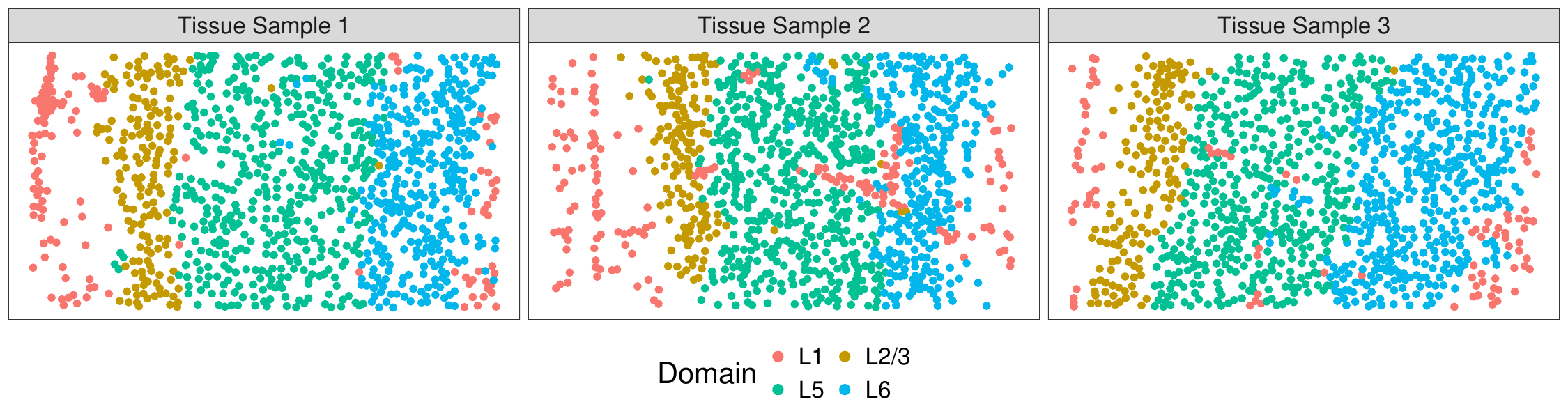}}

    \caption{For the mPFC dataset, spatial clustering results for competing methods. This should be compared with the ground truth in Figure \ref{fig:mpfc_ground_truth_spatial_domains}.}
    \label{fig:mpfc_spatial_domains_clustering_results}
\end{figure}

\begin{figure}

    \centering
    \subfloat[BayesClint\label{fig:mpfc_bayesclint_umap_cell_types}]{\includegraphics[width=0.5\textwidth]{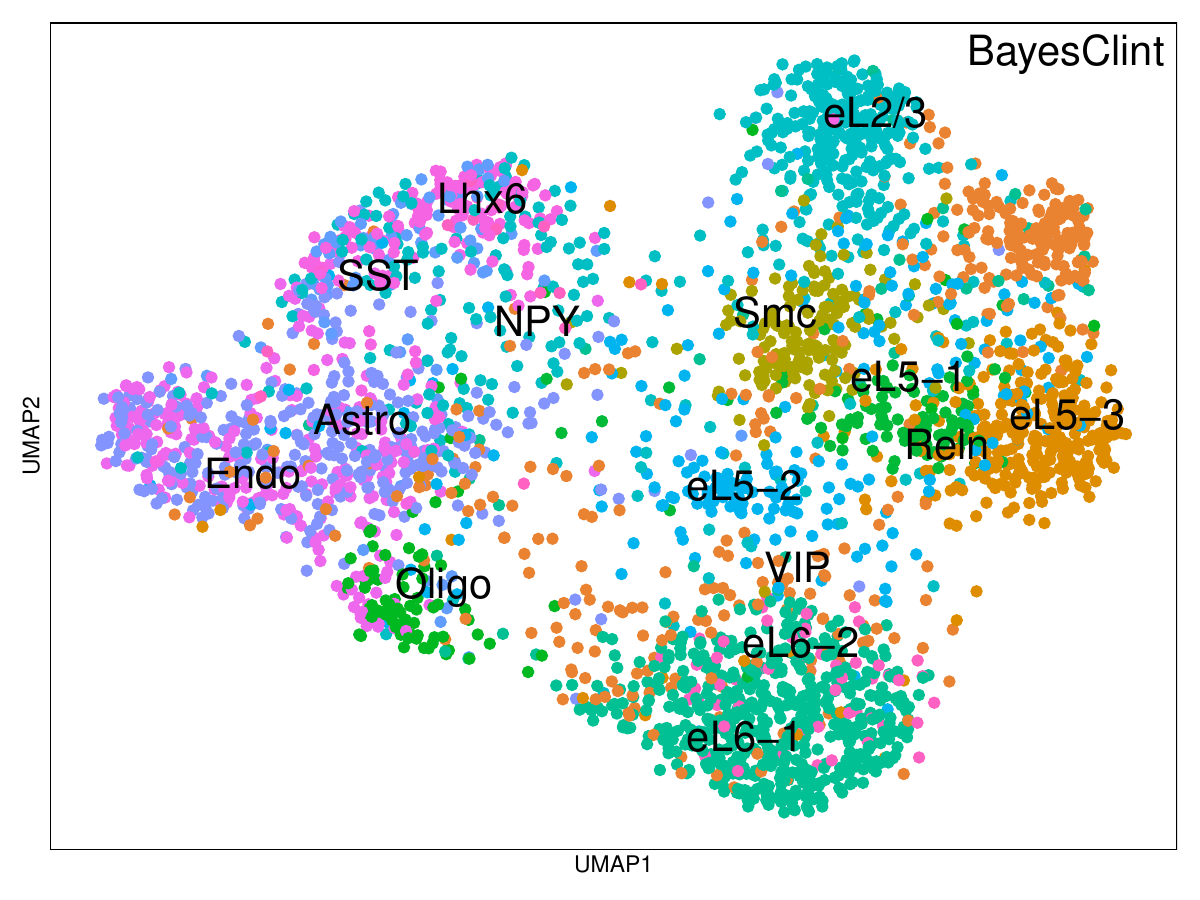}}
    \subfloat[SPC+$k$-means\label{fig:mpfc_spc_kmeans_umap_cell_types}]{\includegraphics[width=0.5\textwidth]{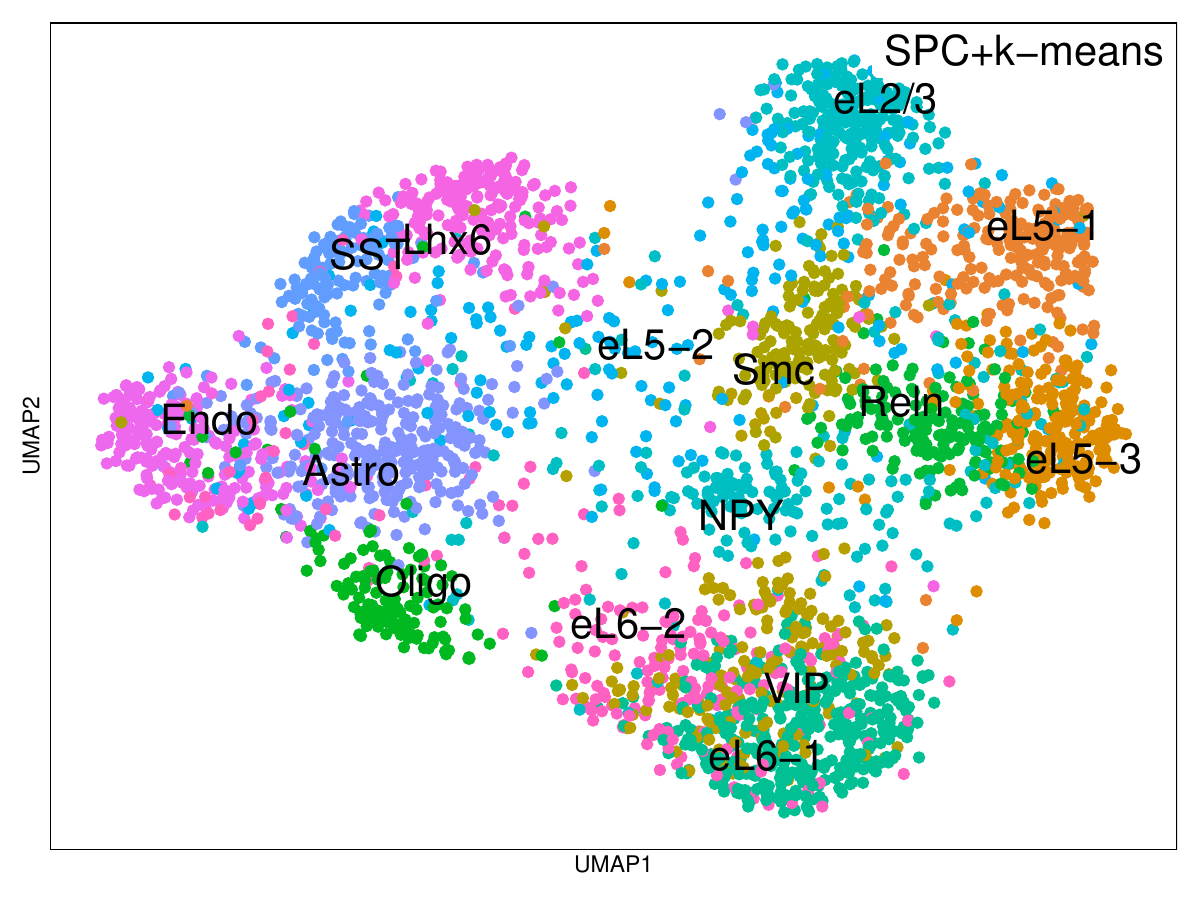}}

    \subfloat[SPC+mclust\label{fig:mpfc_spc_mclust_umap_cell_types}]{\includegraphics[width=0.5\textwidth]{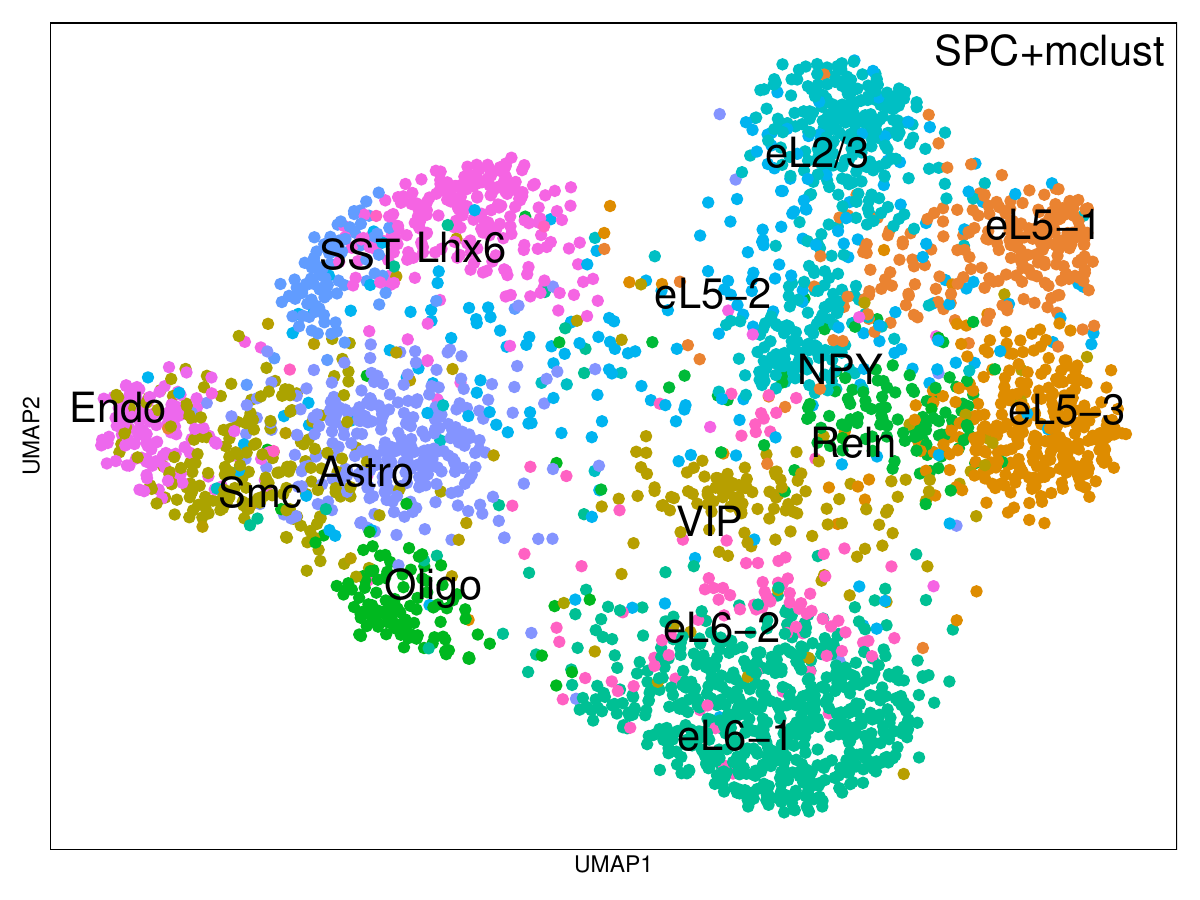}}
    \subfloat[PRECAST\label{fig:mpfc_precast_umap_cell_types}]{\includegraphics[width=0.5\textwidth]{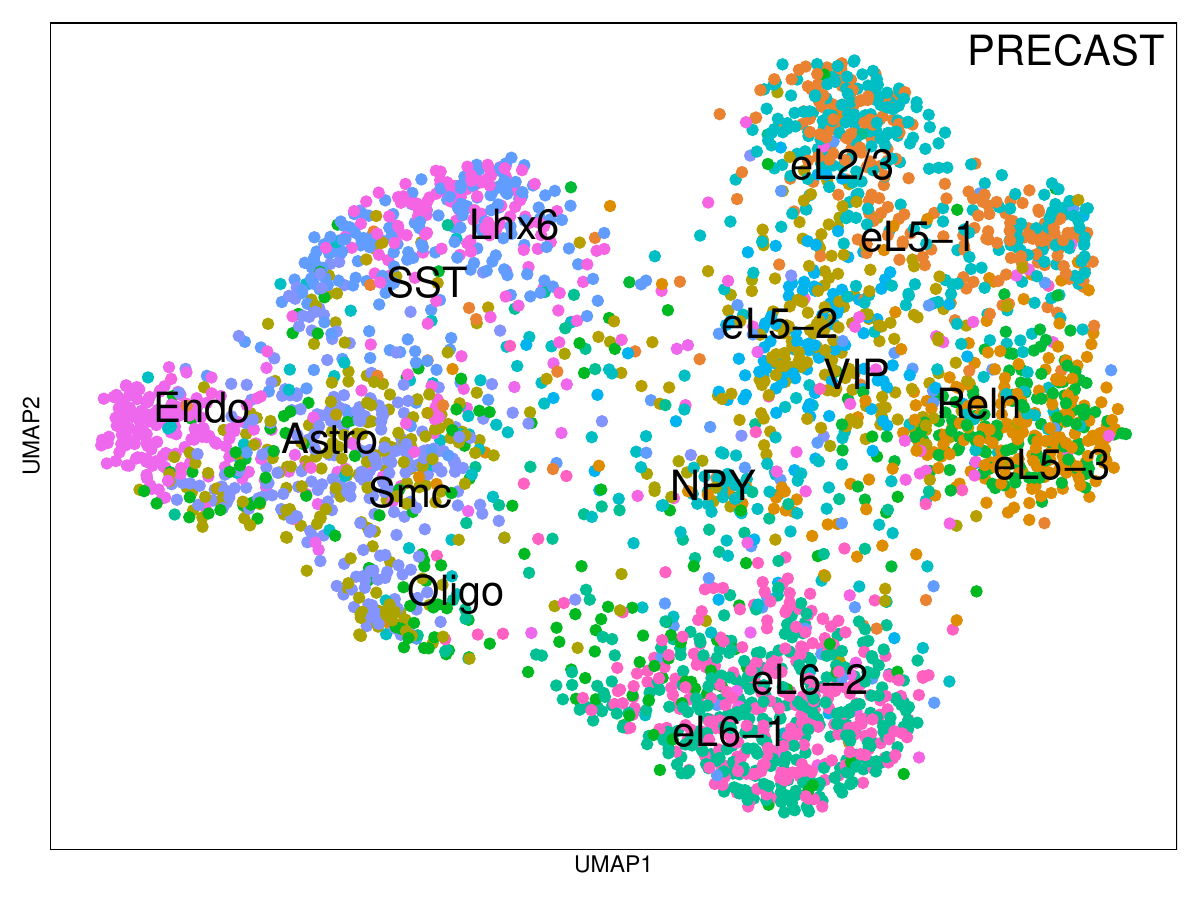}}

    \subfloat[BASS\label{fig:mpfc_bass_umap_cell_types}]{\includegraphics[width=0.5\textwidth]{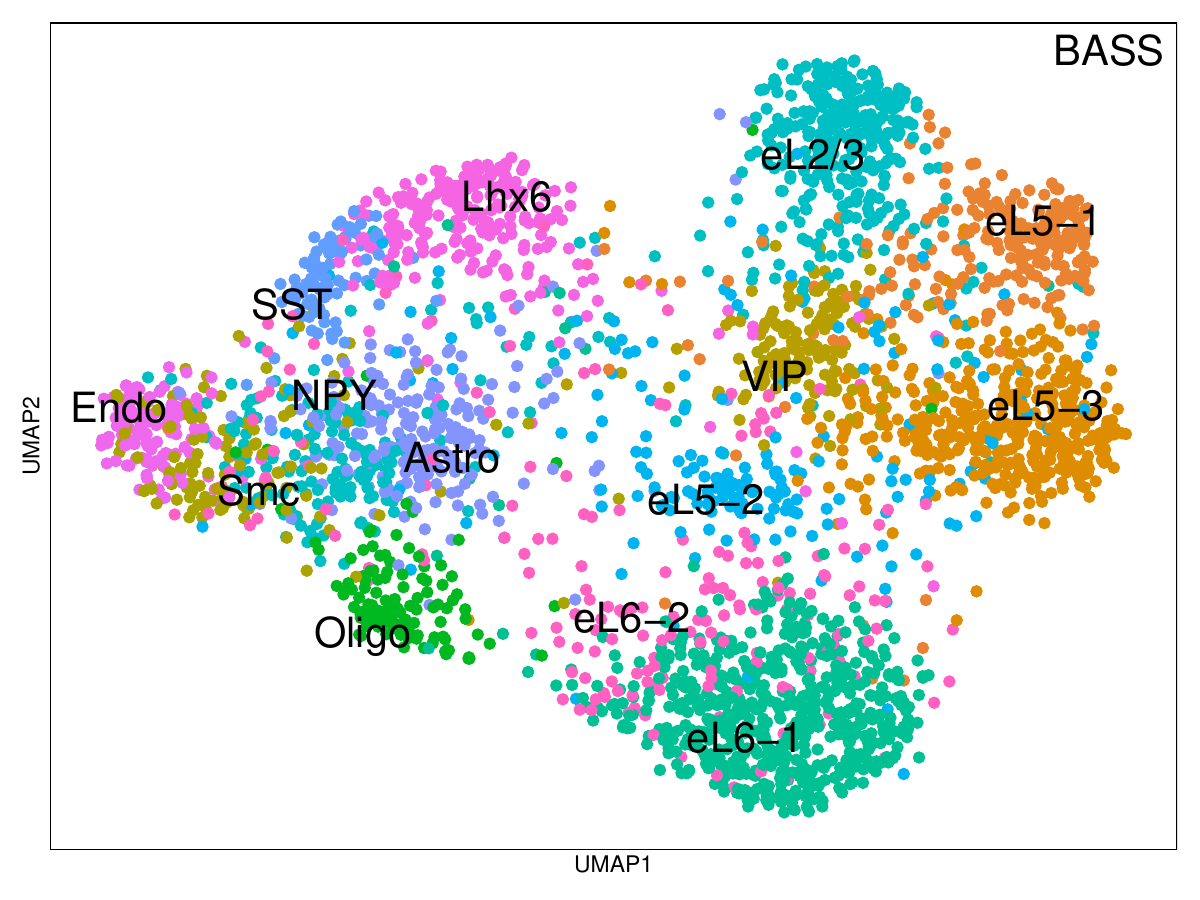}}

    \caption{For the mPFC dataset, cell type clustering results. This should be compared with the ground truth in Figure \ref{fig:mpfc_ground_truth_cell_types}.}
    \label{fig:mpfc_cell_types_clustering_results}
\end{figure}

\begin{figure}

    \centering
    \subfloat[BayesClint\label{fig:mpfc_bayesclint_cell_compositions}]{\includegraphics[width=\textwidth]{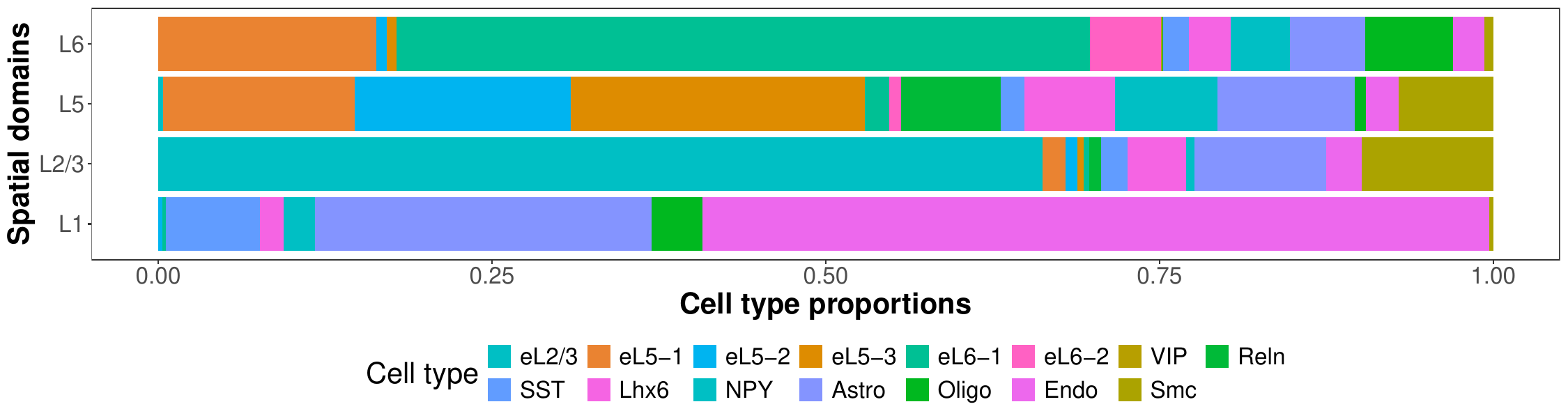}}

    \subfloat[BASS\label{fig:mpfc_bass_cell_compositions}]{\includegraphics[width=\textwidth]{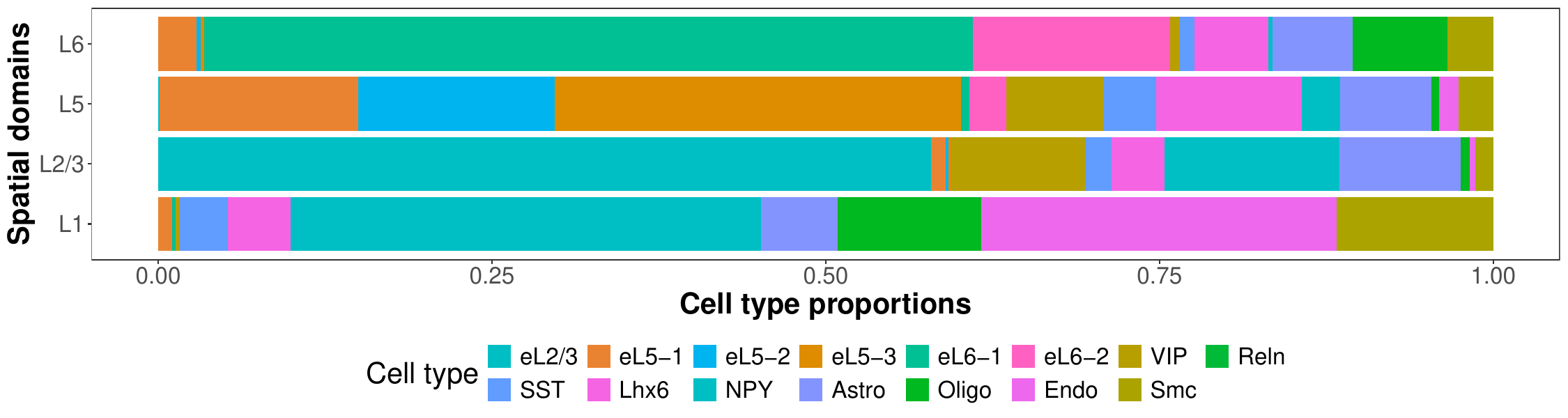}}

    \caption{For the mPFC dataset, cell type composition results. This should be compared with the ground truth in Figure \ref{fig:mpfc_ground_truth_cell_compositions}.}
    \label{fig:mpfc_cell_type_composition_results}
\end{figure}

\begin{figure}
    \centering
    \includegraphics[width=0.75\linewidth]{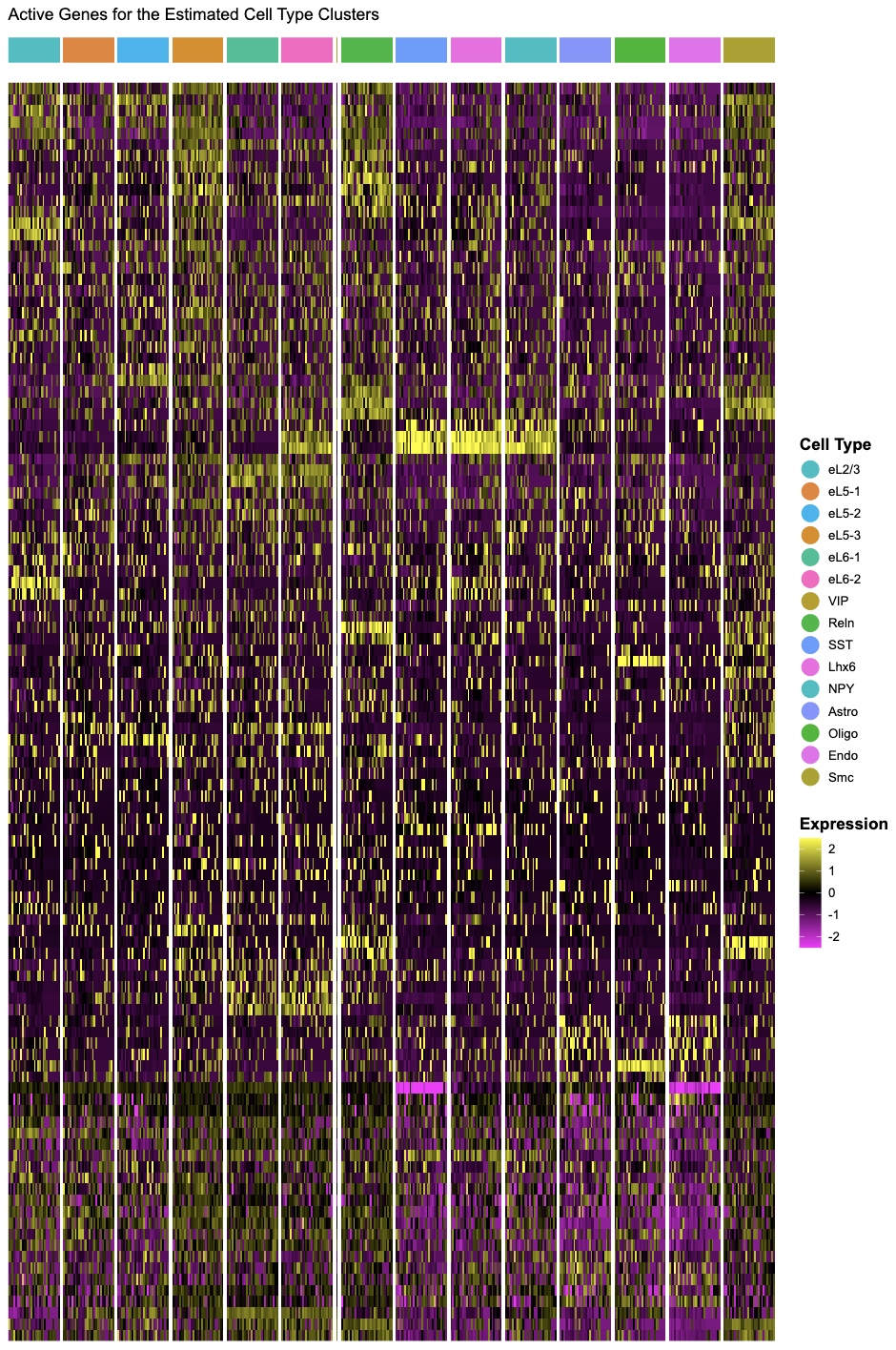}
    \caption{For the mPFC dataset, gene expressions for the active genes selected by BayesClint. To create the heatmap, we first hierarchically cluster the genes' log-normalized expression values to group them. The row-wise ordering of the genes in the pair of heatmaps for mPFC are made consistent for the set of active genes and the subset of differentiating genes. We cluster the cells according to the cell type clusters returned from BayesClint. We randomly downsample to 30 cells per cell type, for visualization purposes. Finally, we used doHeatmap() function from the PRECAST R package on the scaled values (to display the relative expression level centered at zero).}
    \label{fig:mpfc_bayesclint_active}
\end{figure}

\begin{figure}
    \centering
    \includegraphics[width=0.75\linewidth]{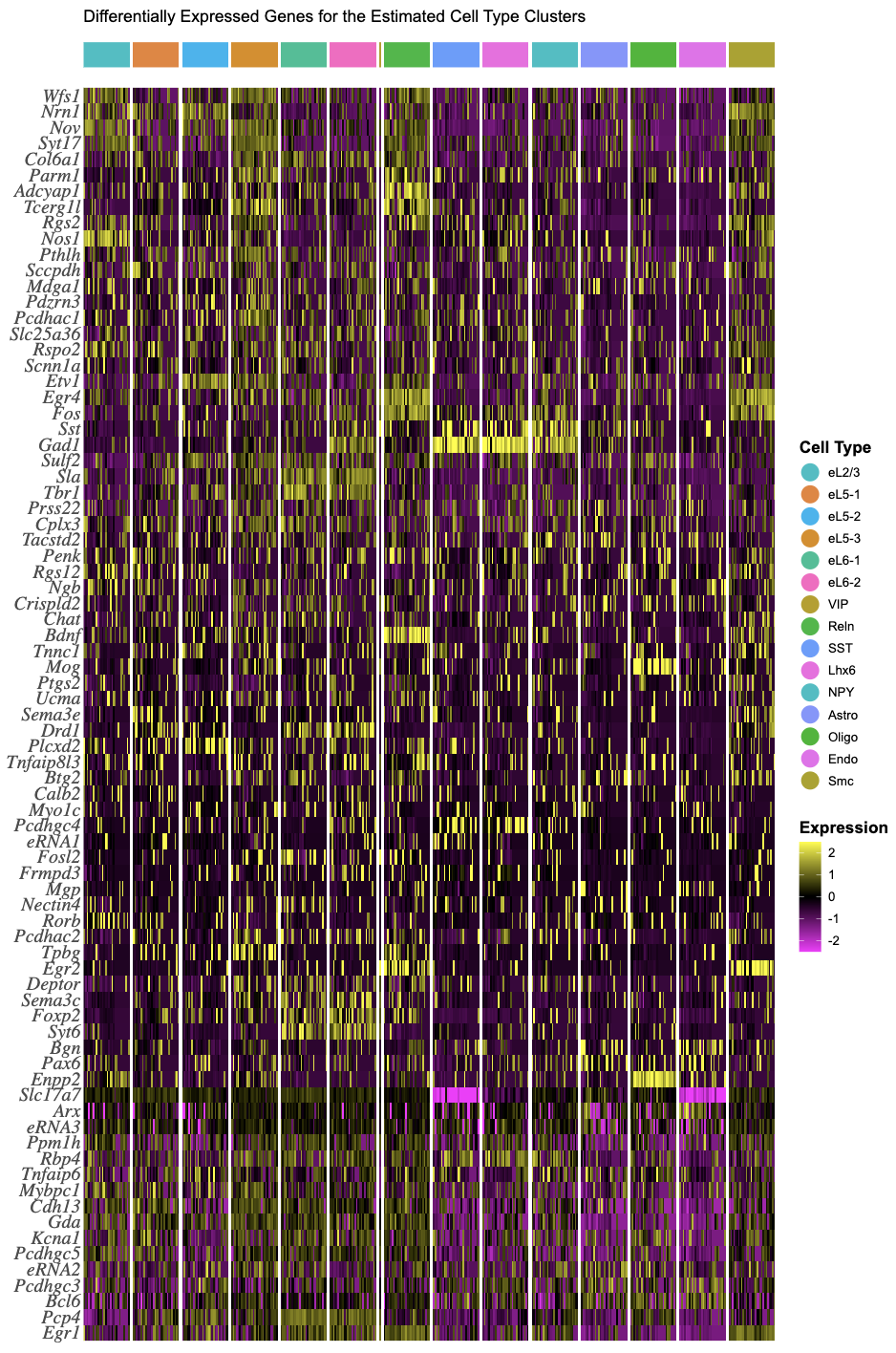}
    \caption{For the mPFC dataset, gene expressions for the subset of 79 genes associated with the top-ranked component (i.e., the component with the minimum combined $p$-value), selected by BayesClint. To create the heatmap, we first hierarchically cluster the genes' log-normalized expression values to group them. The row-wise ordering of the genes in the pair of heatmaps for mPFC are made consistent for the set of active genes and the subset of genes associated with the top-ranked component. We cluster the cells according to the cell type clusters returned from BayesClint. We randomly downsample to 30 cells per cell type, for visualization purposes. Finally, we used doHeatmap() function from the PRECAST R package on the scaled values (to display the relative expression level centered at zero).}
    \label{fig:mpfc_bayesclint_differentiating}
\end{figure}

\clearpage

\section{List of Notation}
\label{sec:list_of_notation}

\begin{itemize}
    \item $\bX_m = (\bx_{m1}, \ldots, \bx_{mn_m})^\intercal \in \mathbb{R}^{n_m \times P}$
    \item $\Upsilon_m = \textrm{diag}(\upsilon^2_{m1}, \ldots, \upsilon^2_{mP})$
    \item $\bU_m = (\bu_{m1}, \ldots, \bu_{mn_m})^\intercal \in \mathbb{R}^{n_m \times r}, m = 1, \ldots, N$: the factors
    \item $\bA \in \mathbb{R}^{r \times P}$: factor loadings matrix
    \item $\brho \in \mathbb{R}^r$: component selection indicator 
    \item $\bGamma \in \mathbb{R}^{r \times P}$: genetic feature selection indicator
    \begin{itemize}
        \item $\bgamma_j$ is the $r$-length vector $(\gamma_{1j}, \ldots, \gamma_{rj})^\intercal$
    \end{itemize}
    \item $\zeta_{mi}$: cell type label
    \item $\kappa_{mi}$: spatial domain label
    \item $\bTheta = (\btheta_1, \ldots, \btheta_K) \in \mathbb{R}^{C \times K}$
    \item $\bmu_{c}$, $r$-length vector $(\mu_{1c}, \ldots, \mu_{rc})^\intercal$: cell type means
    \item $\Sigma \in \mathbb{R}^{r \times r}$: Variance-covariance matrix for the factors $\bu_{mi}$
    \item $\beta_m$: Potts smoothing parameter
    \item $d(B_m, \beta_m)$: Potts normalizing constant, or partition function
    \item $B_m$: neighborhood graph based on $k$-nearest neighbors
    \item $I_r$: Identity matrix of size $r \times r$
    \item $\bI(\zeta_{mi} = c)$: Indicator function that equals 1 if the cell type for cell $i$ in tissue sample $m$ is $c$, and 0 otherwise.
    \item $q_\gamma$: the prior probability of selecting features for a given component.
    \item $t = 1, \ldots, T$: index of the MCMC iterations after burn-in.
    \item $\textrm{PPI}_j$: posterior probability of inclusion for gene $j$ in the set of active genes. 
    \item $\textrm{PPI}_{d,j}$: posterior probability of inclusion for gene $j$ in the set of genes that differentiate across cell types.
    \item $p_{lc}$: Bayesian $p$-value for the null hypothesis that cell type mean $\mu_{lc}$ is zero. 
    \item $\tau$: threshold for inclusion in the set of selected genes, e.g., $\{ \bI(\textrm{PPI}_1 \geq \tau), \ldots, \bI(\textrm{PPI}_P \geq \tau) \}$.
    \item $\tau'$: threshold for inclusion in the set of selected genes which controls the BFDR, e.g., $\{ \bI(1 - \textrm{PPI}_1 < \tau'), \ldots, \bI(1 - \textrm{PPI}_P < \tau') \}$.
\end{itemize}

\bibliographystyle{abbrvnat} 
\bibliography{references}